\documentclass[a4paper,12pt,english]{article}
\usepackage[english]{babel}

\usepackage[T1]{fontenc}          
\usepackage[utf8]{inputenc}       
\usepackage{lmodern}			  
\usepackage[dcucite]{harvard}
\usepackage{mathtools}
\usepackage{amsfonts}
\usepackage{amsmath}
\usepackage{amssymb}
\usepackage{amsthm}
\usepackage{graphics}
\usepackage{bbm}
\usepackage{rotating}
\usepackage{booktabs} 			  
\usepackage{adjustbox}
\usepackage[flushleft]{threeparttable}
\usepackage{float}
\usepackage{subcaption}
\usepackage{color}
\usepackage[toc,page]{appendix}
\usepackage{pdfpages}
\usepackage{xspace}
\usepackage{enumitem}
\usepackage[onehalfspacing]{setspace}
\usepackage[paper=a4paper,top=1in,bottom=1in,left=1in,right=1in]{geometry}
\usepackage{hyperref}

\newcommand{\mR}{\ensuremath{{\mathbb R}}}
\newcommand{\mZ}{\ensuremath{{\mathbb Z}}}

\newcommand{\mN}{\ensuremath{{\mathbb N}}}
\newcommand{\mE}{\ensuremath{{\mathbb E}}}
\newcommand{\mP}{\ensuremath{{\mathbb P}}}

\newcommand{\ve}{\ensuremath{\varepsilon}}
\newcommand{\ie}{\mbox{i.\,e.}\xspace}

\newcommand{\eg}{\mbox{e.\,g.}\xspace}
\newcommand{\Eg}{\mbox{E.\,g.}\xspace}

\newcommand{\floor}[1]{\lfloor #1 \rfloor}

\newtheorem{remark}{Remark}
\newtheorem{assumption}{Assumption}

\newtheorem{proposition}{Proposition}
\newtheorem{lemma}{Lemma}
\newtheorem{theorem}{Theorem}

\usepackage{authblk}
\author[a,b]{Karsten Reichold\thanks{Correspondence to: Karsten Reichold, Department of Statistics, TU Dortmund University, Vogelpothsweg 87, D-44227 Dortmund, Germany.\\ E-mail addresses: \href{mailto:reichold@statistik.tu-dortmund.de}{reichold@statistik.tu-dortmund.de}, \href{mailto:jentsch@statistik.tu-dortmund.de}{jentsch@statistik.tu-dortmund.de}\\ \texttt{MATLAB} code for empirical applications is available on \href{https://github.com/kreichold/CointSelfNorm}{www.github.com/kreichold/CointSelfNorm}.}}

\author[a]{Carsten Jentsch}

\affil[a]{\small{Department of Statistics, TU Dortmund University, 44221, Dortmund, Germany}}

\affil[b]{\small{Department of Economics, University of Klagenfurt, 9020, Klagenfurt, Austria}}

\title{A Bootstrap-Assisted Self-Normalization Approach to Inference in Cointegrating Regressions}

\date{\today}

\begin{document}
	
\numberwithin{equation}{section}
	
\maketitle

\begin{abstract} 
	 \noindent {\footnotesize Traditional inference in cointegrating regressions requires tuning parameter choices to estimate a long-run variance parameter. Even in case these choices are ``optimal'', the tests are severely size distorted. We propose a novel self-normalization approach, which leads to a nuisance parameter free limiting distribution without estimating the long-run variance parameter directly. This makes our self-normalized test tuning parameter free and considerably less prone to size distortions at the cost of only small power losses. In combination with an asymptotically justified vector autoregressive sieve bootstrap to construct critical values, the self-normalization approach shows further improvement in small to medium samples when the level of error serial correlation or regressor endogeneity is large. We illustrate the usefulness of the bootstrap-assisted self-normalized test in empirical applications by analyzing the validity of the Fisher effect in Germany and the United States.}
	
	
	
\end{abstract}

\newpage

\section{Introduction}\label{sec:Int}
Tuning parameter choices complicate statistical inference in cointegrating regressions and affect finite sample distributions of test statistics. Even in case these choices are ``optimal'', hypothesis tests are often severely size distorted. In this paper we address this issue by proposing a novel self-normalized test statistic for general linear hypotheses, that itself is completely tuning parameter free.\footnote{The term ``self-normalization'' was coined in \citeasnoun{Sh10}, who extends the method of \citeasnoun{Lo01} to inference for a general parameter in a stationary time series setting. The distinctive feature of ``self-normalization'' is that it leads to asymptotically pivotal test statistics without requiring the applied researcher to make tuning parameter choices.} Its limiting distribution is non-standard but simulating asymptotically valid critical values is straightforward. In addition, we provide a resampling procedure to construct bootstrap critical values and justify their use by proving a bootstrap invariance principle that may be of independent interest.

Cointegration methods have been and are widely used to analyze long-run relationships between stochastically trending variables in many areas such as macroeconomics, environmental economics and finance, see, \eg, \citeasnoun{BLNW20}, \citeasnoun{Wa15} and \citeasnoun{RLF16} for recent examples. In addition to these classical fields of application, cointegration methods have recently proven to be useful to describe phenomena also in other contexts. For instance, \citeasnoun{DKN18} describe the close connection between cointegration and the theory of phase synchronization in physics and \citeasnoun{PLS20} apply cointegration-based methods to estimate Earth's climate sensitivity.

Although the OLS estimator is consistent in cointegrating regressions, its limiting distribution is usually contaminated by second order bias terms, reflecting the correlation structure between regressors and errors. This makes the OLS estimator infeasible for conducting inference based on (simulated) quantile tables of (non)standard distributions. The literature provides several estimators which overcome this difficulty at the cost of tuning parameter choices: the number of leads and lags for the dynamic OLS (D-OLS) estimator of \citeasnoun{PhLo91}, \citeasnoun{Sa91} and \citeasnoun{StWa93}, kernel and bandwidth choices for the fully modified OLS (FM-OLS) estimator of \citeasnoun{PhHa90} and the canonical cointegrating regression (CCR) estimator of \citeasnoun{Pa92}, or type and number of basis functions for the trend instrument variable (TIV) estimator of \citeasnoun{Ph14}. In addition, traditional hypothesis tests based on these estimators require kernel and bandwidth choices to estimate a long-run variance parameter. Such tuning parameters are often difficult to choose in practice and the finite sample performance of the estimators and tests based upon them often reacts sensitively to their choices. In particular, corresponding tests often suffer from severe size distortions.

In contrast to the aforementioned approaches, the integrated modified OLS (IM-OLS) estimator of \citeasnoun{VoWa14} avoids the choice of tuning parameters. However, standard asymptotic inference based on the IM-OLS estimator still involves estimating a long-run variance parameter. To capture the effects of the required kernel and bandwidth choices in finite samples, \citeasnoun{VoWa14} propose fixed-$b$ theory for obtaining critical values. However, their simulation results reveal that when endogeneity and/or error serial correlation is strong, a large sample size is needed for the procedure to yield reasonable sizes. Moreover, for small to medium samples, test performance still seems to be sensitive to the choice of $b$. Similarly, \citeasnoun{HwSu18} develop ``partial'' fixed-$K$ theory for the TIV estimator, which captures the choice of the number of basis functions but ignores the impact of the basis functions itself.

Instead of trying to capture finite sample effects of tuning parameter choices, we propose a novel IM-OLS based test statistic for general linear hypotheses, which is completely tuning parameter free. The test statistic is based on a self-normalization approach that is similar in spirit to but different from the approach of \citeasnoun{KVB00} for stationary data. The limiting null distribution of the self-normalized test statistic is nonstandard but simulating asymptotic critical values is straightforward. The idea of inference based on self-normalized test statistics has been proven to be widely applicable in stationary time series analysis, see \citeasnoun{Sh15} for a detailed review.\footnote{Self-normalization has also been applied to testing for a unit root in univariate time series, monitoring cointegrating relationships and the analysis of high-dimensional stationary time series, see \citeasnoun{Br02}, \citeasnoun{KWG21} and \citeasnoun{WaSh20}, respectively.}

To further improve the performance of the self-normalized test in small to medium samples, we combine the self-normalization approach with a residual-based vector autoregressive (VAR) sieve resampling procedure to construct critical values.\footnote{The VAR sieve bootstrap is frequently used in related literature: \citeasnoun{Ps01}, inspired by the seminal work of \citeasnoun{LiMa97}, shows the usefulness of the sieve bootstrap in cointegrating regressions and \citeasnoun{Pa02} provides its asymptotic justification by proving an underlying invariance principle result. Subsequently, \citeasnoun{ChPa03} and \citeasnoun{CPS06} apply the sieve bootstrap to unit root testing and to conduct D-OLS based inference in cointegrating regressions, respectively. Note, however, that the approach of \citeasnoun{CPS06} still requires the choice of leads and lags for estimation and additional kernel and bandwidth choices for inference. Moreover, \citeasnoun{PSU10} use the sieve bootstrap in the context of testing for cointegration in conditional error correction models.} The VAR sieve bootstrap captures the second-order dependence structure of the original process in a simple manner. In particular, it only requires the selection of the order of the VAR, which is a straightforward and well understood task in practice.

To show consistency of the VAR sieve bootstrap in our context, we establish a bootstrap invariance principle result under relatively general conditions that may be of independent interest. Our framework allows for so-called weak white noises that are uncorrelated, but not necessarily independent, and also for various concepts to quantify such weak forms of dependence of the innovation process. In particular, we do not impose the assumption of a causal linear process with i.i.d.~innovations as used in, \eg, \citeasnoun{Pa02}.

The theoretical analysis is complemented by a simulation study to assess the performance of the proposed methods benchmarked against competing approaches, including the traditional D-OLS and FM-OLS based Wald-type tests. The main result is that the traditional tests show severe size distortions, whereas our novel approach, which combines the concepts of self normalization and bootstrapping, proves to hold the prescribed level approximately at the expense of only small power losses. Given that large size distortions of hypothesis tests are the rule rather than the exception in the cointegrating literature, the small power losses may be accepted by the applied researcher.

We illustrate the usefulness of the bootstrap-assisted self-normalized test in empirical applications by analyzing the validity of the Fisher effect in Germany and the United States. The Fisher effect is backed by several theoretical models, but many empirical studies reject the hypothesis that inflation and the short-term nominal interest rate cointegrate with the slope of inflation being equal to one. The errors in this regression are likely to be highly persistent even in case cointegration between the two variables prevails. As a consequence, the Fisher effect might be rejected solely due to the adverse effects of highly persistent errors on the performance of the estimators and tests chosen by the applied researcher, see, \eg, \citeasnoun{CaPi04} and \citeasnoun{We08} for a detailed discussion. The bootstrap-assisted self-normalized test remedies these shortcomings, as it leads to reliable inference even in the presence of highly persistent errors.

The rest of the paper is organized as follows: Section~\ref{sec:Model} introduces the model and its underlying assumptions. Section~\ref{sec:Tests} constructs the self-normalized test statistic and derives its limiting null distribution. Section~\ref{sec:Boot} presents the bootstrap procedure and derives its asymptotic validity. Section~\ref{sec:FiniteSample} assesses the finite sample performance of the proposed methods and Section~\ref{sec:Illustration} illustrates the usefulness of the bootstrap-assisted self-normalized test in empirical applications. Section~\ref{sec:Conclusion} concludes. The Online Appendix contains additional theoretical and finite sample results, tabulated critical values and proofs.

We use the following notation. The integer part of a real number $x$ is denoted by $\floor{x}$. For a real matrix $A$ we denote its transpose by $A'$ and its Frobenius norm by $\vert A \vert_F = (\text{tr}(A'A))^{1/2}$, where $\text{tr}(\cdot)$ denotes the trace and for a vector the Frobenius norm becomes the Euclidean norm. The $k$-dimensional identity matrix is denoted by $I_k$ and $0_{j\times k}$ (or simply $0$) denotes a ($j\times k$)-dimensional matrix of zeros. With $\text{diag}(\cdot)$ we denote a (block) diagonal matrix with diagonal elements specified throughout. Equality in distribution is signified by $\overset{d}{=}$. With $\overset{w}{\longrightarrow}$ and $\overset{p}{\longrightarrow}$ we denote weak convergence and convergence in probability, respectively. Adding the superscript ``$*$'' signifies convergence in the bootstrap probability space. The corresponding probability measure is denoted by $\mP^*$ and $\mE^*(\cdot)$ denotes the expectation with respect to $\mP^*$. For notational simplicity, a Brownian motion $\{M(r),\ 0\leq r \leq 1\}$ is denoted by $M(r)$.

\section{The Model and Assumptions}\label{sec:Model}
We consider the cointegrating regression model
\begin{align}
y_t &= x_t'\beta + u_t,\label{eq:y}\\
x_t &= x_{t-1} + v_t, \label{eq:x}
\end{align}
$t=1,\ldots,T$, where $(y_t)_{t=1,\ldots,T}$ is a scalar time series and $(x_t)_{t=1,\ldots,T}$ is an $m\times 1$ vector of time series.\footnote{We exclude deterministic regressors from~\eqref{eq:y} to ease exposition of the main arguments. However, it is straightforward to incorporate, \eg, the leading case of an intercept and polynomial time trends, $d_t = [1,t,\ldots,t^p]'$, $p\geq 1$. Please note that the accompanying \texttt{MATLAB} code allows to handle this more general case.} For brevity we set $x_0=0$. For $\{w_t\}_{t\in\mZ} \coloneqq \{[u_t,v_t']'\}_{t\in \mZ}$ we assume the following:
\begin{assumption}\label{ass:w}
	Let $\{w_t\}_{t \in \mZ}$ be an $\mR^{1+m}$-valued, strictly stationary and purely nondeterministic stochastic process of full rank\footnote{The process $\{w_t\}_{t \in \mZ}$ is of full rank, if the components of its innovation process are linearly independent. For more details we refer to \citeasnoun[p.\,379]{MeKr15}, who study the range of validity of the VAR sieve bootstrap.} with $\mE(w_t)=0$ and\linebreak $\mE(\vert w_t \vert_F^a)<\infty$, for some $a>2$. The autocovariance matrix function $\Gamma(\cdot)$ of $\{w_t\}_{t \in \mZ}$ fulfills $\sum_{h=-\infty}^\infty (1+\vert h \vert)^k \vert \Gamma (h)\vert_F < \infty$ for some $k\geq 3/2$. 
	For the spectral density matrix $f(\cdot)$ of $\{w_t\}_{t \in \mZ}$ we assume that there exists a constant $c>0$ such that $\min \sigma(f(\lambda))\geq c$ for all frequencies $\lambda \in (-\pi,\pi]$, where $\sigma(f(\lambda))$ denotes the spectrum of $f(\cdot)$ at frequency $\lambda$.
\end{assumption}
The short memory condition $\sum_{h=-\infty}^\infty (1+\vert h \vert)^k \vert \Gamma (h)\vert_F < \infty$ for some $k\geq 3/2$ in Assumption~\ref{ass:w} implies a continuously differentiable spectral density $f$, which is particularly bounded from below and from above, uniformly for all frequencies $\lambda \in (-\pi,\pi]$. As shown in \citeasnoun{MeKr15}, a process fulfilling Assumption~\ref{ass:w} does always possess the one-sided representations
\begin{align}
\Phi(L)w_t &= \ve_t, \label{eq:inv}\\
w_t &= \Psi(L)\ve_t, \label{eq:caus}
\end{align}
where $\{\ve_t\}_{t\in \mZ}$ is a strictly stationary uncorrelated -- but not necessarily independent -- white noise process with positive definite covariance matrix $\Sigma$, $\Phi(z)\coloneqq I_{m+1}-\sum_{j=1}^\infty \Phi_j z^j$ and $\Psi(z)\coloneqq I_{m+1}+\sum_{j=1}^\infty \Psi_j z^j$, with $\sum_{j=1}^\infty (1+j)^k \vert \Phi_j\vert_F < \infty$ and $\sum_{j=1}^\infty (1+j)^k \vert \Psi_j\vert_F < \infty$ for the $k\geq 3/2$ from Assumption~\ref{ass:w}. Moreover, it holds that $\text{det}(\Phi(z)) \neq 0$ and $\text{det}(\Psi(z)) \neq 0$ for all $\vert z \vert \leq 1$.
\begin{assumption}\label{ass:cumulants}
	The process $\{w_t\}_{t \in \mZ}$ has absolutely summable cumulants up to order four. More precisely, we have for all $j=2,\ldots,4$ and $\mathbf{a}=[a_1,\ldots,a_j]'$, with $a_1,\ldots,a_j\in\{1,\ldots,m+1\}$, that
	\begin{align*}
		\sum_{h_2,\ldots,h_j=-\infty}^\infty \vert \text{cum}_{\mathbf{a}}(0,h_2,\ldots,h_j)\vert <\infty,	
	\end{align*}
	where $\text{cum}_{\mathbf{a}}(0,h_2,\ldots,h_j)$ denotes the $j$-th joint cumulant of $w_{0,a_1},w_{h_2,a_2},\ldots,w_{h_j,a_j}$ and $w_{t,i}$ denotes the $i$-th element of $w_t$.
\end{assumption}
Let $\Omega$ denote the long-run covariance matrix of $\{w_t\}_{t\in\mZ}$, \ie, 
\begin{align*}
\Omega =
\begin{bmatrix}
\Omega_{uu}&\Omega_{uv}\\
\Omega_{vu}&\Omega_{vv}
\end{bmatrix} = 2\pi f(0) = \sum_{h=-\infty}^\infty\Gamma(h) =\Psi(1)\Sigma \Psi(1)'. 
\end{align*}
From $\Sigma>0$ and $\text{det}(\Psi(1)) \neq 0$ it follows that $\Omega>0$. In particular, positive definiteness of $\Omega_{vv}$ rules out cointegration among the elements of $\{x_t\}_{t\in\mZ}$. As typical in the cointegration literature, we assume that $\{w_t\}_{t \in \mZ}$ fulfills an invariance principle.
\begin{assumption}\label{ass:FCLT}
	Let $\{w_t\}_{t \in \mZ}$ fulfill
	\begin{align}\label{eq:FCLT}
	B_T(r)\coloneqq  T^{-1/2} \sum_{t=1}^{\floor{rT}} w_t \overset{w}{\longrightarrow} B(r) = \Omega^{1/2} W(r),\quad 0\leq r \leq 1,
	\end{align}
	as $T\rightarrow \infty$, where $W(r)=[W_{u\cdot v}(r),W_v(r)']'$ is an $(1+m)$-dimensional vector of independent standard Brownian motions and 
	\begin{align*}
		\Omega^{1/2} =\begin{bmatrix}
		\Omega_{u \cdot v}^{1/2} & \Omega_{uv}(\Omega_{vv}^{-1/2})'\\
		0 & \Omega_{vv}^{1/2}
		\end{bmatrix},
	\end{align*}
	where $\Omega_{u\cdot v} \coloneqq  \Omega_{uu} - \Omega_{uv} \Omega_{vv}^{-1}\Omega_{vu}$, such that $\Omega^{1/2}(\Omega^{1/2})'=\Omega$. For later usage note that $\Omega_{u\cdot v}$ is a scalar and partition $B(r)=[B_u(r),B_v(r)']'$.
\end{assumption}

We emphasize that Assumption~\ref{ass:w} does explicitly not ask for invertibility or causality of the process $\{w_t\}_{t\in\mZ}$ with respect to an \emph{independent} white noise process, in contrast to the assumptions in related literature, compare, \eg, \citeasnoun{Pa02}, \citeasnoun{CPS06} and \citeasnoun{PSU10}. In particular, in this paper, the innovation process $\{\ve_t\}_{t\in \mZ}$ resulting from the representations in \eqref{eq:inv} and \eqref{eq:caus} will generally be uncorrelated but not necessarily independent. Assumption~\ref{ass:cumulants} is of technical nature and satisfied if, \eg, $\{w_t\}_{t \in \mZ}$ is $\alpha$-mixing with strong-mixing coefficients $\alpha(j)$ such that $E(\vert w_t\vert _F^{4+\delta})<\infty$ and $\sum_{j=1}^\infty j^2\alpha(j)^{\delta/(4+\delta)}<\infty$ for some $\delta>0$, see, \eg, \citeasnoun[p.\,221]{Sh10}. In particular, Assumption~\ref{ass:cumulants} requires the existence of fourth moments of $\{w_t\}_{t \in \mZ}$. To establish meaningful asymptotic theory, Assumptions~\ref{ass:w} and~\ref{ass:cumulants} have to be complemented by an invariance principle in Assumption~\ref{ass:FCLT}. This general formulation of an invariance principle allows for various concepts of choice to quantify weak forms of dependence of the innovation process $\{w_t\}_{t\in\mZ}$. These include classical approaches sufficing to prove invariance principles, \eg, several variants of mixing properties, mixingale-type sequences, linear processes including all-pass filters and their multivariate extensions, or (Bernoulli) shift processes, see \citeasnoun{MPU06} for an overview.\footnote{All-pass filters, discussed for univariate times series in, \eg, \citeasnoun{ADB07}, lead to uncorrelated, but dependent white noise processes. \citeasnoun{LaSa13} propose their multivariate extensions based on non-causal and non-invertible vector-valued time series models.} In addition, the general formulation also allows for more modern approaches that cover the general notion of weakly dependent stationary time series discussed in \citeasnoun{DoWi07} or physical dependence proposed by \citeasnoun{Wu05} and employed in \citeasnoun{Wu07} to prove (strong) invariance principles. Finally, note that Assumption~\ref{ass:FCLT} suffices to derive the limiting distribution of the self-normalized test statistic defined in Section~\ref{sec:Tests}. Assumptions~\ref{ass:w} and~\ref{ass:cumulants} are required only to show bootstrap consistency in Section~\ref{sec:Boot}.

\section{Testing General Linear Hypotheses}\label{sec:Tests}
In cointegrating regressions the OLS estimator is consistent, but its limiting distribution is contaminated by second order bias terms. The bias terms reflect the correlation structure between the regressors and the errors and make the OLS estimator unsuitable for conducting asymptotic inference using (simulated) quantile tables of (non)standard distributions. The literature provides several modified estimators that allow for standard asymptotic inference, compare the discussion in the introduction. For our purposes, we choose the tuning parameter free and easy to implement IM-OLS approach of \citeasnoun{VoWa14}.

\subsection{The IM-OLS Estimator Revisited} 
\citeasnoun{VoWa14} propose to compute the partial sum of both sides of~\eqref{eq:y}, then to add $x_t$ as a regressor to the partial sum regression and finally to estimate the regression coefficients by OLS.\footnote{Adding $x_t$ to the partial sums regression serves as an endogeneity correction, which is similar to the leads and lags augmentation in D-OLS estimation. Although similar in spirit, it is considerably simpler as it avoids choosing the numbers of leads and lags.} That is, by computing the OLS estimator in the augmented partial sums regression 
\begin{align}\label{eq:Sy}
S_t^y =S_t^{x\prime}\beta + x_t'\gamma + S_t^u = Z_t'\theta + S_t^u,
\end{align}
with $S_t^y \coloneqq  \sum_{s=1}^t y_s$, $S_t^x \coloneqq  \sum_{s=1}^t x_s$, $S_t^u \coloneqq  \sum_{s=1}^t u_s$ and the $2m$-dimensional vector $Z_t\coloneqq [S_t^{x\prime},x_t']'$, the IM-OLS estimator $\hat\theta_{\scriptstyle\text{IM}}\coloneqq [\hat\beta_{\scriptstyle\text{IM}}',\hat\gamma_{\scriptstyle\text{IM}}']'$ for $\theta\coloneqq [\beta',\gamma']'$ in~\eqref{eq:Sy} is obtained. As shown in \citeasnoun[Theorem~2]{VoWa14} it holds under Assumption~\ref{ass:FCLT} that the limiting distribution of $\hat\theta_{\scriptstyle\text{IM}}$ is given by
\begin{align}\label{eq:IM}
\begin{bmatrix}
T\left(\hat\beta_{\scriptstyle\text{IM}} - \beta\right)\\
\hat\gamma_{\scriptstyle\text{IM}} - \Omega_{vv}^{-1}\Omega_{vu}
\end{bmatrix}
\overset{w}{\longrightarrow}\Omega_{u \cdot v}^{1/2} \left(\Pi'\right)^{-1}\mathcal{Z},
\end{align}
as $T\rightarrow \infty$, where $\Pi\coloneqq \text{diag}\left(\Omega_{vv}^{1/2},\Omega_{vv}^{1/2}\right)$ and 
\begin{align}\label{eq:Z}
\mathcal{Z}\coloneqq \left(\int_0^1 g(r) g(r)'dr\right)^{-1}\int_0^1\left[G(1)-G(r)\right]dW_{u\cdot v}(r),
\end{align}
with $g(r)\coloneqq [\int_0^r W_v(s)'ds, W_v(r)']'$ and $G(r)\coloneqq \int_0^r g(s)ds$.\footnote{Since both $x_t$ and $S_t^u$ are I(1) processes, all correlation -- between $B_v(r)$ and $B_u(r)$ -- is soaked up in the long-run population regression vector $\Omega_{vv}^{-1}\Omega_{vu}$. Therefore, the correct centering parameter for $\hat\gamma_{\scriptstyle\text{IM}}$ in case of endogeneity is $\Omega_{vv}^{-1}\Omega_{vu}$ rather than the population value $\gamma = 0$. For more details see \citeasnoun[p.\,746]{VoWa14}.} Conditional upon $W_v(r)$, the asymptotic distribution in~\eqref{eq:IM} is normal with zero-mean and covariance matrix $\Omega_{u\cdot v}V$, where
\begin{align}\label{eq:V}
V &\coloneqq \left(\Pi'\right)^{-1}\left(\int_0^1 g(r)g(r)'dr\right)^{-1}\left(\int_0^1\left[G(1)-G(r)\right]\left[G(1)-G(r)\right]'dr\right)\notag\\
&\ \times \left(\int_0^1 g(r)g(r)'dr\right)^{-1}\Pi^{-1}\\
&= \left(\Pi'\right)^{-1}\tilde{V}\Pi^{-1},
\end{align} 
with $\tilde V$ implicitly defined by the last equality. Let $A_T \coloneqq  \text{diag}\left(T^{-1}I_m,I_m\right)$ and define
\begin{align}\label{eq:Vhat}
\hat V_T &\coloneqq  \left(\sum_{t=1}^T Z_t Z_t'\right)^{-1}\left(\sum_{t=1}^T c_t c_t'\right)\left(\sum_{t=1}^T Z_t Z_t'\right)^{-1},
\end{align}
where $c_1 \coloneqq  S_T^{Z}$ and $c_t\coloneqq  S_T^{Z} - S_{t-1}^Z$, with $S_{t}^{Z}\coloneqq \sum_{j=1}^t Z_j$, for $t=2,\ldots,T$. Then,
\begin{align}\label{eq:VIM}
A_T^{-1}\hat V_TA_T^{-1}
\overset{w}{\longrightarrow} V,
\end{align}
as $T\rightarrow \infty$, compare \citeasnoun[Proof of Theorem 3]{VoWa14}.

\begin{remark}
	The IM-OLS estimator is rate-$T$ consistent, but its conditional asymptotic covariance matrix is always larger than (or equal to) the conditional asymptotic covariance matrix of the rate-$T$ consistent D- and FM-OLS estimators \citeaffixed[Proposition~2]{VoWa14}{see}. This stems from the fact that the error in the augmented partial sum regression is integrated rather than stationary. However, this comparison may be misleading, as the asymptotic covariance matrix of the D- and FM-OLS estimators does not account for the finite sample effects of tuning parameter choices required for the D- and FM-OLS estimators but not for the IM-OLS estimator. Simulation results in \citeasnoun{VoWa14} and in  Section~\ref{app:FiniteSampleEstimation} in Online Appendix~\ref{app:finitesample} demonstrate that the IM-OLS estimator performs well relative to the two competitors. The IM-OLS estimator thus serves as a useful starting point for developing a self-normalized test statistic.
\end{remark}

\subsection{The Self-Normalized Test Statistic}
The zero mean Gaussian mixture limiting distribution of the IM-OLS estimator in conjunction with \eqref{eq:VIM} forms the basis for standard asymptotic inference based on the traditional Wald-type hypothesis test. To be more precise, for testing $s\leq m$ linearly independent restrictions on $\beta\in \mR^m$ in~\eqref{eq:y}, we consider the hypotheses
\begin{align}\label{eq:H0}
\text{H}_0:R_1\beta=r_0 \quad \text{versus} \quad \text{H}_1:R_1\beta\neq r_0,
\end{align}
where $R_1 \in\mR^{s\times m}$ has full row rank $s$ and $r_0\in \mR^s$. For deriving the limiting null distribution of the corresponding Wald-type test statistic, it is convenient to rewrite the null hypothesis in terms of the correct centering parameter for $\hat\theta_{\scriptstyle\text{IM}}$, given by $[\beta',\left(\Omega_{vv}^{-1}\Omega_{vu}\right)']'$. To this end, we define $R_2 \coloneqq  [R_1,0_{s\times m}]\in\mR^{s\times 2m}$ such that the null hypothesis in~\eqref{eq:H0} reads as $R_1\beta = R_2 [\beta',\left(\Omega_{vv}^{-1}\Omega_{vu}\right)']'=r_0$.\footnote{Note that the auxiliary coefficient vector $\gamma$ is not restricted under the null hypothesis and, in particular, $\Omega_{vv}^{-1}\Omega_{vu}$ does not have to be estimated.} 

Under Assumption~\ref{ass:FCLT} it follows from \citeasnoun[Theorem~3]{VoWa14} that the limiting distribution of the Wald-type test statistic
\begin{align*}
\tau_{\scriptstyle\text{IM}} \coloneqq  \left(R_2 \hat\theta_{\scriptstyle\text{IM}} - r_0\right)'\left[R_2 \hat V_T R_2'\right]^{-1}\left(R_2 \hat\theta_{\scriptstyle\text{IM}} - r_0\right)
\end{align*}
converges under the null hypothesis in distribution to
\begin{align}\label{eq:G1}
\mathcal{G}_{\Omega} \coloneqq  \left(R_2\Omega_{u\cdot v}^{1/2} \left(\Pi'\right)^{-1}\mathcal{Z}\right)'\left(R_2 V R_2'\right)^{-1}\left(R_2\Omega_{u\cdot v}^{1/2} \left(\Pi'\right)^{-1}\mathcal{Z}\right)\overset{d}{=} \Omega_{u\cdot v} \chi_s^2,
\end{align}
as $T\rightarrow \infty$, where $\mathcal{Z}$ is defined in~\eqref{eq:Z} and $\chi_s^2$ denotes a chi-square distribution with $s$ degrees of freedom.\footnote{In practical applications it might be more convenient to express this -- and the following -- test statistic(s) in terms of $\hat\beta_{\scriptstyle\text{IM}}$ and $R_1$, only. This can be achieved by noting that $R_2\hat\theta_{\scriptstyle\text{IM}} = R_1 \hat\beta_{\scriptstyle\text{IM}}$ and $R_2 \hat V_T R_2'= R_1 \hat V_T(1,1)R_1'$, where $\hat V_T(1,1)$ denotes the upper left $(m\times m)$-dimensional block of the $(2m\times 2m)$-dimensional matrix $\hat V_T$.} The limiting null distribution of $\tau_{\scriptstyle\text{IM}}$ is contaminated by a nuisance parameter, $\Omega_{u\cdot v}$.\footnote{As pointed out in \citeasnoun[Remark~4.6(d)]{Ph95}, $\Omega_{u\cdot v}$ is the long-run variance of the regression error $u_t$ corrected for its conditional long-run mean given $v_t$. The dependence of the limiting distribution on $\Omega_{u\cdot v}$ rather than on the long-run variance $\Omega_{uu}$ of the regression errors $u_t$ follows from the endogeneity correction within the IM-OLS procedure.} The presence of the long-run variance parameter makes the limiting distribution highly case dependent and thus infeasible for inference based on tabulated critical values.\footnote{Analogous results also hold for the Wald-type tests based on the D-OLS, FM-OLS and CCR estimators.} To remedy this problem, the literature suggests to plug-in a consistent estimator of $\Omega_{u\cdot v}$, $\hat\Omega_{u\cdot v}$ say, such that the nuisance parameter is scaled out in the limit. For this purpose, we denote
\begin{align}\label{eq:T_k}
\tau_{\scriptstyle\text{IM}}(\kappa) \coloneqq  \left(R_2 \hat\theta_{\scriptstyle\text{IM}} - r_0\right)'\left[R_2 \kappa\hat V_T R_2'\right]^{-1}\left(R_2 \hat\theta_{\scriptstyle\text{IM}} - r_0\right),
\end{align}
with $\kappa$ denoting a one-dimensional quantity. For $\kappa=1$ we thus have $\tau_{\scriptstyle\text{IM}}(1)=\tau_{\scriptstyle\text{IM}}$. Hence, the literature typically considers Wald-type test statistics of the form
\begin{align}\label{eq:TOmega}
\tau_{\scriptstyle\text{IM}}(\hat\Omega_{u\cdot v}) = \left(R_2 \hat\theta_{\scriptstyle\text{IM}} - r_0\right)'\left[R_2 \hat\Omega_{u\cdot v}\hat V_T R_2'\right]^{-1}\left(R_2 \hat\theta_{\scriptstyle\text{IM}} - r_0\right),
\end{align}
which converge under the null hypothesis in distribution to
\begin{align*}
\mathcal{G}&\coloneqq  \left(R_2\left(\Pi'\right)^{-1}\mathcal{Z}\right)'\left(R_2 V R_2'\right)^{-1}\left(R_2\left(\Pi'\right)^{-1}\mathcal{Z}\right) \overset{d}{=} \chi_s^2,
\end{align*}
as $T\rightarrow \infty$. As $\mathcal{G}$ is nuisance parameter free, it allows for standard asymptotic inference based on tabulated critical values. However, estimation of $\Omega_{u\cdot v}$ is cumbersome, as it is typically based on non-parametric kernel estimators of the form
\begin{align}\label{eq:hatOmegaudotv}
	\hat\Omega_{u\cdot v}=\hat\Omega_{u\cdot v}\left(\mathcal{K},b_T\right)\coloneqq\hat\Omega_{uu}\left(\mathcal{K},b_T\right)-\hat\Omega_{uv}\left(\mathcal{K},b_T\right)\hat\Omega_{vv}^{-1}\left(\mathcal{K},b_T\right)\hat\Omega_{vu}\left(\mathcal{K},b_T\right),
\end{align}
where
\begin{align}\label{eq:hatOmega}
	\begin{bmatrix}
		\hat\Omega_{uu}\left(\mathcal{K},b_T\right)&\hat\Omega_{uv}\left(\mathcal{K},b_T\right)\\
		\hat\Omega_{vu}\left(\mathcal{K},b_T\right)&\hat\Omega_{vv}\left(\mathcal{K},b_T\right)
	\end{bmatrix} &\coloneqq T^{-1} \sum_{i=1}^T \sum_{j=1}^T\mathcal{K}\left(\frac{\vert i-j\vert}{b_T}\right) \begin{bmatrix}
	\hat u_i^{\text{\tiny OLS}}\\
	v_i
\end{bmatrix}\begin{bmatrix}
\hat u_j^{\text{\tiny OLS}}\\
v_j
\end{bmatrix}',
\end{align}
with $\hat u_t^{\text{\tiny OLS}}$ and $v_t$ denoting the OLS residuals in~\eqref{eq:y} and the first differences of $x_t$, respectively. The kernel function $\mathcal{K}\left(\cdot\right)$ and the bandwidth parameter $b_T$ have to fulfill some common technical assumptions to ensure consistency of $\hat \Omega_{u \cdot v}$, see, \eg, \citeasnoun{An91}, \citeasnoun{NeWe94} and \citeasnoun{Ja02} for details. As these kernel and bandwidth choices affect the finite sample distribution of $\tau_{\scriptstyle\text{IM}}(\hat\Omega_{u\cdot v})$, but are completely ignored in the conventional asymptotic framework, corresponding tests are usually prone to large size distortions. This is in particular the case, when the level of endogeneity and/or error serial correlation is large or the sample size is small.\footnote{\citeasnoun{VoWa14} suggest to use fixed-$b$ critical values to capture the effects of tuning parameter choices. However, their simulation results reveal that in small to medium samples the observed test performance turns out to be sensitive to the choice of $b$, and worsens as endogeneity and/or error serial correlation increases.}

To avoid \emph{any} tuning parameter choices, we propose a novel test statistic based on self-normalization. Instead of plugging-in a consistent estimator $\hat\Omega_{u\cdot v}$ of $\Omega_{u\cdot v}$ in \eqref{eq:T_k}, we insert a quantity that is asymptotically proportional to $\Omega_{u\cdot v}$ but does not rely on any tuning parameters and can be directly computed from the data. To this end, we define the OLS residuals in the augmented partial sum regression given in~\eqref{eq:Sy} as $\hat{S}_t^u \coloneqq  S_t^y - Z_t'\hat\theta_{\scriptstyle\text{IM}}$, $t=1,\ldots,T$. For $t=2,\ldots,T$ let $\Delta\hat{S}_t^u\coloneqq \hat{S}_t^u - \hat{S}_{t-1}^u$ and define the \emph{self-normalizer} as
\begin{align}\label{self_normalizer}
\hat \eta_T \coloneqq  T^{-2} \sum_{t=2}^T \left( \sum_{s=2}^t \Delta\hat{S}_s^u \right)^2.
\end{align}
The proof of Theorem~\ref{thm:SN} below reveals that under the null hypothesis the self-normalizer converges weakly to $\Omega_{u \cdot v} \int_0^1 \left(W_{u\cdot v}(r) - g(r)'\mathcal{Z}\right)^2 dr$, \ie, its limiting distribution is scale dependent on $\Omega_{u \cdot v}$. Choosing $\kappa=\hat \eta_T$ thus removes the nuisance parameter $\Omega_{u \cdot v}$ asymptotically, without estimating it directly. Therefore, we introduce our \emph{self-normalized test statistic} as
\begin{align}\label{self_normalized_test}
\tau_{\scriptstyle\text{IM}}(\hat \eta_T) = \left(R_2 \hat\theta_{\scriptstyle\text{IM}} - r_0\right)'\left[R_2 \hat \eta_T \hat V_T R_2'\right]^{-1}\left(R_2 \hat\theta_{\scriptstyle\text{IM}} - r_0\right).
\end{align}
Its limiting null distribution is given in the following theorem.
\begin{theorem}\label{thm:SN}
	Let $(y_t)_{t=1}^T$ and $(x_t)_{t=1}^T$ be generated by~\eqref{eq:y} and~\eqref{eq:x}, respectively and let $\{w_t\}_{t\in \mZ}$ satisfy Assumption~\ref{ass:FCLT}. Then it holds under the null hypothesis given in~\eqref{eq:H0} that
	\begin{align}\label{eq:Geta}
	\tau_{\scriptstyle\text{IM}}(\hat \eta_T)\overset{w}{\longrightarrow}\mathcal{G}_{\scriptstyle\text{SN}}\coloneqq & \frac{\left(R_2\left(\Pi'\right)^{-1}\mathcal{Z}\right)'\left(R_2 V R_2'\right)^{-1}\left(R_2\left(\Pi'\right)^{-1}\mathcal{Z}\right)}{\int_0^1 \left(W_{u\cdot v}(r) - g(r)'\mathcal{Z}\right)^2 dr} \notag\\
	=& \frac{\chi_s^2}{\int_0^1 \left(W_{u\cdot v}(r) - g(r)'\mathcal{Z}\right)^2 dr},
	\end{align}
	as $T\rightarrow \infty$.
\end{theorem}
The limiting null distribution of $\tau_{\scriptstyle\text{IM}}(\hat \eta_T)$ is nonstandard but free of \emph{any} nuisance parameters and only depends on the number of restrictions under the null hypothesis and the number of integrated regressors. Although the $\chi_s^2$ distributed random variable in the numerator is correlated with the denominator (as both are driven by $W_{u\cdot v}(r)$ and $W_v(r)$), simulating critical values is straightforward. Table~\ref{tab:critvals} in Online Appendix~\ref{app:critvalsdeter} provides critical values for various choices of $m$ and $s$ as well as for different deterministic regressors in~\eqref{eq:y}.

\begin{remark}
	We have already mentioned in the introduction that our self-normalization approach is similar in spirit to but different from the approach of \citeasnoun{KVB00} in the stationary time series literature. As \citeasnoun{KiVo02} show that the approach of \citeasnoun{KVB00} is exactly equivalent to using HAC standard errors based on the Bartlett kernel with bandwidth equal to sample size, it is tempting to assume that the traditional kernel estimator of $\Omega_{u \cdot v}$ defined in~\eqref{eq:hatOmegaudotv} with the Bartlett kernel ($\mathcal{K}_\text{\scriptsize Bartlett}$) and $b_T=T$ could also serve as a self-normalizer in our context. However, as $\hat\Omega_{u\cdot v}\left(\mathcal{K}_\text{\scriptsize Bartlett},T\right)$ depends on the OLS residuals in~\eqref{eq:y}, several nuisance parameters other than $\Omega_{u\cdot v}$ enter the limiting distribution of $\hat\Omega_{u\cdot v}\left(\mathcal{K}_\text{\scriptsize Bartlett},T\right)$. Thus, the resulting test statistic is not nuisance parameter free.\footnote{In particular, this implies that our self-normalization approach is not a special case of standard long-run variance estimation based on the Bartlett kernel with bandwidth equal to sample size. Instead, in Online Appendix~\ref{app:OtherChoices}, we show that $\hat \eta_T$ is related to a a \textit{seemingly} natural but \textit{inconsistent} kernel estimator of $\Omega_{u \cdot v}$ based on the first differences of the residuals in the augmented partial sum regression.} One way around this problem is to replace the OLS residuals in the construction of $\hat\Omega_{u\cdot v}\left(\mathcal{K}_\text{\scriptsize Bartlett},T\right)$ with FM-OLS residuals \citeaffixed{JPS06}{see}. However, the tuning parameter choices required for FM-OLS estimation may have adverse effects on the performance of the corresponding test in finite samples. Moreover, this approach does not fit to the tuning parameter free flavor of self-normalization. The discussion in Online Appendix~\ref{app:OtherChoices} reveals some other possible choices of $\kappa$ leading to a nuisance parameter free limiting distribution of $\tau_{\scriptstyle\text{IM}}(\kappa)$. A large scale simulation study comparing all mentioned approaches might be interesting but beyond the scope of this paper.
\end{remark}

\begin{remark}
	In the stationary time series literature local asymptotic power of self-normalized tests is often slightly smaller than local asymptotic power of traditional tests, see, \eg, \citeasnoun{KVB00} and \citeasnoun{Sh15}. Focusing on the single regressor case, we derive local asymptotic power of the self-normalized test $\tau_{\scriptstyle\text{IM}}(\hat \eta_T)$ and compare it with local asymptotic power of the traditional test $\tau_{\scriptstyle\text{IM}}(\hat \Omega_{u\cdot v})$ in Online Appendix~\ref{app:LocalPower}. We find that local asymptotic power of the self-normalized test is similar to, but slightly below, local asymptotic power of the traditional test. Interestingly, the corresponding largest relative local asymptotic power loss is smaller than the largest relative local asymptotic power loss of a classical self-normalized test for a restriction on the mean of a stationary time series relative to the corresponding $t$-type test \citeaffixed[p.\,1800]{Sh15}{compare}.
\end{remark}

\section{Bootstrap Inference}\label{sec:Boot}
In small to medium samples asymptotic critical values might not serve as good approximations of the quantiles of the self-normalized test statistic's finite sample distribution. In this section, we combine our self normalization approach with a VAR sieve based resampling procedure to obtain more suitable critical values. We first describe the bootstrap scheme in detail in Section~\ref{sec:Scheme} and then show its consistency in Section~\ref{sec:Consistency}.

\subsection{Bootstrap Method}\label{sec:Scheme}
The representation given in~\eqref{eq:inv} suggests to approximate $\{w_t\}_{t\in\mZ}=\{[u_t,v_t']'\}_{t\in\mZ}$ by a sequence of VAR processes with increasing order $q=q_T\rightarrow\infty$ as $T\rightarrow\infty$. These VAR approximations can be bootstrapped using the vector autoregressive sieve bootstrap. Applying the VAR sieve bootstrap in our context requires to fit a finite order VAR to $w_t=[u_t,v_t']'$, $t=1,\ldots,T$. However, while $v_t=x_t-x_{t-1}$ is simply given by the first difference of $x_t$, the regression error in~\eqref{eq:y}, $u_t$, is unknown. We therefore fit a finite order VAR to $\hat w_t \coloneqq  [\hat u_t,v_t']'$, $t=1,\ldots,T$, instead, where $\hat u_t \coloneqq  y_t - x_t'\hat\beta_{\scriptstyle\text{IM}}$ denotes the IM-OLS residual in~\eqref{eq:y}. In the following, let $\hat \Phi_1(q),\ldots,\hat \Phi_q(q)$ denote the solution of the sample Yule-Walker equations in the regression of $\hat w_{t}$ on $\hat w_{t-1},\ldots,\hat w_{t-q}$, $t=q+1,\ldots,T$, and denote the corresponding residuals by $\hat \ve_t(q) \coloneqq  \hat w_t - \sum_{j=1}^q \hat \Phi_j(q) \hat w_{t-j}$, $t=q+1,\ldots,T$.\footnote{The Yule-Walker estimator is a natural choice, as any finite order VAR estimated by the Yule-Walker estimator is causal and invertible in finite samples. This will be particularly important in the proof of Theorem~\ref{Thm:Bstar} below.}

\noindent\textbf{Bootstrap Scheme:}
\begin{enumerate}[wide]
	\item[$\left.\text{\bf Step~1}\right)$] Obtain the bootstrap sample $\left(\ve_t^*\right)_{t=1}^T$ by randomly drawing $T$ times with replacement from the centered residuals $\left(\hat \ve_t(q) - \bar{\hat{\ve}}_T(q)\right)_{t=q+1}^T$, where $\bar{\hat{\ve}}_T(q)\coloneqq (T-q)^{-1}\sum_{t=q+1}^T \hat{\ve}_{t}(q)$, and construct $\left(w_t^*\right)_{t=1}^T$ recursively as $w_t^* = \hat\Phi_1(q) w_{t-1}^* + \ldots + \hat\Phi_q(q) w_{t-q}^* + \ve_{t}^*$, given initial values $w_{1-q}^*,\ldots,w_0^*$.\footnote{\label{fn:init_values}Though irrelevant for developing asymptotic theory, it is advantageous in practical applications to eliminate the dependencies of the results on the initial values of $w_s^*$, $1-q\leq s\leq 0$, to obtain a stationary sample. We suggest to generate a sufficiently large number of $w_t^*$'s and keep the last $T+q$ of them, only.} Partition $w_t^*=[u_t^*,v_t^{*\prime}]'$ analogously to $w_t$ and define $x_t^* \coloneqq  \sum_{s=1}^t v_s^*$.
	
	\item[$\left.\text{\bf Step~2}\right)$] To generate data under the null hypothesis given in~\eqref{eq:H0}, define $y_t^* \coloneqq  x_t^{*\prime}\hat\beta_{\scriptstyle\text{IM}}^r + u_t^*$, where the restricted IM-OLS estimator of $\beta$ is given by
	\begin{align}
		\hat\beta_{\scriptstyle\text{IM}}^r \coloneqq J
	\left(\hat\theta_{\scriptstyle\text{IM}} - \left(\sum_{t=1}^T Z_t Z_t'\right)^{-1} R_2'\left[R_2 \left(\sum_{t=1}^T Z_t Z_t'\right)^{-1} R_2'\right]^{-1} \left(R_2\hat\theta_{\scriptstyle\text{IM}}-r_0\right)\right),
	\end{align}
	with $J\coloneqq [I_{m},0_{m\times m}]$, such that $R_1\hat\beta_{\scriptstyle\text{IM}}^r = r_0$.
	
	\item[$\left.\text{\bf Step~3}\right)$] Compute the OLS estimator in the bootstrap augmented partial sum regression
	\begin{align}\label{eq:Systar}
	S_t^{y^*} = S_t^{x^*\prime}\beta + x_t^{*\prime} \gamma + S_t^{u^*} = Z_t^{*\prime}\theta + S_t^{u^*},
	\end{align}
	where $S_t^{y^*} \coloneqq  \sum_{s=1}^t y_s^*$, $S_t^{x^*} \coloneqq  \sum_{s=1}^t x_s^*$, $S_t^{u^*} \coloneqq  \sum_{s=1}^t u_s^*$ and $Z_t^*\coloneqq [S_t^{x^*\prime},x_t^{*\prime}]'$, to obtain the bootstrap IM-OLS estimator $\hat\theta_{\scriptstyle\text{IM}}^*\coloneqq [\hat\beta_{\scriptstyle\text{IM}}^{*\prime},\hat\gamma_{\scriptstyle\text{IM}}^{*\prime}]'$ of $\theta$ in~\eqref{eq:Systar}. Define the corresponding residuals $\hat{S}_t^{u^*} \coloneqq  S_t^{y^*} - Z_t^{*\prime}\hat\theta_{\scriptstyle\text{IM}}^*$, $t=1,\ldots,T$, and let $\Delta\hat{S}_t^{u^*}\coloneqq \hat{S}_t^{u^*} - \hat{S}_{t-1}^{u^*}$, $t=2,\ldots,T$, denote their first differences. Define
	\begin{align*}
	\hat V_T^*\coloneqq  \left(\sum_{t=1}^T Z_t^* Z_t^{*\prime}\right)^{-1}\left(\sum_{t=1}^T c_t^* c_t^{*\prime}\right)\left(\sum_{t=1}^T Z_t^* Z_t^{*\prime}\right)^{-1},
	\end{align*}
	where $c_1^* \coloneqq  S_T^{Z^*}$ and $c_t^*\coloneqq  S_T^{Z^*} - S_{t-1}^{Z^*}$, with $S_{t}^{Z^*}\coloneqq \sum_{j=1}^t Z_j^*$, for $t=2,\ldots,T$.
	
	\item[$\left.\text{\bf Step~4}\right)$] Define the bootstrap version of the test statistic $\tau_{\scriptstyle\text{IM}}(\hat \eta_T)$ as
	\begin{align*}
	\tau_{\scriptstyle\text{IM}}^*(\hat \eta_T^*) \coloneqq  \left(R_2 \hat\theta_{\scriptstyle\text{IM}}^* - r_0\right)'\left[R_2 \hat\eta_T^*\hat V_T^* R_2'\right]^{-1}\left(R_2 \hat\theta_{\scriptstyle\text{IM}}^* - r_0\right),
	\end{align*}
	where
	\begin{align*}
	\hat \eta_T^* \coloneqq  T^{-2} \sum_{t=2}^T \left(\sum_{s=2}^t \Delta\hat{S}_s^{u^*} \right)^2.
	\end{align*}

	\item[$\left.\text{\bf Step~5}\right)$] Let $\alpha$ denote the desired nominal size of the test. Repeat $\text{Steps}\left. 1\right) \text{to} \left. 4\right)$ $B$ times, where $B$ is large and $(B+1)(1-\alpha)$ is an integer, to obtain $B$ realizations of the bootstrap test statistic $\tau_{\scriptstyle\text{IM}}^*(\hat \eta_T^*)$. Reject the null hypothesis in~\eqref{eq:H0} if the test statistic based on the original observations, $\tau_{\scriptstyle\text{IM}}(\hat \eta_T)$, is greater than the $(B+1)(1-\alpha)$-th largest realization of the bootstrap test statistic.
\end{enumerate}

Note that we impose the null hypothesis when generating the bootstrap sample $(y_t^*)_{t=1}^T$ by using the restricted IM-OLS estimator, $\hat\beta_{\scriptstyle\text{IM}}^r$, instead of the unrestricted IM-OLS estimator, $\hat\beta_{\scriptstyle\text{IM}}$. However, we use the unrestricted residuals, $\hat u_t = y_t - x_t'\hat\beta_{\scriptstyle\text{IM}}$, instead of the restricted residuals, $y_t - x_t'\hat\beta_{\scriptstyle\text{IM}}^r$, in the definition of $\hat w_t$. It is advantageous to use the restricted residuals in the definition of $\hat w_t$ when the null hypothesis is true. However, the empirical distribution function of the restricted residuals will generally fail to mimic the population distribution under the alternative. This leads to a loss of power of the test relative to the case where $\hat w_t$ is based on the unrestricted residuals, see, \eg, \citeasnoun{GiKi02} and \citeasnoun{PaPo05} for a detailed discussion.

\subsection{Bootstrap Consistency}\label{sec:Consistency}
We now show the asymptotic validity of the testing approach proposed in the previous subsection. To this end, we first prove an invariance principle result to hold for the bootstrap innovations $\{\ve_t^*\}_{t\in\mZ}$, which then enables us to show that an invariance principle result also holds for $\{w_t^*\}_{t\in\mZ}$. To derive asymptotic results, the rate at which $q_T$ goes to infinity needs to fulfill a technical assumption.
\begin{assumption}\label{ass:q}
	Let $q_T\rightarrow \infty$ and $q_T = O((T/\ln(T))^{1/3})$ as $T\rightarrow \infty$.
\end{assumption}
For notational brevity, we suppress, as before, the dependence of $q_T$ on $T$ also in the following. We are now in the position to prove the following invariance principle for the bootstrap innovations.
\begin{proposition}\label{prop:Wstar}
	It holds under Assumptions~\ref{ass:w}, \ref{ass:cumulants} and \ref{ass:q} that 
	\begin{align*}
	W_T^*(r)\coloneqq T^{-1/2} \sum_{t=1}^{\floor{rT}} \ve_t^* \overset{w^*}{\longrightarrow} \Sigma^{1/2} W(r), \quad 0\leq r \leq 1, \quad \text{in $\mP$},
	\end{align*}
	as $T\rightarrow \infty$, with $\Sigma^{1/2}(\Sigma^{1/2})'=\Sigma$.
\end{proposition}
The preceding result together with the Beveridge-Nelson decomposition \cite{PhSo92} allows us to prove an invariance principle for $\{w_t^*\}_{t\in \mZ}$ that may be of independent interest.
\begin{theorem}\label{Thm:Bstar}
	It holds under Assumptions~\ref{ass:w}, \ref{ass:cumulants} and \ref{ass:q} that
	\begin{align*}
	B_T^*(r) \coloneqq  T^{-1/2} \sum_{t=1}^{\floor{rT}} w_t^* \overset{w^*}{\longrightarrow} \Psi(1)\Sigma^{1/2}W(r), \quad 0\leq r \leq 1, \quad \text{in $\mP$},
	\end{align*}
	as $T\rightarrow \infty$. If, in addition, Assumption~\ref{ass:FCLT} is fulfilled, it holds that $\Psi(1)\Sigma^{1/2}W(r) \overset{d}{=} B(r)$, with $B(r)$ introduced in~\eqref{eq:FCLT}. In particular, $\Psi(1)\Sigma^{1/2}W(r)$ has covariance matrix $\Omega$.
\end{theorem}
Theorem~\ref{Thm:Bstar} extends the bootstrap invariance principle of \citeasnoun[Theorem~3.3]{Pa02} in the sense that the innovations $\{\ve_{t}\}_{t\in\mZ}$ in \eqref{eq:inv} and~\eqref{eq:caus} have to be uncorrelated but not necessarily independent. Nevertheless, generating the bootstrap quantities $\left(\ve_t^*\right)_{t=1}^T$ by drawing independently with replacement from the centered residuals $\left(\hat \ve_t(q) - \bar{\hat{\ve}}_T(q)\right)_{t=q+1}^T$ still allows to capture the entire second order dependence structure of $\{w_{t}\}_{t\in\mZ}$, which is in our context both necessary \textit{and} sufficient for the bootstrap to be consistent. This stems from the fact that the dependence structure in the limiting null distribution of $\tau_{\scriptstyle\text{IM}}(\hat \eta_T)$ depends only on the second moments of $\{w_t\}_{t\in\mZ}$ and, with respect to second moments, independence and uncorrelatedness are indistinguishable.

The invariance principle for $\{w_t^*\}_{t\in \mZ}$ is the key ingredient in showing that the bootstrap IM-OLS estimator $\hat\theta_{\scriptstyle\text{IM}}^*$ in~\eqref{eq:Systar} has, conditional upon the original sample, the same limiting distribution as the IM-OLS estimator $\hat\theta_{\scriptstyle\text{IM}}$ in \eqref{eq:Sy}.
\begin{theorem}\label{Thm:IMstar}
	Let the bootstrap quantities be generated as described in Section~\ref{sec:Scheme}. Then it holds under Assumptions~\ref{ass:w}, \ref{ass:cumulants}, \ref{ass:FCLT} and~\ref{ass:q} that
	\begin{align}
	\begin{bmatrix}
	T\left(\hat\beta_{\scriptstyle\text{IM}}^* - \hat\beta_{\scriptstyle\text{IM}}^r\right)\\
	\hat\gamma_{\scriptstyle\text{IM}}^* - \Omega_{vv}^{-1}\Omega_{vu}
	\end{bmatrix}
	\overset{w^*}{\longrightarrow}\Omega_{u \cdot v}^{1/2} \left(\Pi'\right)^{-1}\mathcal{Z}\quad \text{in $\mP$},\label{eq:IMstar}
	\end{align}
	as $T\rightarrow \infty$, with $\mathcal{Z}$ as defined in~\eqref{eq:Z}. Moreover, it holds under the null hypothesis given in~\eqref{eq:H0} that
	\begin{align*}
     \tau_{\scriptstyle\text{IM}}^*(\hat \eta_T^*) &\overset{w^*}{\longrightarrow} \mathcal{G}_{\scriptstyle\text{SN}}\quad \text{in $\mP$},
	\end{align*}
	as $T\rightarrow \infty$, with $\mathcal{G}_{\scriptstyle\text{SN}}$ as defined in~\eqref{eq:Geta}.
\end{theorem}
Theorem~\ref{Thm:IMstar} shows that the VAR sieve bootstrap described in Section~\ref{sec:Scheme} is consistent for the limiting distributions of the IM-OLS estimator and the self-normalized test statistic based upon it under the null hypothesis.\footnote{Following the bootstrap literature, it would be more common to use $\hat \Omega_{vv}^{-1}\hat \Omega_{vu}$ rather than $\Omega_{vv}^{-1}\Omega_{vu}$ as the centering coefficient vector in~\eqref{eq:IMstar}. However, both versions lead to the same limiting distribution. As (an estimate of) $\Omega_{vv}^{-1}\Omega_{vu}$ is not needed to construct the bootstrap samples, we use $\Omega_{vv}^{-1}\Omega_{vu}$ as the centering coefficient vector in~\eqref{eq:IMstar} to stress that estimating $\Omega$ is not necessary for our procedure.} Moreover, by construction, the bootstrap is consistent for the limiting null distribution of the self-normalized test statistic even under deviations from the null hypothesis. In particular, this implies that local asymptotic power of the bootstrap-assisted self-normalized test coincides with local asymptotic power of the asymptotic-version of the self-normalized test, compare the discussion in Remark~\ref{rem:LocalPowerBoot} in Online Appendix~\ref{app:LocalPower}. 

\begin{remark}
	The VAR sieve bootstrap scheme allows to capture the second-order dependence structure of $\{w_t\}_{t\in \mZ}$ in a simple manner. In particular, it only requires the selection of the order of the VAR, which is a straightforward and well understood task in practice. In contrast, choosing the block size for the residual-based block bootstrap proposed in \citeasnoun{PaPo03} for unit root testing seems to be difficult in applications.\footnote{Although \citeasnoun{PoWh04} and \citeasnoun{PPW09} propose estimators of the optimal block size, these are tailor-made for the sample mean of a univariate time series.} The dependent wild bootstrap (DWB), originally proposed in \citeasnoun{Sh10} in a stationary setting and recently extended in \citeasnoun{RhSh19} to univariate unit root testing, might serve as a more non-parametric alternative to the VAR sieve bootstrap. However, it requires the choice of a kernel and a bandwidth parameter to capture dependencies across time. Extending the DWB to our setting and providing a -- possibly data-driven -- rule for selecting the bandwidth parameter for a given kernel seems to be non-trivial. We leave these interesting questions for future research.
\end{remark}

\section{Finite Sample Performance}\label{sec:FiniteSample}
We generate data according to \eqref{eq:y} and \eqref{eq:x} with $m=2$ regressors, \ie,
\begin{align} 
y_t &= x_{1t}\beta_1 + x_{2t}\beta_2 + u_t,\label{eq:linMy}\\
x_{it} &= x_{i,t-1} + v_{it},\quad x_{i0}=0,\quad i=1,2,\label{eq:linMx}
\end{align}
for $t=1,\ldots,T$. The regression errors $u_t$ and the first differences of the stochastic regressors $v_{it}$ are generated as
\begin{align*}
u_t &= \rho_1 u_{t-1} + e_t +\phi e_{t-1} + \rho_2 (\nu_{1t} + \nu_{2t}),\quad u_{-100}=0,\\
v_{it} & = \nu_{it} + 0.5 \nu_{i,t-1},\quad \nu_{i,-100}=0, \quad i=1,2,
\end{align*}
for $t=-99,\ldots,0,1,\ldots,T$. The period $t=-99,\ldots,0$ serves as a burn-in period to ensure stationarity of $u_t$ and $v_{it}$. The parameters $\rho_1$ and $\rho_2$ control the level of serial correlation in the regression errors and the extent of endogeneity, respectively. For $\phi\neq 0$ the error process contains a first order moving average component. To construct $e_t$, $\nu_{1t}$ and $\nu_{2t}$ we first generate three independent univariate stationary GARCH(1,1) processes
\begin{align*}
	\xi_{jt} &= \sigma_{jt}\ve_{jt},\\
	\sigma_{jt}^2 &= a_0 + a_1 \xi_{j,t-1}^2+b_1\sigma_{j,t-1}^2, \quad \xi_{j,-100}^2=1, \quad  \sigma_{j,-100}^2=1,  
\end{align*}
$j=1,2,3$, where $[\ve_{1t},\ve_{2t},\ve_{3t}]'\sim \mathcal{N}\left(0,I_3\right)$ i.i.d.~across $t$, with $a_1,b_1\geq 0$, $a_1+b_1<1$ and $a_0 \coloneqq 1-a_1-b_1$, such that $\mE(\xi_{jt})=0$ and $\mE(\xi_{jt}^2)=\mE(\sigma_{jt}^2)=1$. We then set $[e_t,\nu_{1t},\nu_{2t}]'= L [\xi_{1t},\xi_{2t},\xi_{3t}]'$, where $L$ is the lower triangular matrix of the Cholesky decomposition of 
\begin{align*}
	P \coloneqq \begin{bmatrix}
		1&\rho_3&\rho_3\\
		\rho_3&1&\rho_3\\
		\rho_3&\rho_3&1
	\end{bmatrix},
\end{align*} 
such that $LL'=P$. The parameter $\rho_3$ thus controls the level of correlation between the univariate GARCH(1,1) processes $e_t$, $\nu_{1t}$ and $\nu_{2t}$. In the following we set $\rho_3=0.2$ to impose weak correlation between the three process. We set $\beta_1=\beta_2=1$ and choose the order $1\leq q \leq \floor{T^{1/3}} =: q_{\max}$ of the VAR sieve as the one that minimizes the Akaike information criterion (AIC) computed on the evaluation period $t=q_{\max}+1,\ldots,T$, as suggested by \citeasnoun[p.\,56]{KiLu17}.\footnote{Results based on the Bayesian information criterion are similar and therefore not reported.} We consider results for $T\in\{75,100,250,500\}$, $\rho_1=\rho_2\in\{0,0.3,0.6,0.9\}$ and $\phi\in\{0,0.3,0.9\}$. To mimic typical empirical GARCH patterns, we set $a_1=0.05$ and $b_1=0.94$ \citeaffixed[p.\,77]{BJT16}{compare}.\footnote{The performance advantages of the (bootstrap-assisted) self-normalized test over the traditional tests, that will be apparent below, also prevail for other choices of $a_1$, $b_1$ and $\rho_3$. In particular, we observe similar results in case $e_t$, $\nu_{1t}$ and $\nu_{2t}$ are i.i.d.~standard normal and independent of each other ($a_1=b_1=\rho_3=0$). We provide corresponding results in Table~\ref{tab:sizes_iid} and Figure~\ref{fig:power_iid} in Online Appendix~\ref{app:finitesampleiid}.} In all cases, the number of Monte Carlo and bootstrap replications is $3{,}000$ and $1{,}499$, respectively.

We present simulation results under the null hypothesis in Section~\ref{sec:FiniteSampleTest} and under deviations from the null hypothesis in Section~\ref{sec:FiniteSamplePower}. Section~\ref{app:FiniteSampleEstimation} in Online Appendix~\ref{app:finitesample} compares the estimation performance of the IM-OLS estimator with the performance of the D- and FM-OLS estimators. Finally, Section~\ref{app:FiniteSampleJohansen} in Online Appendix~\ref{app:finitesample} compares the performance of the \textit{non-parametric} (bootstrap-assisted) self-normalized test with the performance of \citename{Jo95}'s~\citeyear{Jo95} \textit{parametric} likelihood ratio test.

\subsection{Test Performance Under the Null Hypothesis}\label{sec:FiniteSampleTest}
We start with the performance of the (bootstrap-assisted) self-normalized test under the null hypothesis $\text{H}_0: \beta_1=1,\ \beta_2=1$. The results are benchmarked against the traditional Wald-type tests based on the D-, FM- and IM-OLS estimators, in the following denoted by $\tau_{\scriptstyle\text{D}}(\hat{\Omega}_{u\cdot v}), \tau_{\scriptstyle\text{FM}}(\hat{\Omega}_{u\cdot v})$ and $\tau_{\scriptstyle\text{IM}}(\hat{\Omega}_{u\cdot v})$, respectively. The traditional tests rely on a kernel estimator of the long-run variance parameter $\Omega_{u\cdot v}$ as defined in~\eqref{eq:hatOmegaudotv}. We analyze results for the Bartlett kernel and the QS kernel together with the corresponding data-dependent bandwidth selection rules of \citeasnoun{An91}. In addition, we analyze the performance of the IM-OLS based test statistic that is neither divided by $\hat{\Omega}_{u\cdot v}$ nor by $\hat \eta_T$, given by $\tau_{\scriptstyle\text{IM}}(1)$. In conjunction with corresponding bootstrap critical values the test is in the following denoted by $\tau_{\scriptstyle\text{IM}}^{*}(1)$.\footnote{We obtain bootstrap critical values for $\tau_{\scriptstyle\text{IM}}^*(1)$ using the scheme described in Section~\ref{sec:Scheme}, with obvious modifications.} We analyze the performance of $\tau_{\scriptstyle\text{IM}}^{*}(1)$ to assess whether the bootstrap is able to approximate the limiting distribution $\Omega_{u \cdot v} \chi_s^2$ of the tuning parameter free test statistic $\tau_{\scriptstyle\text{IM}}(1)$ adequately. Moreover, we also present results of the IM-OLS based Wald-type test that relies on bootstrap rather than asymptotically valid chi-square critical values. We denote this test in the following as $\tau_{\scriptstyle\text{IM}}^*(\hat{\Omega}_{u\cdot v})$ and refer to it as the IM-OLS based Wald-type bootstrap test.\footnote{Again, we obtain bootstrap critical values for the IM-OLS based Wald-type statistic using the scheme described in Section~\ref{sec:Scheme}, with obvious modifications. In particular, in each bootstrap iteration, the estimate of the long-run variance $\Omega_{u\cdot v}$ is based on the corresponding bootstrap sample rather than on the original sample, which has turned out to be beneficial in preliminary simulations.} Table~\ref{tab:sizes} displays the empirical null rejection probabilities -- in the following referred to as (empirical) sizes -- of the tests. 

Overall, size distortions of the tests are the larger the larger $\rho_1,\rho_2$ or $\phi$ and decrease as $T$ increases. In line with \citeasnoun{VoWa14}, we find that the traditional Wald-type test based on the IM-OLS estimator performs better than the Wald-type tests based on the D- and FM-OLS estimators, but is still severely size distorted. In addition, the following three key observations emerge: First, the self-normalized test $\tau_{\scriptstyle\text{IM}}(\hat{\eta}_{T})$ performs much better than the traditional Wald-type tests for all sample sizes considered. The performance advantage of self-normalization over the traditional approach is the more pronounced the larger $\rho_1,\rho_2$ or $\phi$. Second, the self-normalized test $\tau_{\scriptstyle\text{IM}}(\hat{\eta}_{T})$ has only minor size-distortions for $\rho_1,\rho_2\in\{0,0.3,0.6\}$ even for small sample sizes. However, for $\rho_1,\rho_2=0.9$ size distortions of $\tau_{\scriptstyle\text{IM}}(\hat{\eta}_{T})$ -- although considerably smaller than size distortions of the traditional tests -- are less satisfactory in small to medium samples. In these cases, bootstrapping reduces the size distortions considerably. Third, for $\rho_1,\rho_2\in\{0,0.3,0.6\}$ the bootstrap-assisted self-normalized test $\tau_{\scriptstyle\text{IM}}^*(\hat{\eta}_{T})$ and the IM-OLS based Wald-type bootstrap test $\tau_{\scriptstyle\text{IM}}^*(\hat{\Omega}_{u\cdot v})$ perform similarly. In case $\rho_1,\rho_2=0.9$, however, $\tau_{\scriptstyle\text{IM}}^*(\hat{\eta}_{T})$ outperforms $\tau_{\scriptstyle\text{IM}}^*(\hat{\Omega}_{u\cdot v})$, with the performance advantage being the more pronounced the larger $\phi$ and the smaller $T$.

It is worth emphasizing that for $\rho_1,\rho_2\in\{0,0.3,0.6\}$ the non-bootstrapped self-normalized test $\tau_{\scriptstyle\text{IM}}(\hat{\eta}_{T})$ outperforms the IM-OLS based Wald-type bootstrap $\tau_{\scriptstyle\text{IM}}^*(\hat{\Omega}_{u\cdot v})$. In case $\rho_1,\rho_2=0.9$, however, $\tau_{\scriptstyle\text{IM}}^*(\hat{\Omega}_{u\cdot v})$ has smaller size distortions than $\tau_{\scriptstyle\text{IM}}(\hat{\eta}_{T})$, but the difference decreases considerably as $T$ or $\phi$ increase. This demonstrates that self-normalization itself, \ie, without bootstrap assistance, is already a useful tool to reduce size distortions observed for the traditional Wald-type tests in cointegrating regression. It is also interesting to note that for $T\in\{75,100\}$ and $\rho_1,\rho_2=0.9$ the bootstrap version of the traditional IM-OLS Wald-type test based on the QS kernel has larger size distortions than the version based on the Bartlett kernel. This indicates that the bootstrap is unable to capture the effects of tuning parameter choices on the performance of the traditional tests in small to medium samples. Finally, we note that both the self-normalized test and the bootstrap-assisted self-normalized test outperform $\tau_{\scriptstyle\text{IM}}^{*}(1)$. The performance advantage of self-normalization over the traditional Wald-type tests based on $\hat{\Omega}_{u\cdot v}$ thus seems to result from the ability of the self-normalizer to converge in distribution to a limit that is scale dependent on the true long-run variance parameter rather than from avoiding estimating $\Omega_{u\cdot v}$ completely.

\begin{table}[!ht]
\adjustbox{max width=\textwidth}{\begin{threeparttable}
\centering
\caption{Empirical sizes of the tests for $\text{H}_0: \beta_1=1,\ \beta_2=1$ at $5\%$ level}
\label{tab:sizes}
\begin{tabular}{ccccccccccccc}
	\toprule[1pt]\midrule[0.3pt]
	\multicolumn{2}{c}{}&\multicolumn{3}{c}{}&\multicolumn{8}{c}{Traditional Wald-type tests}\\
	\cmidrule(lr){6-13}
	\multicolumn{3}{c}{}&\multicolumn{2}{c}{Self-normalized tests}&\multicolumn{4}{c}{Bartlett kernel}&\multicolumn{4}{c}{QS kernel}\\
	\cmidrule(lr){4-5}
	\cmidrule(lr){6-9}
	\cmidrule(lr){10-13}
	$\phi$&$\rho_1,\rho_2$ & $\tau_{\tiny \text{IM}}^*(1)$ &
	$\tau_{\scriptstyle\text{IM}}(\hat{\eta}_{T})$ & 
	$\tau_{\scriptstyle\text{IM}}^*(\hat{\eta}_{T})$ &  $\tau_{\scriptstyle\text{D}}(\hat{\Omega}_{u\cdot v})$ & $\tau_{\scriptstyle\text{FM}}(\hat{\Omega}_{u\cdot v})$ & $\tau_{\scriptstyle\text{IM}}(\hat{\Omega}_{u\cdot v})$ &
	$\tau_{\tiny \text{IM}}^*(\hat{\Omega}_{u\cdot v})$ & $\tau_{\scriptstyle\text{D}}(\hat{\Omega}_{u\cdot v})$ & $\tau_{\scriptstyle\text{FM}}(\hat{\Omega}_{u\cdot v})$ & $\tau_{\scriptstyle\text{IM}}(\hat{\Omega}_{u\cdot v})$ & 
	$\tau_{\tiny \text{IM}}^*(\hat{\Omega}_{u\cdot v})$ \\
	\midrule
	\multicolumn{13}{l}{Panel A: $T=75$}\\
	\midrule
	$0$&0 & 0.10 & 0.03 & 0.07 & 0.16 & 0.15 & 0.11 & 0.07 & 0.19 & 0.20 & 0.14 & 0.07 \\
	&0.3 & 0.12 & 0.05 & 0.08 & 0.18 & 0.21 & 0.14 & 0.08 & 0.19 & 0.23 & 0.14 & 0.08 \\
	&0.6 & 0.16 & 0.08 & 0.08 & 0.34 & 0.40 & 0.19 & 0.09 & 0.33 & 0.43 & 0.18 & 0.09 \\
	&0.9 & 0.53 & 0.36 & 0.19 & 0.69 & 0.83 & 0.69 & 0.25 & 0.77 & 0.88 & 0.77 & 0.28 \\\midrule
	$0.3$&0 & 0.11 & 0.04 & 0.07 & 0.18 & 0.18 & 0.13 & 0.07 & 0.20 & 0.21 & 0.15 & 0.07 \\
	&0.3 & 0.13 & 0.06 & 0.07 & 0.22 & 0.23 & 0.15 & 0.08 & 0.22 & 0.25 & 0.15 & 0.08 \\
	&0.6 & 0.18 & 0.09 & 0.09 & 0.39 & 0.41 & 0.22 & 0.10 & 0.40 & 0.44 & 0.23 & 0.09 \\
	&0.9 & 0.54 & 0.35 & 0.20 & 0.72 & 0.81 & 0.69 & 0.25 & 0.82 & 0.89 & 0.80 & 0.29 \\\midrule
	$0.9$&0 & 0.16 & 0.05 & 0.09 & 0.23 & 0.19 & 0.15 & 0.10 & 0.24 & 0.23 & 0.16 & 0.08 \\
	&0.3 & 0.19 & 0.07 & 0.10 & 0.29 & 0.26 & 0.18 & 0.11 & 0.31 & 0.30 & 0.19 & 0.09 \\
	&0.6 & 0.24 & 0.10 & 0.11 & 0.46 & 0.41 & 0.26 & 0.13 & 0.50 & 0.49 & 0.31 & 0.11 \\
	&0.9 & 0.55 & 0.33 & 0.20 & 0.77 & 0.79 & 0.70 & 0.27 & 0.88 & 0.90 & 0.83 & 0.31 \\
	\midrule
	\multicolumn{13}{l}{Panel B: $T=100$}\\
	\midrule
	$0$&0 & 0.08 & 0.04 & 0.07 & 0.13 & 0.13 & 0.10 & 0.07 & 0.15 & 0.17 & 0.12 & 0.06 \\
	&0.3 & 0.10 & 0.05 & 0.07 & 0.14 & 0.18 & 0.13 & 0.08 & 0.14 & 0.20 & 0.13 & 0.07 \\
	&0.6 & 0.13 & 0.07 & 0.07 & 0.26 & 0.35 & 0.16 & 0.09 & 0.25 & 0.35 & 0.16 & 0.08 \\
	&0.9 & 0.44 & 0.29 & 0.15 & 0.61 & 0.77 & 0.59 & 0.19 & 0.68 & 0.82 & 0.65 & 0.21 \\\midrule
	$0.3$&0 & 0.09 & 0.04 & 0.06 & 0.14 & 0.16 & 0.12 & 0.07 & 0.15 & 0.18 & 0.13 & 0.06 \\
	&0.3 & 0.11 & 0.06 & 0.07 & 0.16 & 0.20 & 0.14 & 0.08 & 0.16 & 0.21 & 0.14 & 0.07 \\
	&0.6 & 0.15 & 0.08 & 0.08 & 0.29 & 0.35 & 0.19 & 0.10 & 0.29 & 0.38 & 0.19 & 0.09 \\
	&0.9 & 0.44 & 0.29 & 0.16 & 0.64 & 0.76 & 0.59 & 0.20 & 0.73 & 0.84 & 0.69 & 0.22 \\\midrule
	$0.9$&0 & 0.13 & 0.05 & 0.08 & 0.16 & 0.18 & 0.13 & 0.09 & 0.17 & 0.19 & 0.14 & 0.08 \\
	&0.3 & 0.16 & 0.06 & 0.09 & 0.19 & 0.23 & 0.16 & 0.10 & 0.20 & 0.25 & 0.17 & 0.09 \\
	&0.6 & 0.20 & 0.08 & 0.10 & 0.33 & 0.36 & 0.22 & 0.11 & 0.35 & 0.42 & 0.25 & 0.11 \\
	&0.9 & 0.47 & 0.27 & 0.17 & 0.69 & 0.74 & 0.60 & 0.22 & 0.81 & 0.86 & 0.74 & 0.24 \\
	\midrule
	\multicolumn{13}{l}{Panel C: $T=250$}\\
	\midrule
	$0$&0 & 0.06 & 0.04 & 0.06 & 0.09 & 0.09 & 0.07 & 0.06 & 0.09 & 0.10 & 0.08 & 0.06 \\
	&0.3 & 0.07 & 0.05 & 0.06 & 0.10 & 0.12 & 0.09 & 0.06 & 0.09 & 0.11 & 0.08 & 0.06 \\
	&0.6 & 0.08 & 0.06 & 0.06 & 0.14 & 0.21 & 0.10 & 0.07 & 0.13 & 0.19 & 0.10 & 0.06 \\
	&0.9 & 0.19 & 0.12 & 0.08 & 0.30 & 0.58 & 0.26 & 0.09 & 0.34 & 0.60 & 0.29 & 0.10 \\\midrule
	$0.3$&0 & 0.06 & 0.05 & 0.06 & 0.10 & 0.10 & 0.08 & 0.06 & 0.09 & 0.10 & 0.08 & 0.06 \\
	&0.3 & 0.07 & 0.06 & 0.06 & 0.11 & 0.13 & 0.10 & 0.06 & 0.10 & 0.12 & 0.09 & 0.06 \\
	&0.6 & 0.09 & 0.06 & 0.06 & 0.15 & 0.21 & 0.11 & 0.07 & 0.14 & 0.21 & 0.11 & 0.07 \\
	&0.9 & 0.20 & 0.13 & 0.08 & 0.35 & 0.56 & 0.29 & 0.10 & 0.40 & 0.60 & 0.34 & 0.10 \\\midrule
	$0.9$&0 & 0.08 & 0.05 & 0.07 & 0.11 & 0.12 & 0.09 & 0.07 & 0.10 & 0.11 & 0.09 & 0.07 \\
	&0.3 & 0.10 & 0.06 & 0.07 & 0.12 & 0.15 & 0.11 & 0.07 & 0.12 & 0.14 & 0.10 & 0.07 \\
	&0.6 & 0.12 & 0.07 & 0.07 & 0.18 & 0.22 & 0.13 & 0.08 & 0.17 & 0.23 & 0.13 & 0.07 \\
	&0.9 & 0.25 & 0.14 & 0.10 & 0.41 & 0.54 & 0.32 & 0.11 & 0.48 & 0.62 & 0.41 & 0.12 \\
	\midrule
	\multicolumn{13}{l}{Panel D: $T=500$}\\
	\midrule
	$0$&0 & 0.05 & 0.04 & 0.05 & 0.07 & 0.07 & 0.07 & 0.05 & 0.07 & 0.07 & 0.07 & 0.05 \\
	&0.3 & 0.06 & 0.05 & 0.05 & 0.08 & 0.09 & 0.08 & 0.06 & 0.07 & 0.08 & 0.07 & 0.06 \\
	&0.6 & 0.06 & 0.05 & 0.05 & 0.10 & 0.14 & 0.08 & 0.06 & 0.09 & 0.13 & 0.07 & 0.05 \\
	&0.9 & 0.09 & 0.05 & 0.05 & 0.15 & 0.38 & 0.11 & 0.05 & 0.15 & 0.38 & 0.12 & 0.06 \\\midrule
	$0.3$&0 & 0.06 & 0.04 & 0.05 & 0.08 & 0.08 & 0.08 & 0.06 & 0.07 & 0.08 & 0.07 & 0.06 \\
	&0.3 & 0.06 & 0.05 & 0.05 & 0.09 & 0.10 & 0.09 & 0.06 & 0.08 & 0.09 & 0.08 & 0.05 \\
	&0.6 & 0.06 & 0.05 & 0.05 & 0.11 & 0.14 & 0.09 & 0.06 & 0.10 & 0.13 & 0.08 & 0.06 \\
	&0.9 & 0.10 & 0.06 & 0.05 & 0.18 & 0.37 & 0.13 & 0.06 & 0.20 & 0.39 & 0.15 & 0.06 \\\midrule
	$0.9$&0 & 0.06 & 0.04 & 0.05 & 0.08 & 0.09 & 0.08 & 0.06 & 0.07 & 0.08 & 0.07 & 0.06 \\
	&0.3 & 0.07 & 0.05 & 0.05 & 0.09 & 0.11 & 0.09 & 0.06 & 0.09 & 0.10 & 0.08 & 0.06 \\
	&0.6 & 0.07 & 0.05 & 0.05 & 0.12 & 0.15 & 0.10 & 0.06 & 0.11 & 0.14 & 0.09 & 0.06 \\
	&0.9 & 0.13 & 0.08 & 0.06 & 0.23 & 0.36 & 0.17 & 0.07 & 0.26 & 0.40 & 0.20 & 0.07 \\
	\midrule[0.3pt]\bottomrule[1pt]
\end{tabular}
\begin{tablenotes}
	\item Notes: Superscript~``$*$'' signifies the use of bootstrap critical values. The asymptotic critical value for the self-normalized test $\tau_{\scriptstyle\text{IM}}(\hat{\eta}_{T})$ is given in Table~\ref{tab:critvals} in Online Appendix~\ref{app:critvalsdeter} ($167.23$; Panel A, $m=2$, $s=2$).
\end{tablenotes}
\end{threeparttable}}
\end{table}

\subsection{Size-Adjusted Power}\label{sec:FiniteSamplePower}
To analyze the properties of the tests under deviations from the null hypothesis, we generate data for $\beta_1=\beta_2\in\left(1,1.2\right]$ using $20$ values on a grid with mesh size $0.01$. The large differences in sizes (\ie, under the null hypothesis), however, make a meaningful comparison of the performances under the alternative difficult. To enable a ``fair'' comparison, we follow \citeasnoun[p.\,826]{CNR15} and first simulate under the null hypothesis and record for each test the nominal size $\tilde \alpha$ that yields an empirical size equal to the desired $\alpha=0.05$. We then use critical values corresponding to $\tilde \alpha$ in the simulations under deviations from the null hypothesis. 

\begin{figure}[!t]
	\begin{center}
		\begin{subfigure}{0.4\textwidth}
			\centering
			\caption*{$T=100$, $\phi=0$}
			\vspace{-1ex}
			\includegraphics[trim={0cm 0cm 1cm 1cm},width=\textwidth,clip]{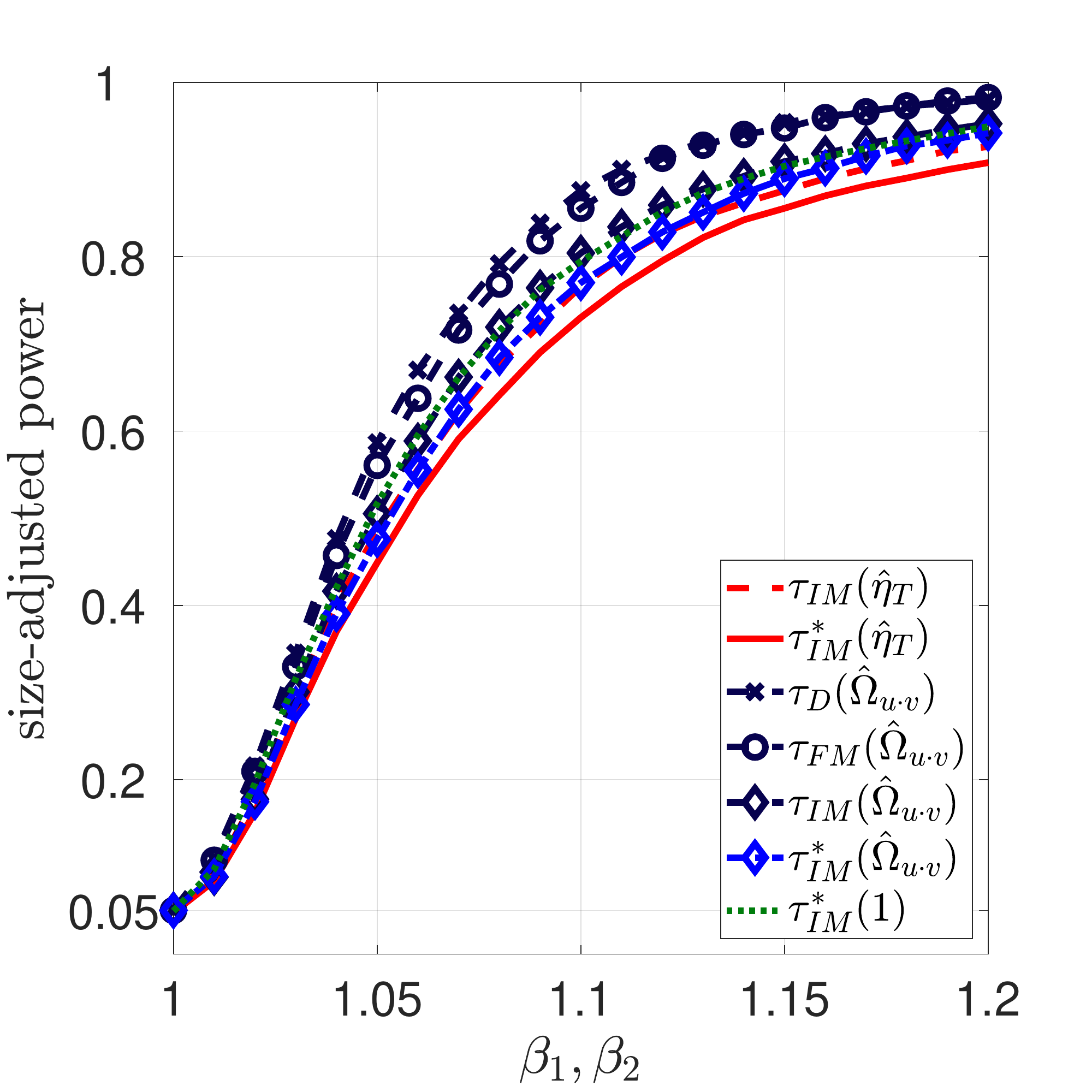}
		\end{subfigure}\begin{subfigure}{0.4\textwidth}
			\centering
			\caption*{$T=100$, $\phi=0.9$}
			\vspace{-1ex}
			\includegraphics[trim={0cm 0cm 1cm 1cm},width=\textwidth,clip]{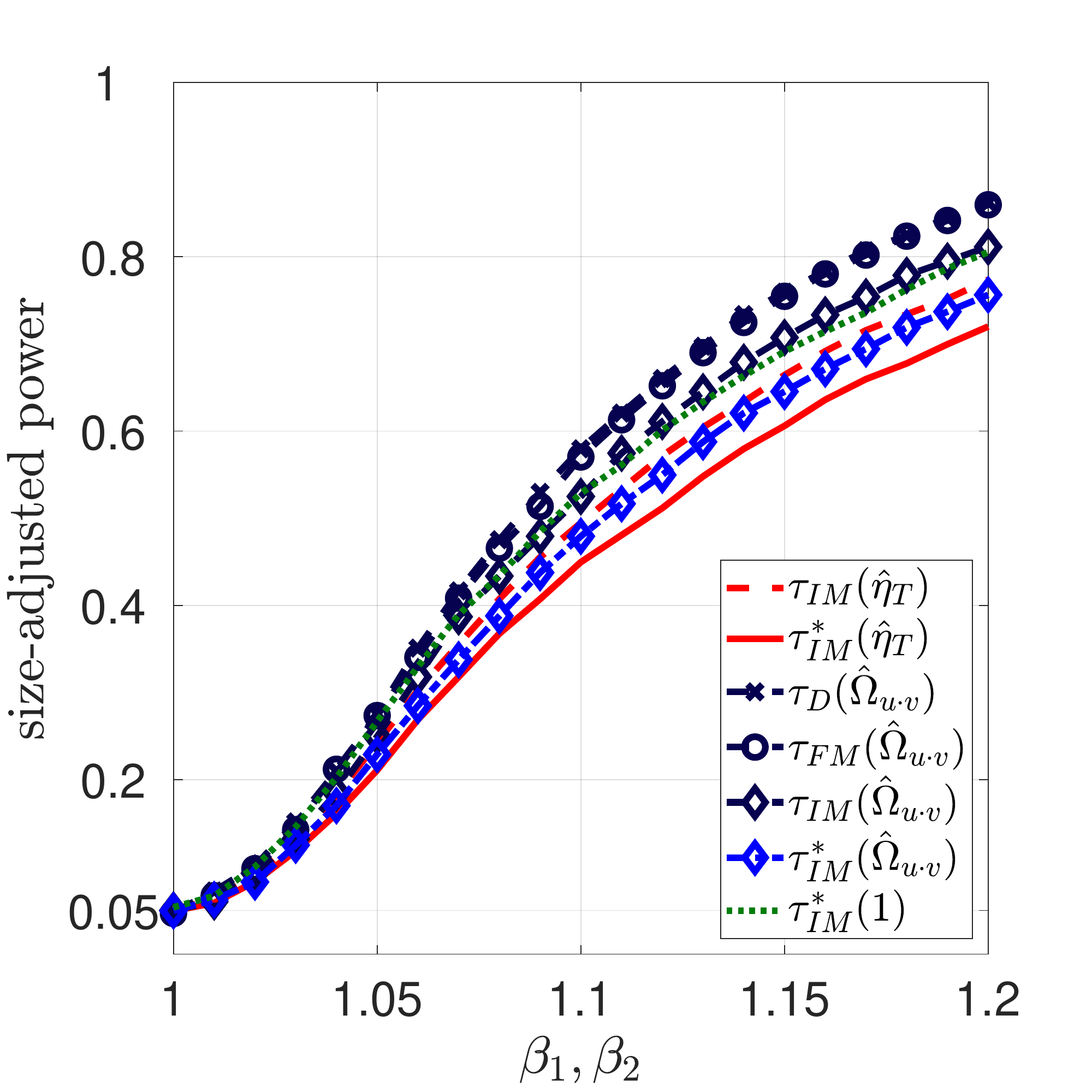}
		\end{subfigure}
	
	\vspace{2ex}
	
		\begin{subfigure}{0.4\textwidth}
			\centering
			\caption*{$T=250$, $\phi=0$}
			\vspace{-1ex}
			\includegraphics[trim={0cm 0cm 1cm 1cm},width=\textwidth,clip]{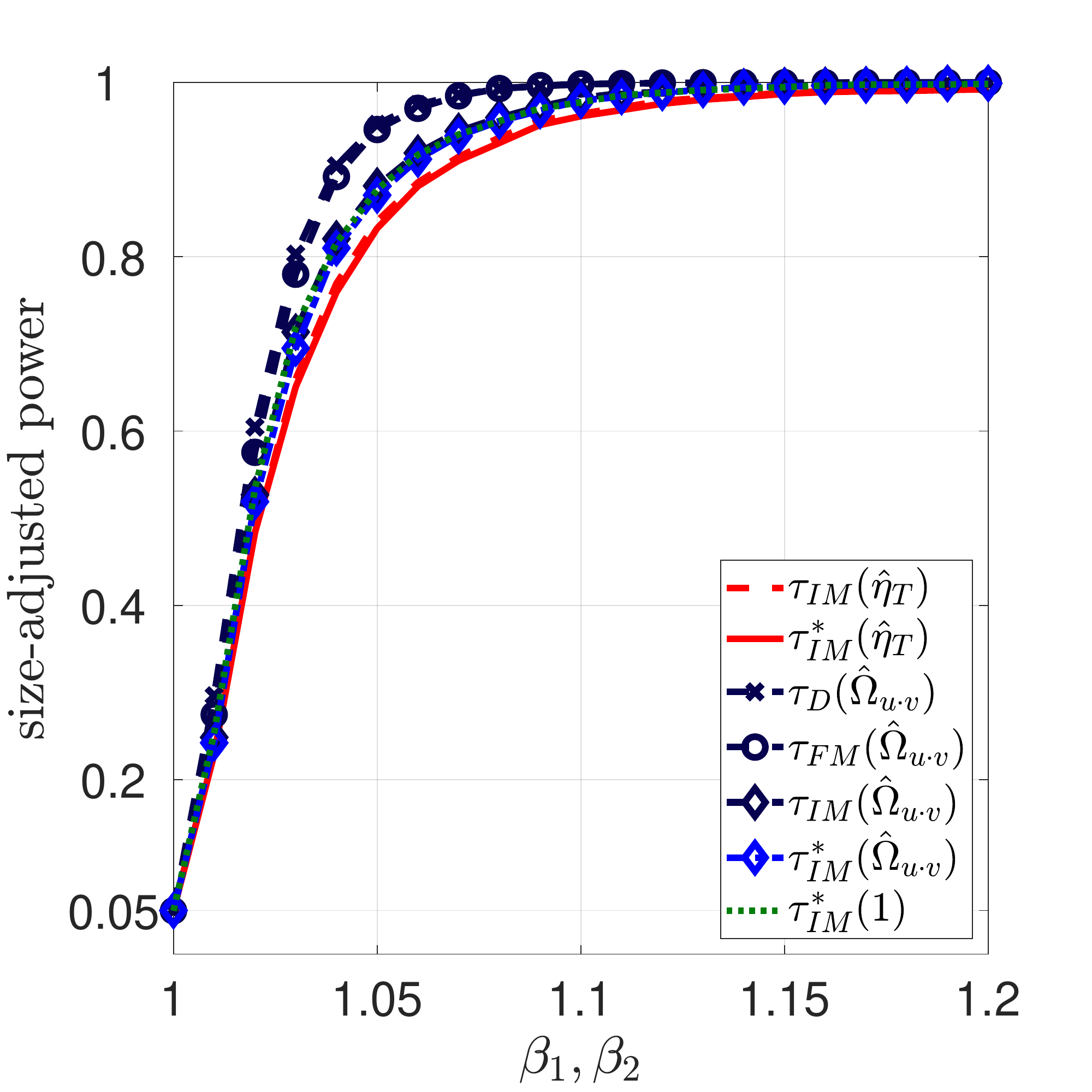}
		\end{subfigure}\begin{subfigure}{0.4\textwidth}
			\centering
			\caption*{$T=250$, $\phi=0.9$}
			\vspace{-1ex}
			\includegraphics[trim={0cm 0cm 1cm 1cm},width=\textwidth,clip]{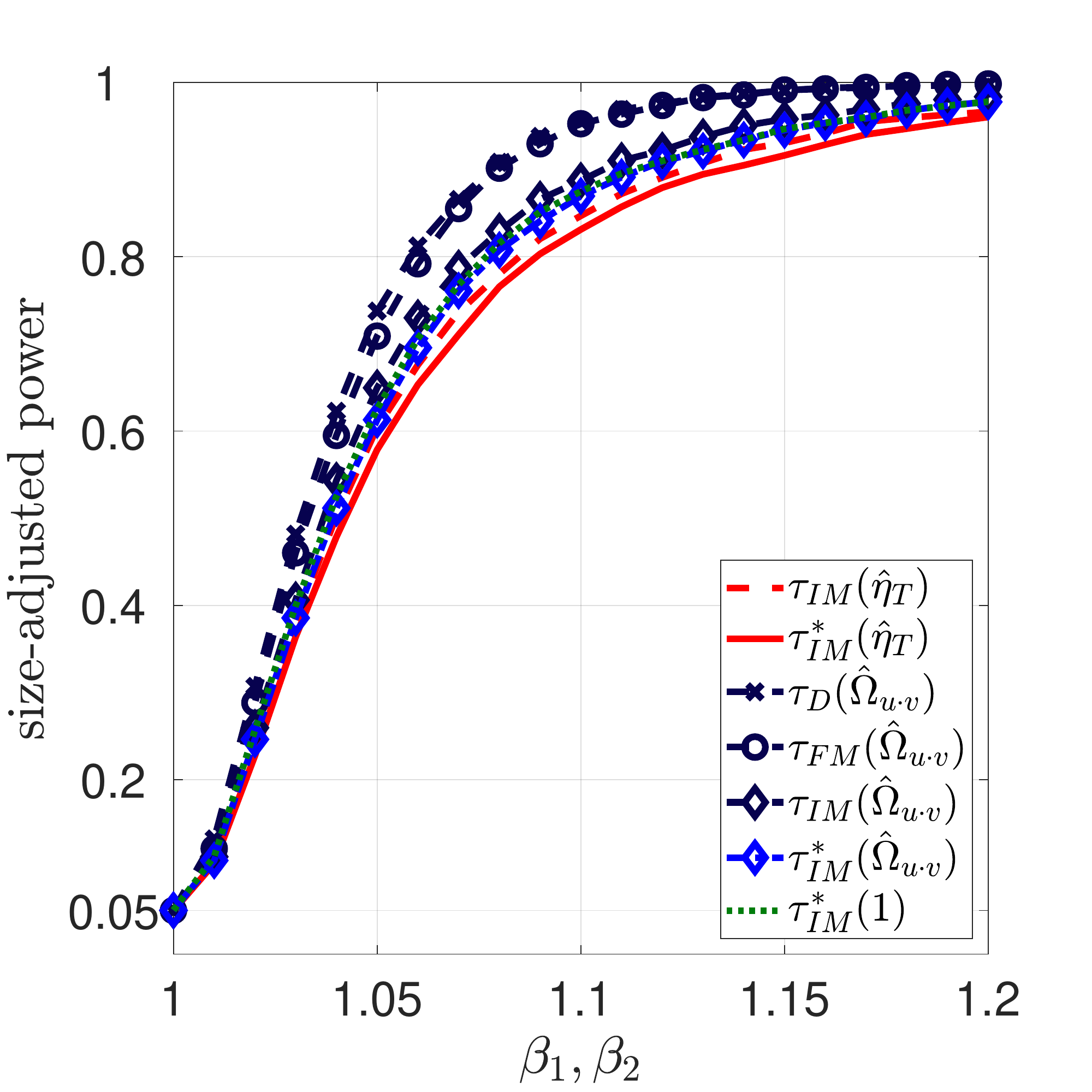}
		\end{subfigure}
	
	\end{center}
	\vspace{-2ex}
	\caption{Size-adjusted power of the tests for $\text{H}_0: \beta_1=1,\ \beta_2=1$ at 5\% level for $\rho_1,\rho_2=0.6$. Notes: Superscript~``$*$'' signifies the use of bootstrap critical values. Whenever necessary, we use the Bartlett kernel to estimate long-run variance parameters.}
	\label{fig:power}
\end{figure}

Figure~\ref{fig:power} displays illustrative results for $T\in\{100,250\}$, $\rho_1,\rho_2=0.6$ and $\phi\in\{0,0.9\}$, where, whenever necessary, we use the Bartlett kernel to estimate long-run variance parameters. Results for the QS kernel and other choices of $T$, $\rho_1,\rho_2$ and $\phi$ are qualitatively similar.\footnote{Because of the enormous size distortions of $\tau_{\tiny \text{IM}}^*(1)$ under the null hypothesis in case $\rho_1,\rho_2=0.9$ and $T\in\{75,100\}$, constructing the size-adjusted power curves for this test requires a much larger number of bootstrap replications.} In general, we find that size-adjusted power of the tests increases with sample size, but larger values of $\rho_1,\rho_2$ and $\phi$ lead to smaller size-adjusted power. In line with \citeasnoun{VoWa14} we find that the traditional Wald-type test based on the IM-OLS estimator has slightly smaller size-adjusted power than the tests based on the D- and FM-OLS estimators, with the difference vanishing as the sample size increases. Moreover, size-adjusted power of $\tau_{\tiny \text{IM}}^*(1)$ is similar to size-adjusted power of $\tau_{\scriptstyle\text{IM}}(\hat{\Omega}_{u\cdot v})$. With respect to self-normalization, we observe that size-adjusted power of $\tau_{\scriptstyle\text{IM}}(\hat{\eta}_T)$ is very similar to size-adjusted power of the traditional tests for small deviations from the null hypothesis, but then becomes slightly lower for larger deviations from the null. This finding is in line with the power properties of self-normalized tests in the stationary time series literature~\citeaffixed{Sh15}{see, \eg,} and with the local asymptotic power properties of the self-normalized test analyzed in detail in Online Appendix~\ref{app:LocalPower}. However, given the enormous size improvement of the self-normalized test with respect to the traditional tests under the null hypothesis, the observed loss in power is difficult to deem relevant. The bootstrap versions of the self-normalized test and the traditional Wald-type test have slightly smaller power than the versions based on asymptotic critical values. In this respect, we note that size-adjusted power of the IM-OLS based Wald-type bootstrap test seems to be slightly larger than size-adjusted power of the bootstrap-assisted self-normalized test. However, there are some exceptions, \eg, for $T=100$ and $\rho_1,\rho_2=0.9$, compare Figure~\ref{fig:powerLR_rho109} in Online Appendix~\ref{app:FiniteSampleJohansen}. Thus, in this case, the bootstrap-assisted self-normalized test performs better than the IM-OLS based Wald-type bootstrap test both under the null hypothesis and in terms of size-adjusted power.

\section{Empirical Illustration: The Fisher Effect}\label{sec:Illustration}
In this section we briefly illustrate the usefulness of the self-normalized test developed in this paper in applications. Many empirical studies suggest that inflation and the short-term nominal interest rate do not cointegrate with the slope of inflation being equal to one, a finding at odds with the so-called \textit{Fisher effect} that is backed by many theoretical models.\footnote{For a brief description of the underlying economic theory we refer to \citeasnoun[pp.\,195]{We08}.} The errors $u_t$ in the Fisher equation
\begin{align}\label{eq:FE}
 i_t = \alpha + \beta \pi_t + u_t,
\end{align}
$t=1,\ldots,T$, are likely to be highly persistent even in case cointegration between inflation $\pi_t$ and the short-term nominal interest rate $i_t$ prevails, \citeaffixed[pp.\,217]{We08}{see, \eg,}. As demonstrated in our simulation study (and in many simulation studies before), highly persistent errors are well known to have adverse effects on the performance of estimators and tests in cointegrating regressions. Consequently, the Fisher effect might be rejected, even if it exists, solely due to the poor performance of the methods chosen by the applied researcher \citeaffixed[for a detailed analysis of this phenomenon]{CaPi04}{see}. \citeasnoun{We08} shows that this problem can be tackled by using suitable panel methods, \ie, at the cost of including a large number of additional countries into the analysis. This may be undesirable for the applied researcher originally interested in analyzing the existence of the Fisher effect for a particular country only. Given its superior finite sample performance demonstrated in Section~\ref{sec:FiniteSample} compared to the traditional tests, the bootstrap-assisted self-normalized test developed in this paper seems to be useful to address this problem in the time series setting.

We investigate the validity of the Fisher effect for Germany and the United States between 1965Q1 and 2021Q1, using quarterly data ($T=225$) obtained from the OECD databases Economic Outlook and Main Economic Indicators.\footnote{The inflation rates and short-term interest rates are available on \href{https://data.oecd.org/price/inflation-cpi.htm}{https://data.oecd.org/price/inflation-cpi.htm} and \href{https://data.oecd.org/interest/short-term-interest-rates.htm}{https://data.oecd.org/interest/short-term-interest-rates.htm}, respectively (Accessed: March 24, 2022).} Before estimating~\eqref{eq:FE} for both countries separately, we first assess whether the short-term interest rate and inflation are indeed non-stationary and cointegrated. To this end, we employ \citename{Br02}'s~\citeyear{Br02} self-normalized variance ratio unit root test and a Shin-type cointegration test based on the IM-OLS residuals allowing for a potentially non-zero mean. To estimate the long-run variance parameter required for the cointegration test, we use the Bartlett kernel and the corresponding data-dependent bandwidth selection rule of \citeasnoun{An91}. All tests in this section are carried out at the nominal $10\%$ level. For both countries, the unit root test fails to reject the null hypothesis of non-stationarity of the interest rate and inflation and the cointegration test fails to reject the null hypothesis of cointegration between the two variables. 

These preliminary results justify to estimate~\eqref{eq:FE} on the individual country level using the IM-OLS estimator and to test whether $\beta$ is indeed equal to one using the bootstrap-assisted self-normalized test developed in this paper. We set the number of bootstrap replications to $1{,}499$ and determine the order of the VAR sieve as described in Section~\ref{sec:FiniteSample}.\footnote{Note again that ready-to-use \texttt{MATLAB} code for empirical applications is available on \href{https://github.com/kreichold/CointSelfNorm}{www.github.com/kreichold/CointSelfNorm}.} For comparison, we also estimate~\eqref{eq:FE} using the FM-OLS estimator and test unity of $\beta$ using the corresponding traditional Wald-type test. For the FM-OLS estimator and the test statistic based upon it we use the  Bartlett kernel and the corresponding data-dependent bandwidth selection rule of \citeasnoun{An91} to estimate the long-run parameters. The results are reported in Table~\ref{tab:application}, which also presents the OLS estimates as a benchmark. 

\begin{table}[!ht]
	\adjustbox{width=\textwidth}{\begin{threeparttable}
			\centering
			\caption{Country specific results}
			\label{tab:application}
			\begin{tabular}{lcccccc}
				\toprule[1pt]\midrule[0.3pt]
				\multicolumn{1}{c}{}&\multicolumn{3}{c}{Estimates of $\beta$}&\multicolumn{2}{c}{Test statistics for $\text{H}_0:\beta=1$} & \multicolumn{1}{c}{Persistence}\\
				\cmidrule(lr){2-4}
				\cmidrule(lr){5-6}
				\cmidrule(lr){7-7}
				Country & OLS & IM-OLS & FM-OLS & $\tau_{\scriptstyle\text{IM}}^*(\hat{\eta}_{T})$& $\tau_{\scriptstyle\text{FM}}(\hat{\Omega}_{u\cdot v})$ & $\hat \rho_1$\\
				\midrule
				Germany & 1.36& 1.68& 1.53&26.72 (102.57) & 2.85 (2.71)& 0.92\\
				United States & 0.99 & 1.39& 1.14 &24.29 (102.32) & 0.47 (2.71)& 0.91\\
				\midrule[0.3pt]\bottomrule[1pt]
			\end{tabular}
			\begin{tablenotes}
				\item Notes: All tests are carried out at the nominal $10\%$ level and reject the null hypothesis if the test statistic is larger than the corresponding critical value.  Bootstrap critical values for the bootstrap-assisted self-normalized test $\tau_{\scriptstyle\text{IM}}^*(\hat{\eta}_{T})$ and chi-square critical values for the traditional test $\tau_{\scriptstyle\text{FM}}(\hat{\Omega}_{u\cdot v})$ in parenthesis. The measure of persistence $\hat \rho_1$ in the regression errors in~\eqref{eq:FE} is obtained by regressing the OLS residuals on their first lag.
			\end{tablenotes}
	\end{threeparttable}}
\end{table}

The OLS estimate of $\beta$ is given by $1.36$ for Germany and $0.99$ for the United States, with the modified estimators yielding slightly larger estimates. The estimated values of $\beta$ compare well with the estimates obtained in \citeasnoun[Table~VII]{We08} for the subperiod 1980Q1--2004Q4. To assess the persistence of the errors in~\eqref{eq:FE}, we regress the OLS residuals on their first lag. The resulting estimates of the AR(1) parameter are $0.92$ for Germany and $0.91$ for the United States, showing that the regression errors are indeed highly persistent. With respect to testing unity of $\beta$, Table~\ref{tab:application} shows that the traditional FM-OLS based test accepts the hypothesized value of one for $\beta$ for the United States, but rejects the null hypothesis for Germany. In contrast, the bootstrap-assisted self-normalized test does not reject the hypothesis that $\beta$ is equal to one for both countries. Interestingly, the realizations of the self-normalized test statistic and the corresponding bootstrap critical values are very similar for Germany and the United States. This indicates that the validity of the Fisher effect is equally likely in both countries, although the estimates of $\beta$ are slightly closer to one for the United States. The traditional test thus gives misleading results.\footnote{It is worth mentioning that the asymptotic critical value for the self-normalized test statistic is given by $64.13$ (see Table~\ref{tab:critvals} in Online Appendix~\ref{app:critvalsdeter}, Panel B, $m=1$, $s=1$). This implies that also the self-normalized test based on the asymptotic critical value does not reject the validity of the Fisher effect for the two countries.}

Although our results show the importance of accounting for highly persistent regression errors, a more careful analysis is needed to assess the validity of the Fisher effect for the full period, as the link between inflation and the short-term interest rate might have weakened in the aftermath of the global financial crisis.

\section{Conclusion}\label{sec:Conclusion}
We propose a novel self-normalized test statistic for general linear restrictions in cointegrating regressions avoiding direct estimation of a long-run variance parameter. Its limiting null distribution is nonstandard, but we provide asymptotic critical values. Combining the self-normalization approach with a VAR sieve bootstrap to construct critical values further improves the performance of the self-normalized test in small to medium samples when the level of error serial correlation or regressor endogeneity is large. Constructing bootstrap critical values requires the choice of a single tuning parameter, the order of a VAR, which is a straightforward and well understood task in practice.

Simulation results and local asymptotic power analyses demonstrate that the bootstrap-assisted self-normalized test is considerably less prone to size distortions than the traditional tests at the cost of only small power losses. From a practical point of view, these small power losses are difficult to deem relevant in cointegrating regressions, given the enormous size improvements under the null hypothesis. An empirical application analyzing the validity of the Fisher effect in Germany and the United States exemplifies the advantage of the bootstrap-assisted self-normalized test over the traditional tests. Given that the bootstrap-assisted self-normalized test is easy to implement, we conclude that it should become a serious competitor to the traditional tests in practice.

Bootstrap-assisted self-normalized inference may also be a promising approach to address the enormous size distortions of hypothesis tests often observed in, \eg, cointegrated panels, cointegrating polynomial regressions and non-linear cointegrating regressions. We leave these interesting extensions for future research.

\section*{Acknowledgements}
Karsten Reichold gratefully acknowledges partial financial support by the German Research Foundation via the Collaborative Research Center SFB~823 \emph{Statistical Modelling of Nonlinear Dynamic Processes}. The authors are grateful to Katharina Hees, Fabian Knorre and participants at the Econometrics Colloquium at the University of Konstanz, the IAAE 2021 Annual Conference, the 2021 Asian Meeting of the Econometric Society and the 2021 North American Summer Meeting of the Econometric Society for helpful comments.

\bibliographystyle{ifac}

\newpage
\clearpage
\thispagestyle{empty}

\begin{NoHyper}
	
	\begin{center}
		\doublespacing
		
		{\LARGE Online Appendix to \\ ``A Bootstrap-Assisted Self-Normalization Approach to Inference in Cointegrating Regressions''}
		
		\vspace{0.5cm}
		
		{\large Karsten Reichold and Carsten Jentsch}
		
		\vspace{0.5cm}
		
		{\large \today}
		
		\vspace{1cm}
		
		\textbf{Equation, Table and Figure numbers not preceded by a letter refer to the main article.}
		
	\end{center}	
	
\newpage
\clearpage

\renewcommand\thesection{\Alph{section}}

\setcounter{page}{1}
\setcounter{section}{0}
\setcounter{footnote}{29}

\setcounter{equation}{0}
\renewcommand\theequation{\Alph{section}.\arabic{equation}}	
\setcounter{table}{0}
\renewcommand\thetable{\Alph{section}.\arabic{table}}
\setcounter{figure}{0}
\renewcommand\thefigure{\Alph{section}.\arabic{figure}}

\section{Other Possible Choices for $\kappa$}\label{app:OtherChoices}
Our particular choice for $\kappa$ in~\eqref{eq:T_k}, the self-normalizer $\hat \eta_T$, is closely related to a \textit{seemingly} natural kernel estimator of $\Omega_{u \cdot v}$ defined as
\begin{align*}
		\widetilde \Omega_{u \cdot v}\left(\mathcal{K},b_T\right) \coloneqq T^{-1} \sum_{i=2}^T \sum_{j=2}^T \mathcal{K}\left(\frac{\vert i-j\vert}{b_T}\right)\Delta\hat{S}_i^u\Delta\hat{S}_j^u,
\end{align*}
where, as before, $\mathcal{K}(\cdot)$ is a kernel function and $b_T$ is a bandwidth parameter. In contrast to the traditional estimator $\hat\Omega_{u \cdot v}\left(\mathcal{K},b_T\right)$ defined in~\eqref{eq:hatOmegaudotv}, $\widetilde \Omega_{u \cdot v}\left(\mathcal{K},b_T\right)$ is \textit{inconsistent} for $\Omega_{u \cdot v}$ under common kernel and bandwidth assumptions. Nevertheless, $\widetilde \Omega_{u \cdot v}\left(\mathcal{K},b_T\right)$ serves as a useful starting point to motivate our choice of $\hat \eta_T$ as the self-normalizer.

We first note that under traditional kernel and bandwidth assumptions $\widetilde \Omega_{u \cdot v}\left(\mathcal{K},b_T\right)\overset{w}{\longrightarrow} \Omega_{u\cdot v}\left(1+\mathcal{Z}(2)'\mathcal{Z}(2)\right)$, as $T\rightarrow \infty$, where $\mathcal{Z}(2)$ denotes the vector of the last $m$ components of the $2m$-dimensional vector $\mathcal{Z}$ defined in~\eqref{eq:Z}, see \citeasnoun[Proof of Theorem~3]{VoWa14}. That is, the limiting distribution of $\widetilde \Omega_{u \cdot v}\left(\mathcal{K},b_T\right)$ is nuisance parameter free up to its scale-dependence on $\Omega_{u \cdot v}$, which leads to a nuisance parameter free limiting distribution of the test statistic $\tau_{\scriptstyle\text{IM}}(\widetilde \Omega_{u \cdot v}\left(\mathcal{K},b_T\right))$. However, $\widetilde \Omega_{u \cdot v}\left(\mathcal{K},b_T\right)$ is not a self-normalizer in the original sense, as its construction requires tuning parameter choices, which may have adverse effects on the performance of $\tau_{\scriptstyle\text{IM}}(\widetilde \Omega_{u \cdot v}\left(\mathcal{K},b_T\right))$ in finite samples. In particular, unreported preliminary simulation results indicate that $\widetilde \Omega_{u \cdot v}\left(\mathcal{K},b_T\right)$ is less successful in reducing the size distortions of the traditional test relative to $\hat \eta_T$.

In the special case where $\mathcal{K}\left(\cdot\right)$ is the Bartlett kernel ($\mathcal{K}_\text{\scriptsize Bartlett}$) and $b_T=T$, it follows from algebraic arguments used in \citeasnoun[Proof of Lemma~1]{CaSh06} in combination with similar arguments as used in the proof of Theorem~\ref{thm:SN} that
\begin{align*}
	&\widetilde \Omega_{u \cdot v}\left(\mathcal{K}_\text{\scriptsize Bartlett},T\right) = \hat \eta_T + T^{-2} \sum_{t=2}^T \left(\sum_{s=2}^t \Delta\hat{S}_s^u - \sum_{s=2}^T \Delta\hat{S}_s^u\right)^2\notag \\
	& \overset{w}{\longrightarrow} \Omega_{u \cdot v} \left(\int_0^1 \left(W_{u\cdot v}(r) - g(r)'\mathcal{Z}\right)^2 dr \right. \notag\\
	& \hspace{2cm}\left. +\int_0^1 \left([W_{u\cdot v}(r) - g(r)'\mathcal{Z}]-[W_{u\cdot v}(1) - g(1)'\mathcal{Z}]\right)^2 dr\right),
\end{align*}
as $T\rightarrow \infty$. The quantity $\widetilde \Omega_{u \cdot v}\left(\mathcal{K}_\text{\scriptsize Bartlett},T\right)$ is thus closely related to $\hat \eta_T$ and its limiting distribution is again nuisance parameter free up to its scale-dependence on $\Omega_{u \cdot v}$. Unreported preliminary simulation results show that $\tau_{\scriptstyle\text{IM}}(\widetilde \Omega_{u \cdot v}\left(\mathcal{K}_\text{\scriptsize Bartlett},T\right))$ performs similarly to $\tau_{\scriptstyle\text{IM}}(\hat \eta_T)$ under the null hypothesis. However, $\tau_{\scriptstyle\text{IM}}(\widetilde \Omega_{u \cdot v}\left(\mathcal{K}_\text{\scriptsize Bartlett},T\right))$ has smaller (local asymptotic) power than $\tau_{\scriptstyle\text{IM}}(\hat \eta_T)$ under the alternative.

\begin{remark}
	There is another interesting relation we want to highlight. If we assume for a moment that $\sum_{s=2}^T \Delta\hat{S}_s^u=\hat S_T^u - \hat S_1^u=0$, which does not hold in general, we obtain $\widetilde \Omega_{u \cdot v}\left(\mathcal{K}_\text{\scriptsize Bartlett},T\right) = 2\hat \eta_T$. This relation fits well to the finding of \citeasnoun{KiVo02} in the stationary time series literature. The authors show that the self-normalization approach of \citeasnoun{KVB00} is exactly equivalent to using HAC standard errors based on the Bartlett kernel with bandwidth equal to sample size. From this perspective, choosing $\hat \eta_T$ as the self-normalizer seems to be the natural extension of the approach of \citeasnoun{KVB00} from the stationary time series literature to cointegrating regressions.
\end{remark}

\section{Local Asymptotic Power}\label{app:LocalPower}
In this section we compare the asymptotic power properties of the self-normalized test $\tau_{\scriptstyle\text{IM}}(\hat \eta_T)$ and the traditional test $\tau_{\scriptstyle\text{IM}}(\hat\Omega_{u\cdot v})$ under local alternatives. The results hold under Assumption~\ref{ass:FCLT}, with $(y_t)_{t=1}^T$ and $(x_t)_{t=1}^T$ generated by~\eqref{eq:y} and~\eqref{eq:x}, respectively. To ease exposition of the main arguments, we restrict attention to the single regressor case ($m=1$), which suffices to illustrate the main similarities and differences between the two tests. For $m=1$ the regression model becomes $y_t = x_t \beta + u_t$, with $\beta$ being a one-dimensional parameter. We are interested in testing the null hypothesis $\text{H}_0:\beta = \beta_0$. In this case, $\tau_{\scriptstyle\text{IM}}(\kappa)$ defined in~\eqref{eq:T_k} simplifies to
\begin{align}
	\tau_{\scriptstyle\text{IM}}(\kappa) = \left(\frac{\left(\hat\beta_{\scriptstyle\text{IM}}-\beta_0\right)}{\sqrt{\kappa \hat V_T(1,1)}}\right)^2,
\end{align}
where $\hat V_T(1,1)$ denotes the upper-left element of the ($2\times 2$)-dimensional matrix $\hat V_T$ defined in~\eqref{eq:Vhat}. Straightforward calculations reveal that under the local alternative $\text{H}_1:\beta = \beta_0 + c\,T^{-1}$, with $c\in\mR$, the limiting distribution of $\tau_{\scriptstyle\text{IM}}(\hat \Omega_{u\cdot v})$ is given by 
\begin{align}\label{eq:local_lim_omega}
	\tau_{\scriptstyle\text{IM}}(\hat \Omega_{u\cdot v}) \overset{w}{\longrightarrow} \mathcal{G}_{c} \coloneqq \left( \frac{c\,\Omega_{vv}^{1/2}}{\Omega_{u\cdot v}^{1/2}\sqrt{\tilde V(1,1)}} + \frac{\mathcal{Z}(1)}{\sqrt{\tilde V(1,1)}}\right)^2,
\end{align}
where $\tilde V(1,1)$ denotes the upper-left element of the nuisance parameter free ($2\times 2$)-dimensional matrix $\tilde V$ defined in~\eqref{eq:V} and $\mathcal{Z}(1)$ denotes the first element in the two-dimensional vector $\mathcal{Z}$ defined in~\eqref{eq:Z}. Analogously, the limiting distribution of $\tau_{\scriptstyle\text{IM}}(\hat \eta_T)$ under the local alternative is given by
\begin{align}\label{eq:local_lim_SN}
	\tau_{\scriptstyle\text{IM}}(\hat \eta_T) & \overset{w}{\longrightarrow} \mathcal{G}_{\scriptstyle\text{SN},c} \coloneqq
	\left( \frac{c\,\Omega_{vv}^{1/2}}{\Omega_{u\cdot v}^{1/2}\sqrt{\int_0^1 \left(W_{u\cdot v}(r) - g(r)'\mathcal{Z}\right)^2 dr}\sqrt{\tilde V(1,1)}}\right.\notag\\
	& \left. \hspace{3cm}+ \frac{\mathcal{Z}(1)}{\sqrt{\int_0^1 \left(W_{u\cdot v}(r) - g(r)'\mathcal{Z}\right)^2 dr}\sqrt{\tilde V(1,1)}}\right)^2.
\end{align}
Naturally, for $c=0$ the limiting distributions in~\eqref{eq:local_lim_omega} and~\eqref{eq:local_lim_SN} coincide with the chi-square distribution with one degree of freedom and the distribution $\mathcal{G}_{\scriptstyle\text{SN}}$ defined in~\eqref{eq:Geta}, respectively. It follows that local asymptotic power of $\tau_{\scriptstyle\text{IM}}(\hat \Omega_{u\cdot v})$ at the nominal $5\%$ level is given by $\mP\left(\mathcal{G}_{c} > \chi^2_{1,0.95}\right)$, where $\chi^2_{1,0.95}$ denotes the $0.95$-quantile of the chi-square distribution with one degree of freedom. Analogously, local asymptotic power of $\tau_{\scriptstyle\text{IM}}(\hat \eta_T)$ at the nominal $5\%$ level is given by $\mP\left(\mathcal{G}_{\scriptstyle\text{SN},c} > q_{1,1,0.95}\right)$, where $q_{1,1,0.95}$ denotes the corresponding $0.95$-quantile for the self-normalized test statistic given in Table~\ref{tab:critvals} in Online Appendix~\ref{app:critvalsdeter} ($56.58$; Panel A, $m=1$, $s=1$). Local asymptotic power of both tests depends on $c\,\Omega_{vv}^{1/2}\Omega_{u\cdot v}^{-1/2}$. For a given $c\neq 0$, local asymptotic power of the tests thus depends on both the long-run variance $\Omega_{vv}$ of the first differences $v_t=\Delta x_t$ of the regressor $x_t$ and the long-run variance $\Omega_{u\cdot v}$ of the regression error $u_t$ corrected for its conditional long-run mean given $v_t$. In particular, it follows from the definition of $\Omega_{u\cdot v}$ that local asymptotic power of the tests decreases as the variability in the regression errors increases. To assess the effect of the location parameter $c$, we plot the two power curves as a function of $c$ in Figure~\ref{fig:local_power} for fixed $\Omega_{vv}$ and $\Omega_{u\cdot v}$ using simulations. In particular, we set $\Omega_{vv}^{1/2}\Omega_{u\cdot v}^{-1/2}=1$ and approximate the stochastic terms using similar methods to those used to generate the critical values in Online Appendix~\ref{app:critvalsdeter}.

\begin{figure}[!ht]
	\begin{center}
		\includegraphics[trim={1.3cm 0.3cm 2.3cm 0.8cm},width=0.7\textwidth,clip]{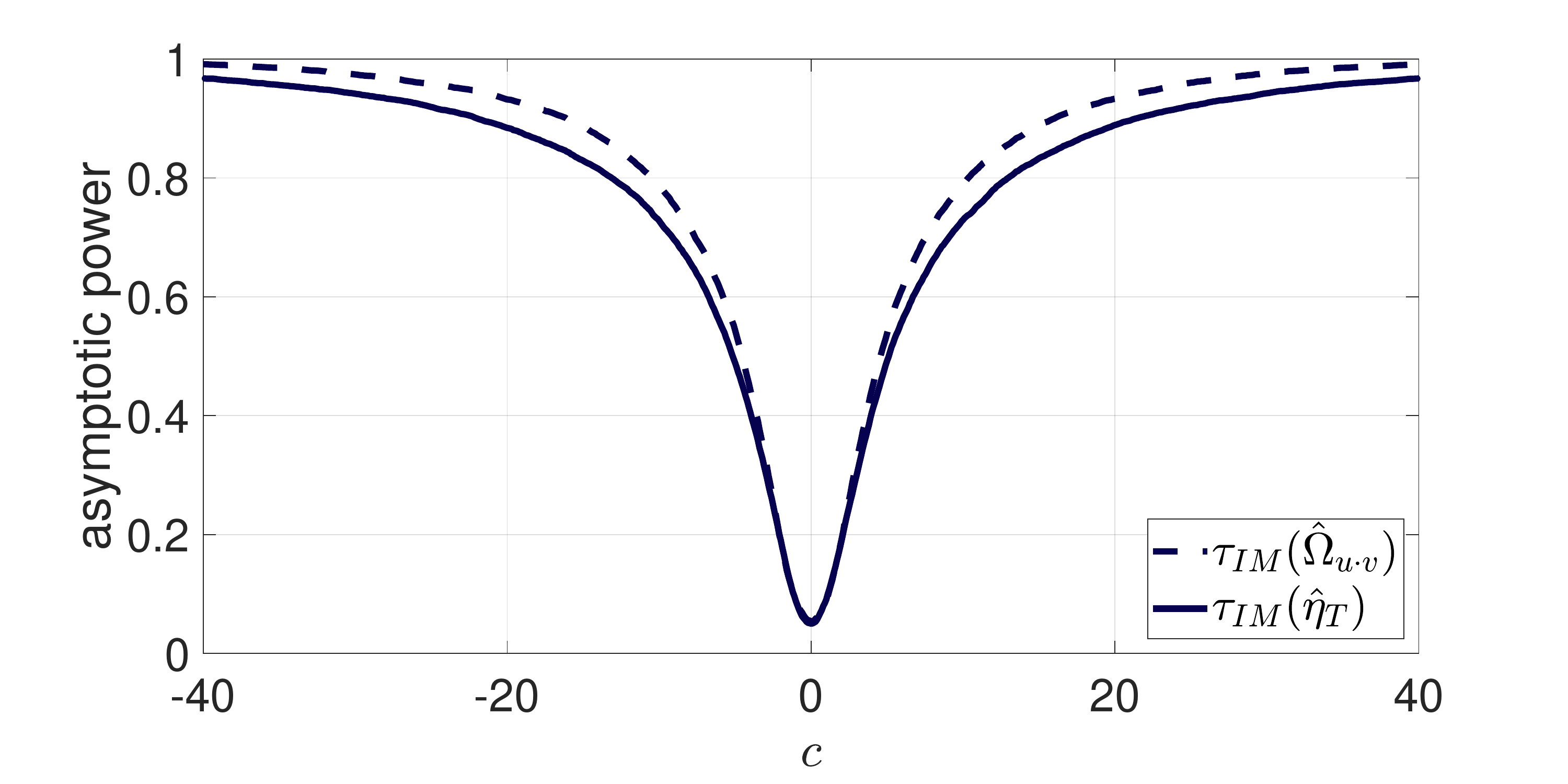}
	\end{center}
	\vspace{-2ex}
	\caption{Asymptotic power of the traditional test and the self-normalized test for $\text{H}_0:\beta = \beta_0$ under the local alternative $\beta = \beta_0 + c\,T^{-1}$. Both tests are carried out at the nominal $5\%$ level.}
	\label{fig:local_power}
\end{figure}

Figure~\ref{fig:local_power} shows that both tests have asymptotic power against local alternatives, with local asymptotic power increasing symmetrically as $c$ moves away from zero. Local asymptotic power of the self-normalized test is similar to, but slightly below, local asymptotic power of the traditional Wald-type test. To quantify the local power loss, we consider two measures: the absolute local power loss, \ie, the difference in local power between the traditional test and the self-normalized test, and the relative local power loss, \ie, the absolute local power loss divided by local power of the traditional test. The largest absolute local power loss is about $0.06$ and occurs at around $c=\pm 10.6$, whereas the largest relative local power loss is about $10.3\%$ and occurs at around $c=\pm 5.1$. The results are consistent with the findings in the stationary time series literature, where local asymptotic power of self-normalized tests is well known to be often slightly below that of traditional tests, see, \eg, \citeasnoun{KVB00} and \citeasnoun{Sh15}.\footnote{For comparison, note that \citeasnoun[p.\,1800]{Sh15} analyzes local asymptotic power of a self-normalized test for testing a hypothesis on the mean of a strictly stationary time series $\{X_t\}_{t\in \mZ}$. He reports that under suitable moment and weak dependence conditions, the largest relative local power loss of the self-normalized test statistic compared to the $t$-type test based on an estimator of the long-run variance of $\{X_t\}_{t\in \mZ}$ is around $28.5\%$. In this light, the largest relative local power loss of our self-normalized test compared to the traditional test seems to be relatively small.}

\begin{remark}\label{rem:LocalPowerBoot}
	Given the results in Theorem~\ref{Thm:IMstar}, it follows from consistency of the IM-OLS estimator of $\beta$ and imposing the null hypothesis when constructing the bootstrap data $y_t^*$ that the VAR sieve bootstrap is consistent for the limiting null distribution of the self-normalized test statistic even when the null hypothesis is incorrect. In the limit, the bootstrap critical value for the self-normalized test statistic thus coincides with the asymptotic critical value  even under deviations from the null hypothesis. Consequently, local asymptotic power of the bootstrap-assisted self-normalized test coincides with local asymptotic power of the asymptotic version of the self-normalized test.
\end{remark}

\section{Additional Finite Sample Results}\label{app:finitesample}

\subsection{Estimator Performance}\label{app:FiniteSampleEstimation}
We briefly compare the IM-OLS estimator with the D- and FM-OLS estimators in terms of bias and root mean squared error (RMSE). Implementing the D-OLS estimator requires choosing the numbers of leads and lags of the first differences of the integrated regressors. To this end, we use the Bayesian information criterion (BIC) analyzed in \citeasnoun{ChKu12}, as it appears to be the most successful criterion -- among those considered by the authors -- in reducing the mean squared error of the D-OLS estimator. The FM-OLS estimator is based on an estimator of the long-run covariance matrix of $\{[u_t,v_t']'\}_{t\in\mZ}$ as defined in~\eqref{eq:hatOmega}. We analyze the results for the Bartlett kernel and the Quadratic Spectral (QS) kernel, together with the corresponding data-dependent bandwidth selection rules of \citeasnoun{An91}. Table~\ref{tab:biasRMSE} displays results for $\beta_1$.\footnote{Results for $\beta_2$ are similar and therefore not reported.} Although asymptotically unbiased, the D-, FM- and IM-OLS estimators are biased in finite samples. Generally, bias and RMSE increase in $\rho_1,\rho_2$ and $\phi$ and decrease in $T$. In terms of bias the IM-OLS estimator performs similar to the D-OLS estimator and clearly outperforms the FM-OLS estimator. In terms of RMSE the results show that overall the D-OLS estimator performs best, followed by the FM-OLS estimator based on the Bartlett kernel. RMSE of the IM-OLS estimator is comparable to the RMSE of the FM-OLS estimator based on the QS kernel, which is often slightly larger than RMSE of the FM-OLS estimator based on the Bartlett kernel. We conclude that the IM-OLS estimator performs relatively well and serves as a good starting point for self-normalized inference in cointegrating regressions.

\begin{table}[!h]
\centering
\adjustbox{totalheight=\textheight}{\begin{threeparttable}
\caption{Bias and RMSE of various estimators of $\beta_1$}
\label{tab:biasRMSE}
\begin{tabular}{ccrrrrrrrrrr}
	\toprule[1pt]\midrule[0.3pt]
	&&\multicolumn{5}{c}{$\text{Bias}\times 100$}&\multicolumn{5}{c}{$\text{RMSE}\times 100$}\\
	\cmidrule(lr){3-7}
	\cmidrule(lr){8-12}
	&&&&\multicolumn{2}{c}{FM-OLS}&&&&\multicolumn{2}{c}{FM-OLS}&\\
	\cmidrule(lr){5-6}
	\cmidrule(lr){10-11}
	$\phi$&$\rho_1,\rho_2$ & IM-OLS &  D-OLS &  Bartlett &  QS & QML & IM-OLS &  D-OLS &  Bartlett &  QS & QML\\
	\midrule
	\multicolumn{12}{l}{Panel A: $T=75$}\\
	\midrule
	$0$&0 & $-$0.25 & $-$0.27 & 0.04 & 0.11 & $-$0.05 & 5.15 & 3.52 & 3.28 & 3.42 & 15.31 \\
	&0.3 & $-$0.25 & $-$0.08 & 0.66 & 0.79 & $-$0.45 & 7.29 & 4.91 & 5.02 & 5.38 & 19.79 \\
	&0.6 & 0.88 & 1.65 & 3.64 & 3.70 & $-$6.82 & 12.72 & 11.68 & 11.12 & 11.97 & 374.48 \\
	&0.9 & 19.63 & 12.69 & 24.93 & 25.00 & $-$606.33 & 58.08 & 46.22 & 48.09 & 55.54 & 31796.03 \\\midrule
	$0.3$&0 & $-$0.28 & $-$0.18 & 0.13 & 0.19 & 0.39 & 6.67 & 4.46 & 4.24 & 4.42 & 20.65 \\
	&0.3 & $-$0.28 & 0.10 & 0.87 & 1.00 & 0.42 & 9.46 & 7.43 & 6.50 & 6.93 & 23.56 \\
	&0.6 & 0.91 & 1.52 & 4.16 & 4.30 & 1.25 & 16.45 & 16.59 & 13.63 & 14.75 & 77.59 \\
	&0.9 & 20.44 & 12.89 & 26.44 & 26.37 & 75.50 & 68.99 & 58.15 & 54.10 & 88.03 & 6141.08 \\\midrule
	$0.9$&0 & $-$0.35 & 0.09 & 0.34 & 0.41 & $-$0.66 & 9.74 & 8.06 & 6.27 & 6.56 & 20.00 \\
	&0.3 & $-$0.33 & 0.35 & 1.37 & 1.51 & $-$0.95 & 13.83 & 12.87 & 9.56 & 10.16 & 76.13 \\
	&0.6 & 0.96 & 1.60 & 5.14 & 5.59 & $-$3.55 & 23.95 & 26.29 & 18.48 & 19.82 & 354.12 \\
	&0.9 & 22.08 & 13.54 & 29.43 & 29.69 & $-$31.68 & 92.30 & 83.42 & 66.97 & 122.62 & 4171.58 \\
	\midrule
	\multicolumn{12}{l}{Panel B: $T=100$}\\
	\midrule
	$0$&0 & $-$0.07 & $-$0.15 & 0.05 & 0.09 & 0.13 & 3.91 & 2.37 & 2.38 & 2.46 & 6.22 \\
	&0.3 & $-$0.03 & 0.03 & 0.45 & 0.51 & $-$0.14 & 5.55 & 3.25 & 3.55 & 3.72 & 9.96 \\
	&0.6 & 0.74 & 1.53 & 2.48 & 2.44 & $-$2.05 & 9.71 & 6.63 & 7.76 & 8.25 & 115.53 \\
	&0.9 & 14.76 & 11.97 & 20.06 & 20.35 & $-$54.65 & 45.30 & 33.92 & 39.03 & 42.27 & 3497.87 \\\midrule
	$0.3$&0 & $-$0.07 & $-$0.07 & 0.12 & 0.15 & 0.02 & 5.07 & 2.97 & 3.09 & 3.18 & 3.97 \\
	&0.3 & $-$0.01 & 0.19 & 0.60 & 0.65 & 0.09 & 7.20 & 4.21 & 4.62 & 4.84 & 11.40 \\
	&0.6 & 0.83 & 1.71 & 2.92 & 2.94 & $-$5.09 & 12.56 & 8.85 & 9.67 & 10.28 & 289.60 \\
	&0.9 & 15.60 & 12.88 & 21.49 & 22.36 & 12.16 & 53.98 & 42.04 & 44.37 & 50.19 & 903.93 \\\midrule
	$0.9$&0 & $-$0.06 & 0.12 & 0.30 & 0.31 & 0.24 & 7.39 & 4.42 & 4.56 & 4.71 & 9.54 \\
	&0.3 & 0.03 & 0.49 & 1.00 & 1.05 & 0.06 & 10.51 & 6.40 & 6.84 & 7.18 & 21.90 \\
	&0.6 & 1.01 & 2.04 & 3.72 & 3.86 & $-$1.24 & 18.27 & 13.33 & 13.42 & 14.37 & 168.09 \\
	&0.9 & 17.27 & 14.19 & 24.28 & 25.81 & 123.50 & 72.40 & 58.58 & 55.89 & 66.60 & 4255.25 \\
	\midrule
	\multicolumn{12}{l}{Panel C: $T=250$}\\
	\midrule
	$0$&0 & 0.01 & $-$0.06 & 0.02 & 0.02 & 0.01 & 1.59 & 0.91 & 0.91 & 0.92 & 0.95 \\
	&0.3 & 0.03 & 0.02 & 0.12 & 0.12 & 0.01 & 2.27 & 1.25 & 1.33 & 1.35 & 1.38 \\
	&0.6 & 0.21 & 0.46 & 0.79 & 0.69 & 0.04 & 3.96 & 2.38 & 2.80 & 2.84 & 2.50 \\
	&0.9 & 5.01 & 4.25 & 8.83 & 8.67 & $-$0.17 & 17.71 & 11.78 & 17.39 & 18.36 & 15.07 \\\midrule
	$0.3$&0 & 0.02 & $-$0.03 & 0.04 & 0.04 & 0.01 & 2.07 & 1.14 & 1.18 & 1.19 & 1.25 \\
	&0.3 & 0.05 & 0.10 & 0.17 & 0.16 & $-$0.00 & 2.95 & 1.61 & 1.74 & 1.77 & 1.83 \\
	&0.6 & 0.24 & 0.55 & 0.93 & 0.84 & 0.01 & 5.14 & 3.06 & 3.54 & 3.63 & 3.33 \\
	&0.9 & 5.32 & 4.98 & 9.64 & 9.82 & $-$1.37 & 21.88 & 14.73 & 19.96 & 21.40 & 36.14 \\\midrule
	$0.9$&0 & 0.04 & 0.08 & 0.09 & 0.09 & 0.02 & 3.02 & 1.67 & 1.74 & 1.77 & 1.91 \\
	&0.3 & 0.07 & 0.25 & 0.30 & 0.29 & 0.07 & 4.30 & 2.41 & 2.58 & 2.65 & 2.87 \\
	&0.6 & 0.31 & 0.75 & 1.21 & 1.15 & 0.03 & 7.50 & 4.44 & 5.01 & 5.19 & 6.09 \\
	&0.9 & 5.95 & 6.43 & 11.20 & 11.85 & 5.25 & 30.54 & 20.66 & 25.18 & 27.23 & 290.19 \\
	\midrule
	\multicolumn{12}{l}{Panel D: $T=500$}\\
	\midrule
	$0$&0 & 0.01 & $-$0.01 & 0.01 & 0.01 & 0.01 & 0.77 & 0.45 & 0.45 & 0.46 & 0.46 \\
	&0.3 & 0.02 & 0.02 & 0.05 & 0.05 & 0.01 & 1.10 & 0.63 & 0.66 & 0.67 & 0.66 \\
	&0.6 & 0.08 & 0.21 & 0.30 & 0.25 & 0.03 & 1.93 & 1.16 & 1.33 & 1.33 & 1.17 \\
	&0.9 & 1.59 & 1.72 & 3.77 & 3.45 & 0.02 & 8.22 & 5.38 & 8.33 & 8.54 & 5.14 \\\midrule
	$0.3$&0 & 0.02 & $-$0.00 & 0.02 & 0.02 & 0.01 & 1.00 & 0.58 & 0.59 & 0.59 & 0.60 \\
	&0.3 & 0.03 & 0.06 & 0.07 & 0.06 & 0.02 & 1.43 & 0.82 & 0.86 & 0.87 & 0.86 \\
	&0.6 & 0.09 & 0.24 & 0.35 & 0.28 & 0.02 & 2.50 & 1.50 & 1.70 & 1.72 & 1.54 \\
	&0.9 & 1.71 & 2.05 & 4.11 & 3.92 & $-$0.07 & 10.41 & 6.82 & 9.64 & 10.07 & 6.86 \\\midrule
	$0.9$&0 & 0.03 & 0.05 & 0.04 & 0.04 & 0.02 & 1.46 & 0.84 & 0.87 & 0.87 & 0.91 \\
	&0.3 & 0.04 & 0.13 & 0.12 & 0.10 & $-$0.01 & 2.09 & 1.21 & 1.28 & 1.29 & 1.33 \\
	&0.6 & 0.12 & 0.31 & 0.45 & 0.39 & $-$0.01 & 3.65 & 2.18 & 2.43 & 2.47 & 2.34 \\
	&0.9 & 1.96 & 2.73 & 4.82 & 4.82 & 0.08 & 14.87 & 9.69 & 12.35 & 13.12 & 10.14 \\
	\midrule[0.3pt]\bottomrule[1pt]
\end{tabular}
\begin{tablenotes}
	\item Notes: The terms ``Bartlett'' and ``QS'' signify the kernel used to apply the FM-OLS estimator. QML denotes the quasi maximum likelihood estimator of \citeasnoun{Jo95}, as described in Section~\ref{app:FiniteSampleJohansen}.
\end{tablenotes}
\end{threeparttable}}
\end{table}

\newpage
\clearpage

\subsection{Results in the i.i.d.~Innovations Case ($a_1=b_1=\rho_3=0$)}\label{app:finitesampleiid}

\begin{table}[!h]
\adjustbox{max width=\textwidth}{\begin{threeparttable}
\centering
\caption{Empirical sizes of the tests for $\text{H}_0: \beta_1=1,\ \beta_2=1$ at $5\%$ level in the i.i.d.~innovations case ($a_1=b_1=\rho_3=0$)}
\label{tab:sizes_iid}
\begin{tabular}{ccccccccccccc}
	\toprule[1pt]\midrule[0.3pt]
	\multicolumn{2}{c}{}&\multicolumn{3}{c}{}&\multicolumn{8}{c}{Traditional Wald-type tests}\\
	\cmidrule(lr){6-13}
	\multicolumn{3}{c}{}&\multicolumn{2}{c}{Self-normalized tests}&\multicolumn{4}{c}{Bartlett kernel}&\multicolumn{4}{c}{QS kernel}\\
	\cmidrule(lr){4-5}
	\cmidrule(lr){6-9}
	\cmidrule(lr){10-13}
	$\phi$&$\rho_1,\rho_2$ & $\tau_{\tiny \text{IM}}^*(1)$ &
	$\tau_{\scriptstyle\text{IM}}(\hat{\eta}_{T})$ & 
	$\tau_{\scriptstyle\text{IM}}^*(\hat{\eta}_{T})$ &  $\tau_{\scriptstyle\text{D}}(\hat{\Omega}_{u\cdot v})$ & $\tau_{\scriptstyle\text{FM}}(\hat{\Omega}_{u\cdot v})$ & $\tau_{\scriptstyle\text{IM}}(\hat{\Omega}_{u\cdot v})$ &
	$\tau_{\tiny \text{IM}}^*(\hat{\Omega}_{u\cdot v})$ & $\tau_{\scriptstyle\text{D}}(\hat{\Omega}_{u\cdot v})$ & $\tau_{\scriptstyle\text{FM}}(\hat{\Omega}_{u\cdot v})$ & $\tau_{\scriptstyle\text{IM}}(\hat{\Omega}_{u\cdot v})$ & 
	$\tau_{\tiny \text{IM}}^*(\hat{\Omega}_{u\cdot v})$ \\
	\midrule
	\multicolumn{13}{l}{Panel A: $T=75$}\\
	\midrule
	$0$&0 & 0.09 & 0.03 & 0.07 & 0.16 & 0.16 & 0.11 & 0.07 & 0.19 & 0.20 & 0.14 & 0.06 \\
	&0.3 & 0.11 & 0.05 & 0.07 & 0.20 & 0.20 & 0.14 & 0.07 & 0.20 & 0.23 & 0.15 & 0.06 \\
	&0.6 & 0.15 & 0.09 & 0.08 & 0.34 & 0.35 & 0.23 & 0.09 & 0.34 & 0.37 & 0.22 & 0.08 \\
	&0.9 & 0.48 & 0.35 & 0.19 & 0.72 & 0.79 & 0.66 & 0.25 & 0.79 & 0.84 & 0.73 & 0.26 \\\midrule
	$0.3$&0 & 0.09 & 0.04 & 0.07 & 0.18 & 0.18 & 0.13 & 0.06 & 0.20 & 0.21 & 0.15 & 0.06 \\
	&0.3 & 0.12 & 0.06 & 0.07 & 0.23 & 0.22 & 0.17 & 0.07 & 0.23 & 0.24 & 0.17 & 0.06 \\
	&0.6 & 0.18 & 0.09 & 0.09 & 0.40 & 0.36 & 0.25 & 0.09 & 0.41 & 0.40 & 0.27 & 0.08 \\
	&0.9 & 0.48 & 0.32 & 0.19 & 0.75 & 0.77 & 0.66 & 0.24 & 0.83 & 0.85 & 0.76 & 0.28 \\\midrule
	$0.9$&0 & 0.16 & 0.05 & 0.10 & 0.23 & 0.20 & 0.15 & 0.09 & 0.24 & 0.22 & 0.16 & 0.08 \\
	&0.3 & 0.19 & 0.06 & 0.10 & 0.32 & 0.26 & 0.19 & 0.11 & 0.33 & 0.29 & 0.21 & 0.09 \\
	&0.6 & 0.26 & 0.10 & 0.12 & 0.48 & 0.38 & 0.29 & 0.13 & 0.53 & 0.45 & 0.34 & 0.11 \\
	&0.9 & 0.51 & 0.30 & 0.19 & 0.78 & 0.74 & 0.68 & 0.25 & 0.88 & 0.87 & 0.82 & 0.28 \\
	\midrule
	\multicolumn{13}{l}{Panel B: $T=100$}\\
	\midrule
	$0$&0 & 0.08 & 0.03 & 0.07 & 0.13 & 0.14 & 0.10 & 0.06 & 0.15 & 0.17 & 0.12 & 0.06 \\
	&0.3 & 0.09 & 0.05 & 0.07 & 0.16 & 0.18 & 0.14 & 0.07 & 0.15 & 0.18 & 0.13 & 0.06 \\
	&0.6 & 0.13 & 0.08 & 0.07 & 0.27 & 0.31 & 0.19 & 0.08 & 0.26 & 0.31 & 0.18 & 0.07 \\
	&0.9 & 0.39 & 0.27 & 0.14 & 0.66 & 0.73 & 0.57 & 0.18 & 0.72 & 0.79 & 0.64 & 0.20 \\\midrule
	$0.3$&0 & 0.08 & 0.05 & 0.07 & 0.15 & 0.15 & 0.12 & 0.06 & 0.15 & 0.17 & 0.13 & 0.06 \\
	&0.3 & 0.10 & 0.06 & 0.07 & 0.18 & 0.19 & 0.15 & 0.07 & 0.17 & 0.20 & 0.14 & 0.07 \\
	&0.6 & 0.14 & 0.09 & 0.08 & 0.29 & 0.32 & 0.22 & 0.09 & 0.29 & 0.33 & 0.22 & 0.08 \\
	&0.9 & 0.40 & 0.27 & 0.15 & 0.68 & 0.71 & 0.58 & 0.19 & 0.77 & 0.81 & 0.69 & 0.22 \\\midrule
	$0.9$&0 & 0.13 & 0.05 & 0.08 & 0.17 & 0.17 & 0.13 & 0.08 & 0.16 & 0.18 & 0.14 & 0.07 \\
	&0.3 & 0.15 & 0.06 & 0.09 & 0.21 & 0.22 & 0.17 & 0.09 & 0.21 & 0.24 & 0.17 & 0.09 \\
	&0.6 & 0.21 & 0.09 & 0.10 & 0.32 & 0.33 & 0.25 & 0.11 & 0.35 & 0.37 & 0.28 & 0.09 \\
	&0.9 & 0.43 & 0.25 & 0.16 & 0.69 & 0.69 & 0.59 & 0.21 & 0.81 & 0.84 & 0.74 & 0.23 \\
	\midrule
	\multicolumn{13}{l}{Panel C: $T=250$}\\
	\midrule
	$0$&0 & 0.06 & 0.04 & 0.06 & 0.08 & 0.09 & 0.07 & 0.06 & 0.09 & 0.10 & 0.08 & 0.05 \\
	&0.3 & 0.07 & 0.05 & 0.06 & 0.10 & 0.12 & 0.09 & 0.06 & 0.09 & 0.11 & 0.09 & 0.06 \\
	&0.6 & 0.07 & 0.06 & 0.06 & 0.17 & 0.19 & 0.11 & 0.06 & 0.15 & 0.17 & 0.10 & 0.06 \\
	&0.9 & 0.18 & 0.13 & 0.07 & 0.38 & 0.52 & 0.28 & 0.09 & 0.40 & 0.55 & 0.31 & 0.09 \\\midrule
	$0.3$&0 & 0.07 & 0.05 & 0.06 & 0.09 & 0.11 & 0.09 & 0.06 & 0.09 & 0.10 & 0.08 & 0.06 \\
	&0.3 & 0.07 & 0.05 & 0.06 & 0.11 & 0.13 & 0.10 & 0.06 & 0.10 & 0.12 & 0.09 & 0.06 \\
	&0.6 & 0.08 & 0.06 & 0.06 & 0.17 & 0.20 & 0.13 & 0.06 & 0.16 & 0.19 & 0.12 & 0.06 \\
	&0.9 & 0.19 & 0.13 & 0.07 & 0.41 & 0.50 & 0.31 & 0.09 & 0.45 & 0.54 & 0.36 & 0.09 \\\midrule
	$0.9$&0 & 0.08 & 0.05 & 0.06 & 0.10 & 0.12 & 0.09 & 0.07 & 0.09 & 0.11 & 0.09 & 0.07 \\
	&0.3 & 0.10 & 0.05 & 0.07 & 0.12 & 0.15 & 0.11 & 0.07 & 0.11 & 0.14 & 0.10 & 0.07 \\
	&0.6 & 0.12 & 0.07 & 0.07 & 0.17 & 0.21 & 0.15 & 0.07 & 0.17 & 0.21 & 0.14 & 0.07 \\
	&0.9 & 0.23 & 0.14 & 0.09 & 0.43 & 0.48 & 0.35 & 0.11 & 0.50 & 0.57 & 0.42 & 0.11 \\
	\midrule
	\multicolumn{13}{l}{Panel D: $T=500$}\\
	\midrule
	$0$&0 & 0.05 & 0.04 & 0.05 & 0.07 & 0.07 & 0.07 & 0.05 & 0.07 & 0.07 & 0.07 & 0.05 \\
	&0.3 & 0.05 & 0.05 & 0.05 & 0.08 & 0.08 & 0.08 & 0.05 & 0.07 & 0.08 & 0.07 & 0.05 \\
	&0.6 & 0.06 & 0.05 & 0.05 & 0.11 & 0.13 & 0.09 & 0.05 & 0.10 & 0.11 & 0.08 & 0.05 \\
	&0.9 & 0.08 & 0.06 & 0.05 & 0.20 & 0.34 & 0.13 & 0.05 & 0.20 & 0.35 & 0.14 & 0.05 \\\midrule
	$0.3$&0 & 0.06 & 0.04 & 0.05 & 0.08 & 0.08 & 0.07 & 0.05 & 0.07 & 0.07 & 0.07 & 0.05 \\
	&0.3 & 0.06 & 0.05 & 0.05 & 0.09 & 0.09 & 0.08 & 0.05 & 0.08 & 0.08 & 0.07 & 0.05 \\
	&0.6 & 0.06 & 0.05 & 0.05 & 0.12 & 0.13 & 0.10 & 0.05 & 0.11 & 0.12 & 0.09 & 0.05 \\
	&0.9 & 0.09 & 0.07 & 0.05 & 0.24 & 0.34 & 0.16 & 0.05 & 0.24 & 0.36 & 0.17 & 0.06 \\\midrule
	$0.9$&0 & 0.07 & 0.04 & 0.05 & 0.08 & 0.08 & 0.08 & 0.06 & 0.07 & 0.08 & 0.07 & 0.06 \\
	&0.3 & 0.07 & 0.05 & 0.05 & 0.09 & 0.10 & 0.09 & 0.06 & 0.08 & 0.10 & 0.08 & 0.05 \\
	&0.6 & 0.08 & 0.05 & 0.05 & 0.13 & 0.14 & 0.11 & 0.06 & 0.12 & 0.13 & 0.10 & 0.06 \\
	&0.9 & 0.12 & 0.08 & 0.06 & 0.28 & 0.33 & 0.20 & 0.06 & 0.30 & 0.36 & 0.21 & 0.06 \\
	\midrule[0.3pt]\bottomrule[1pt]
\end{tabular}
\begin{tablenotes}
\item Notes: Superscript~``$*$'' signifies the use of bootstrap critical values. The asymptotic critical value for the self-normalized test $\tau_{\scriptstyle\text{IM}}(\hat{\eta}_{T})$ is given in Table~\ref{tab:critvals} in Online Appendix~\ref{app:critvalsdeter} ($167.23$; Panel A, $m=2$, $s=2$).
\end{tablenotes}
\end{threeparttable}}
\end{table}

\begin{figure}[H]
	\begin{center}
		\begin{subfigure}{0.4\textwidth}
			\centering
			\caption*{$T=100$, $\phi=0$}
			\vspace{-1ex}
			\includegraphics[trim={0cm 0cm 1cm 1cm},width=\textwidth,clip]{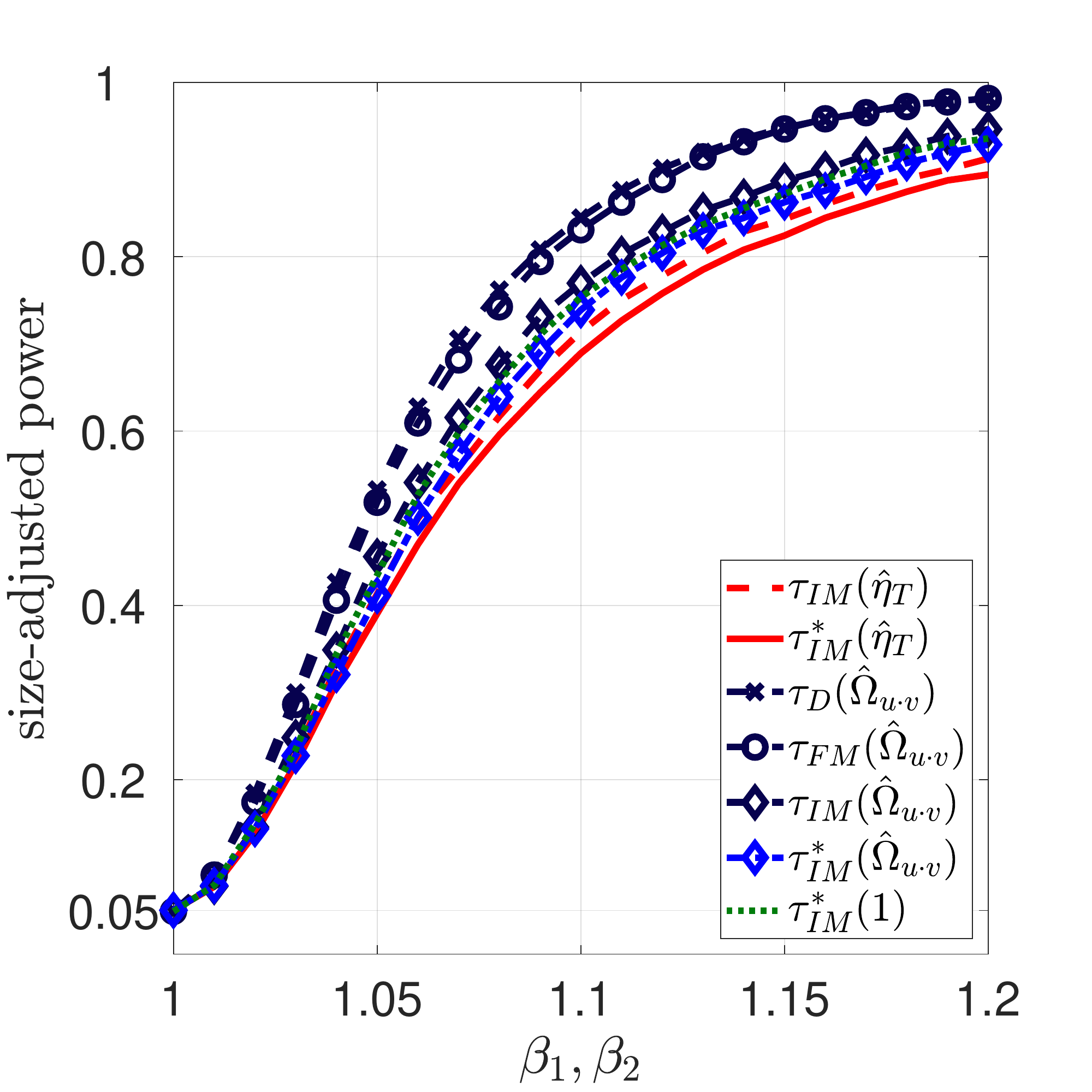}
		\end{subfigure}\begin{subfigure}{0.4\textwidth}
			\centering
			\caption*{$T=100$, $\phi=0.9$}
			\vspace{-1ex}
			\includegraphics[trim={0cm 0cm 1cm 1cm},width=\textwidth,clip]{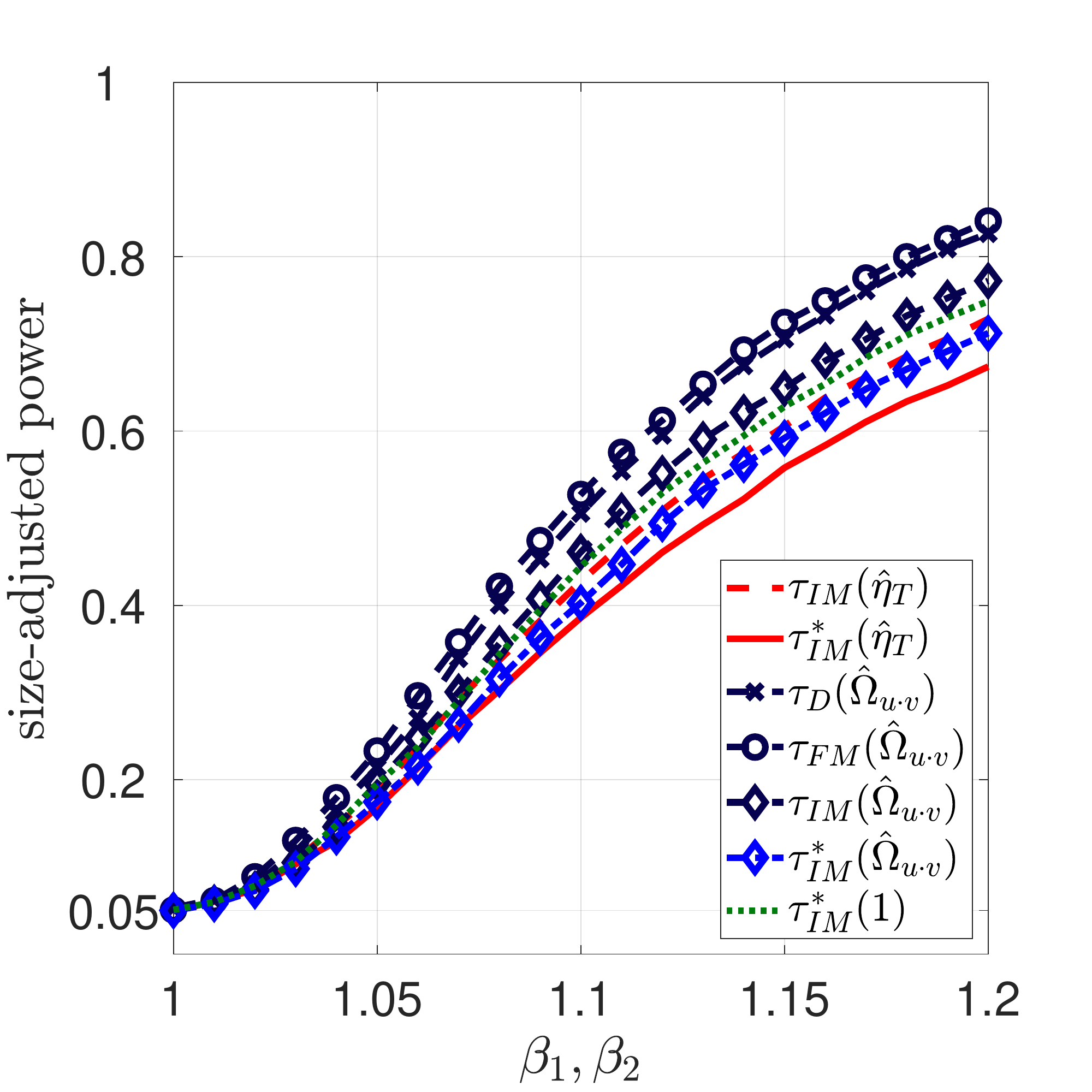}
		\end{subfigure}
		
		\vspace{2ex}
		
		\begin{subfigure}{0.4\textwidth}
			\centering
			\caption*{$T=250$, $\phi=0$}
			\vspace{-1ex}
			\includegraphics[trim={0cm 0cm 1cm 1cm},width=\textwidth,clip]{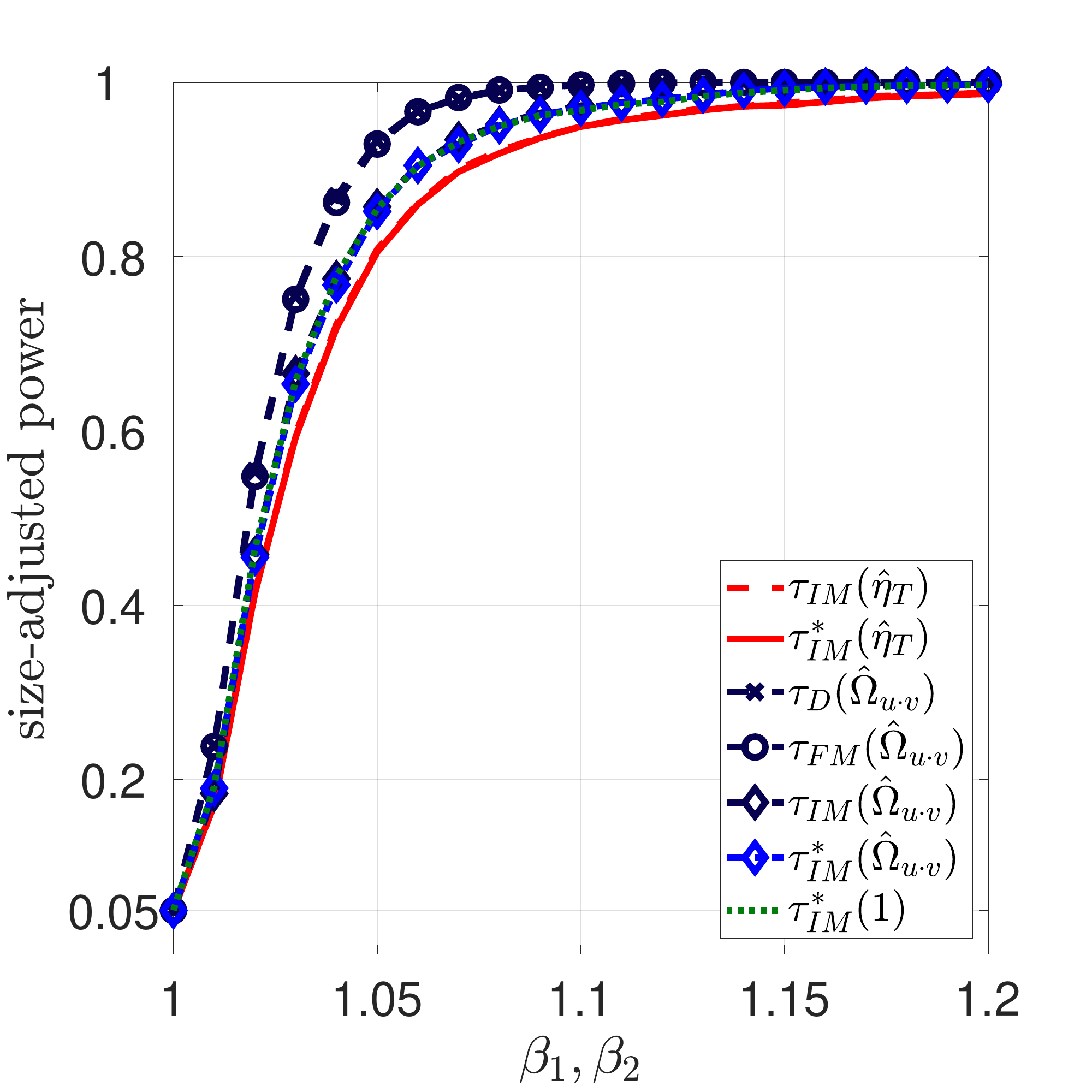}
		\end{subfigure}\begin{subfigure}{0.4\textwidth}
			\centering
			\caption*{$T=250$, $\phi=0.9$}
			\vspace{-1ex}
			\includegraphics[trim={0cm 0cm 1cm 1cm},width=\textwidth,clip]{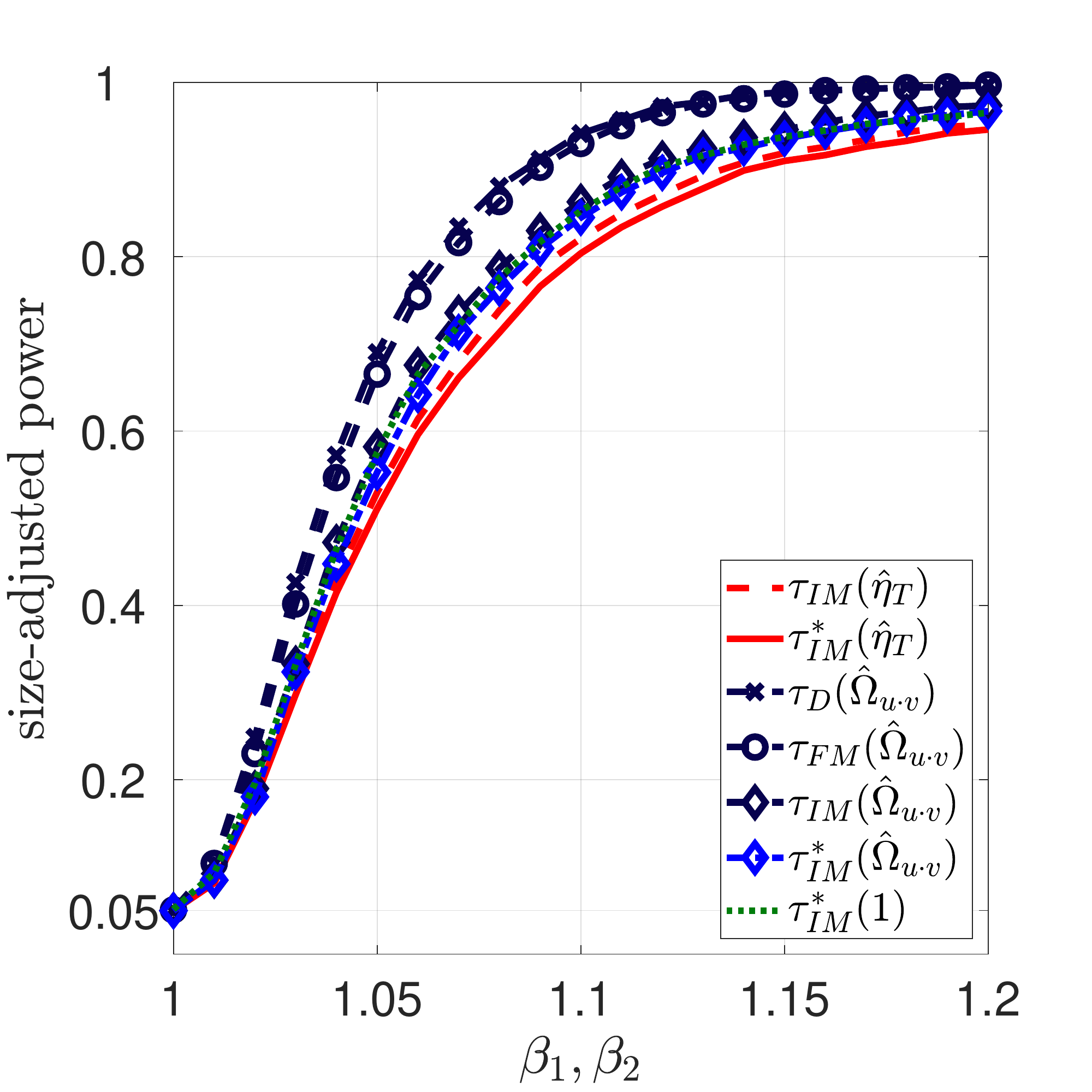}
		\end{subfigure}
	\end{center}
	\vspace{-2ex}
	\caption{Size-adjusted power of the tests for $\text{H}_0: \beta_1=1,\ \beta_2=1$ at 5\% level for $\rho_1,\rho_2=0.6$ in the i.i.d.~innovations case ($a_1=b_1=\rho_3=0$). Notes: Superscript~``$*$'' signifies the use of bootstrap critical values. Whenever necessary, we use the Bartlett kernel to estimate long-run variance parameters.}
	\label{fig:power_iid}
\end{figure}

\newpage
\clearpage

\subsection{Comparison with Johansen's Likelihood Ratio Test}\label{app:FiniteSampleJohansen}
We can rewrite the data generating process (DGP) given in~\eqref{eq:linMy} and~\eqref{eq:linMx} as
\begin{align*}
	\underbrace{\begin{bmatrix}
			1&-\beta_1&-\beta_2\\
			0&1&0\\
			0&0&1
	\end{bmatrix}}_{=:A_0}
	\begin{bmatrix}
		y_t\\
		x_{1t}\\
		x_{2t}
	\end{bmatrix}
	=
	\underbrace{\begin{bmatrix}
			0&0&0\\
			0&1&0\\
			0&0&1
	\end{bmatrix}}_{=:A_1}
	\begin{bmatrix}
		y_{t-1}\\
		x_{1,t-1}\\
		x_{2,t-1}
	\end{bmatrix}
	+
	\begin{bmatrix}
		u_t\\
		v_{1t}\\
		v_{2t}
	\end{bmatrix}
	\Leftrightarrow X_t = \Pi_1 X_{t-1} + \epsilon_t,
\end{align*}
where $X_t \coloneqq [y_t,x_{1t},x_{2t}]'$, $\epsilon_t\coloneqq A_0^{-1}[u_t,v_{1t},v_{2t}]'$ and $\Pi_1 \coloneqq A_0^{-1}A_1$. While the estimators and tests discussed so far ignore the exact characteristics of the DGP, the reduced rank quasi maximum likelihood (QML) estimator of \citeasnoun{Jo95} and the likelihood ratio (LR) test based upon it are tailor-made for this particular class of DGPs. Under this DGP, the performance of the \textit{parametric} LR test thus serves as a natural benchmark for the performance of the \textit{non-parametric} self-normalized test developed in this paper. To fix ideas, we consider the error correction model of order $k\geq 1$ with one cointegrated relation, \ie, 
\begin{align}\label{eq:ECM}
	\Delta X_t = a b' X_{t-1} + \sum_{l=1}^{k-1} \Gamma_l \Delta X_{t-l} + \epsilon_t,
\end{align}
where $\Delta X_s \coloneqq X_s - X_{s-1}$, $a=[-1,0,0]'$ and $b=[1,-\beta_1,-\beta_2]'$. To test general linear hypotheses on the cointegration vector $b$, we first determine the order $k$ in~\eqref{eq:ECM} similarly as described in the beginning of Section~\ref{sec:FiniteSample} for the order of the VAR sieve. We then estimate the vector $b$ using the reduced rank QML estimator of \citeasnoun{Jo95}. Testing the null hypothesis $\text{H}_0: \beta_1=1,\ \beta_2=1$ in~\eqref{eq:linMy} corresponds to testing the null hypothesis $\text{H}_0: b = [1,-1,-1]'$ in~\eqref{eq:ECM}. Under the null hypothesis, the LR test statistic of \citeasnoun{Jo95} is asymptotically chi-squared distributed with two degrees of freedom. \citeasnoun{CNR15} show how to replace the asymptotic chi-square critical values with bootstrap critical values to reduce the size distortions of the LR test in finite samples.

Before we compare the finite sample performance of the self-normalized test based on asymptotic critical values with the LR test based on asymptotic critical values and the bootstrap-assisted self-normalized test with the LR test based on bootstrap critical values generated as described in \citeasnoun[p.\,817]{CNR15}, we should mention one particular drawback of the reduced rank QML estimator. Table~\ref{tab:biasRMSE} in Online Appendix~\ref{app:FiniteSampleEstimation} shows that the QML estimator of $\beta_1$ has a very large bias for $\rho_1,\rho_2=0.9$ and an extremely large RMSE for $\rho_1,\rho_2\in\{0.6,0.9\}$ in small to medium samples.\footnote{Results for $\beta_2$ are similar.} These results are in line with the findings in related literature, where the QML estimator is observed to occasionally produce estimates that are far away from the true parameter values, see, \eg, \citeasnoun{BrLu05} and references therein. 

We now turn to the performance of the tests. Table~\ref{tab:sizes_with_LR} displays the empirical sizes of both the LR test and the LR test based on bootstrap critical values ($\text{LR}^*$). Comparing empirical sizes of the LR test with those of the self-normalized test in Table~\ref{tab:sizes}, we note that that the self-normalized test based on asymptotic critical values clearly outperforms the LR test for almost all combinations of $T$, $\rho_1,\rho_2$ and $\phi$. However, in case $\rho_1,\rho_2=0.9$ and $\phi$ is smaller than $0.9$ size distortions of the self-normalized test are slightly larger than those of the LR test. In terms of size-adjusted power we note that the self-normalized test outperforms the LR test in small to medium samples or in case $\rho_1,\rho_2=0.9$, see Figures~\ref{fig:powerLR_rho106} and~\ref{fig:powerLR_rho109}.

Replacing asymptotic critical values with bootstrap critical values improves the performance of the LR test in small to medium samples considerably. Comparing the empirical sizes of $\text{LR}^*$ in Table~\ref{tab:sizes_with_LR} with those of the bootstrap-assisted self-normalized test in Table~\ref{tab:sizes} reveals that the two bootstrap tests perform similarly for all sample sizes considered, but $\text{LR}^*$ has slight performance advantages over the bootstrap-assisted self-normalized test in small to medium samples in case $\rho_1,\rho_2=0.9$. However, this performance advantage of $\text{LR}^*$ over the bootstrap-assisted self-normalized test under the null hypothesis comes at the cost of some losses in size-adjusted power, as Figure~\ref{fig:powerLR_rho109} reveals. In case $\rho_1,\rho_2$ is large relative to the sample size, we observe that size-adjusted power of $\text{LR}^*$ is considerably smaller than size-adjusted power of the bootstrap-assisted self-normalized test. We also observe that the power difference between these tests can be much larger than the power difference between the bootstrap-assisted self-normalized test and the IM-OLS based Wald-type bootstrap test, compare Figure~\ref{fig:powerLR_rho106} in case $T=100$, $\phi=0.9$ and $\rho_1,\rho_2=0.6$ and Figure~\ref{fig:powerLR_rho109} in case $T=75$, $\phi\in\{0,0.9\}$ and $\rho_1,\rho_2=0.9$. For larger sample sizes, however, $\text{LR}^*$ has slightly larger size-adjusted power than the bootstrap-assisted self-normalized test and the IM-OLS based Wald-type bootstrap test, compare Figure~\ref{fig:powerLR_rho106} in case $T=250$.

We conclude that the (bootstrap-assisted) self-normalized test performs well both under the null hypothesis and under the alternative even when compared to the parametric likelihood-ratio (bootstrap) test tailor made for the DGP under consideration. Therefore, in applications focusing on a particular cointegrating relation between the variables, the IM-OLS estimator and the (bootstrap-assisted) self-normalized test are a good alternative to the QML estimator and the likelihood-ration (bootstrap) test, especially when the QML estimator yields implausible estimates of the cointegrating vector.

\begin{table}[!t]
	\centering
	\adjustbox{max width=\textwidth}{\begin{threeparttable}
			\caption{Empirical sizes of the LR test and the LR bootstrap test for $\text{H}_0: \beta_1=1,\ \beta_2=1$ at $5\%$ level}
			\label{tab:sizes_with_LR}
			\begin{tabular}{cccccccccc}
				\toprule[1pt]\midrule[0.3pt]
				\multicolumn{2}{c}{}&\multicolumn{2}{c}{$T=75$}&\multicolumn{2}{c}{$T=100$}&\multicolumn{2}{c}{$T=250$}&\multicolumn{2}{c}{$T=500$}\\
				\cmidrule(lr){3-4}
				\cmidrule(lr){5-6}
				\cmidrule(lr){7-8}
				\cmidrule(lr){9-10}
				$\phi$&$\rho_1,\rho_2$ & $\text{LR}$ & $\text{LR}^*$ & $\text{LR}$ & $\text{LR}^*$& $\text{LR}$ & $\text{LR}^*$& $\text{LR}$ & $\text{LR}^*$\\
				\midrule
				$0$&0 &  0.13 & 0.06 & 0.10 & 0.06 & 0.07 & 0.06& 0.07 & 0.06\\
				&0.3 &  0.14 & 0.06 & 0.11 & 0.06 & 0.08 & 0.06& 0.07 & 0.06\\
				&0.6 &  0.16 & 0.07 & 0.13 & 0.06& 0.09 & 0.06& 0.07 & 0.06\\
				&0.9 &  0.27 & 0.08 &0.20 & 0.07& 0.11 & 0.07& 0.08 & 0.06\\\midrule
				$0.3$&0 &  0.14 & 0.07 & 0.11 & 0.06& 0.08 & 0.05& 0.07 & 0.06\\
				&0.3 &  0.15 & 0.06 & 0.12 & 0.07& 0.09 & 0.07& 0.07 & 0.06\\
				&0.6 &  0.19 & 0.07 & 0.15 & 0.08& 0.09 & 0.07& 0.07 & 0.05\\
				&0.9 &  0.32 & 0.10 & 0.25 & 0.09& 0.12 & 0.07& 0.08 & 0.06\\\midrule
				$0.9$&0 &  0.18 & 0.09 & 0.14 & 0.08& 0.09 & 0.07& 0.07 & 0.06\\
				&0.3 &  0.20 & 0.08 & 0.15 & 0.08& 0.10 & 0.06& 0.07 & 0.06\\
				&0.6 &  0.23 & 0.09 & 0.18 & 0.07& 0.10 & 0.06& 0.09 & 0.07\\
				&0.9 &  0.39 & 0.09 & 0.31 & 0.09& 0.14 & 0.07& 0.12 & 0.08\\
				\midrule[0.3pt]\bottomrule[1pt]
			\end{tabular}
			\begin{tablenotes}
				\item Note: Superscript~``$*$'' signifies the use of bootstrap critical values.
			\end{tablenotes}
	\end{threeparttable}}
\end{table}

\begin{figure}[!t]
	\begin{center}
		\begin{subfigure}{0.4\textwidth}
			\centering
			\caption*{$T=100$, $\phi=0$}
			\vspace{-1ex}
			\includegraphics[trim={0cm 0cm 1cm 1cm},width=\textwidth,clip]{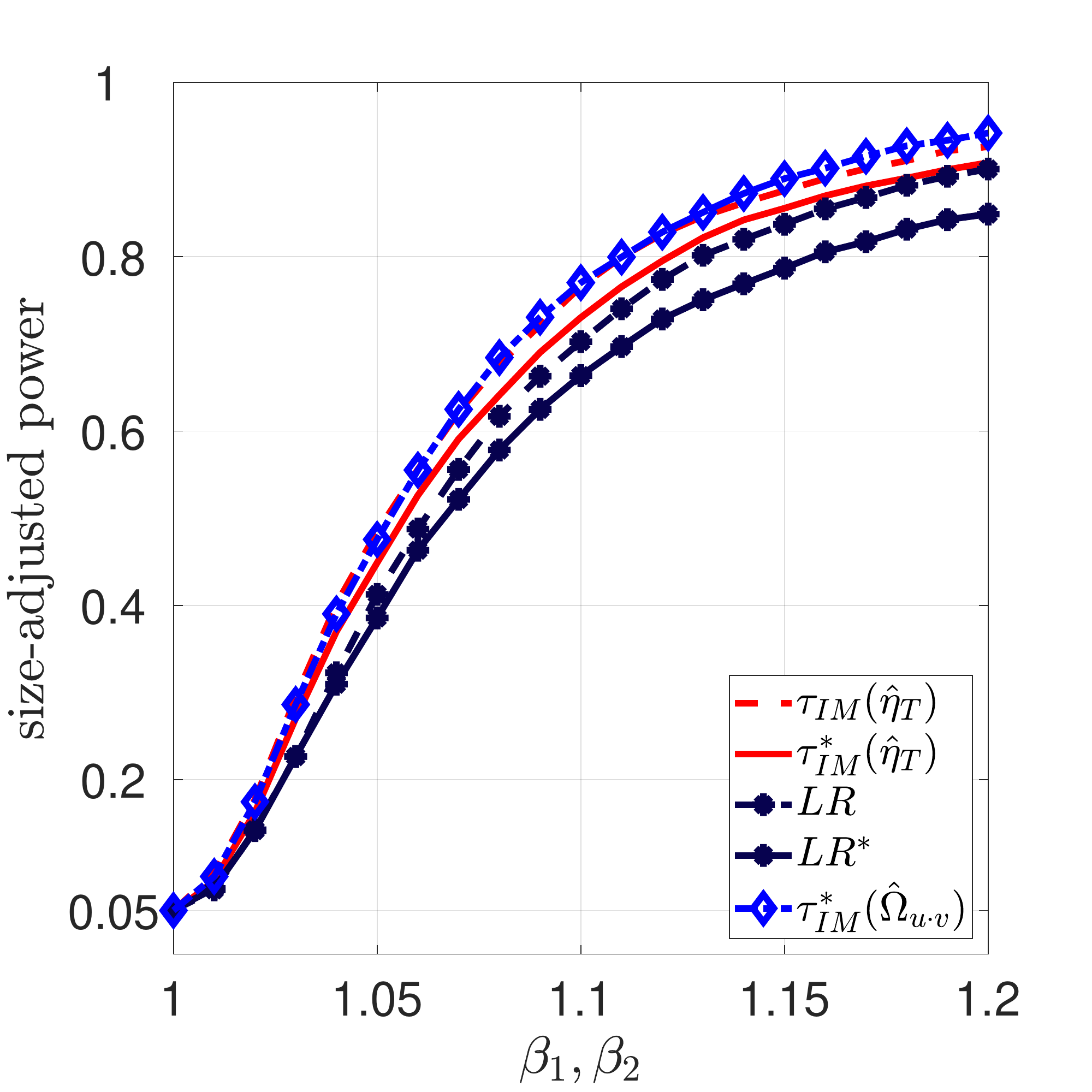}
		\end{subfigure}\begin{subfigure}{0.4\textwidth}
			\centering
			\caption*{$T=100$, $\phi=0.9$}
			\vspace{-1ex}
			\includegraphics[trim={0cm 0cm 1cm 1cm},width=\textwidth,clip]{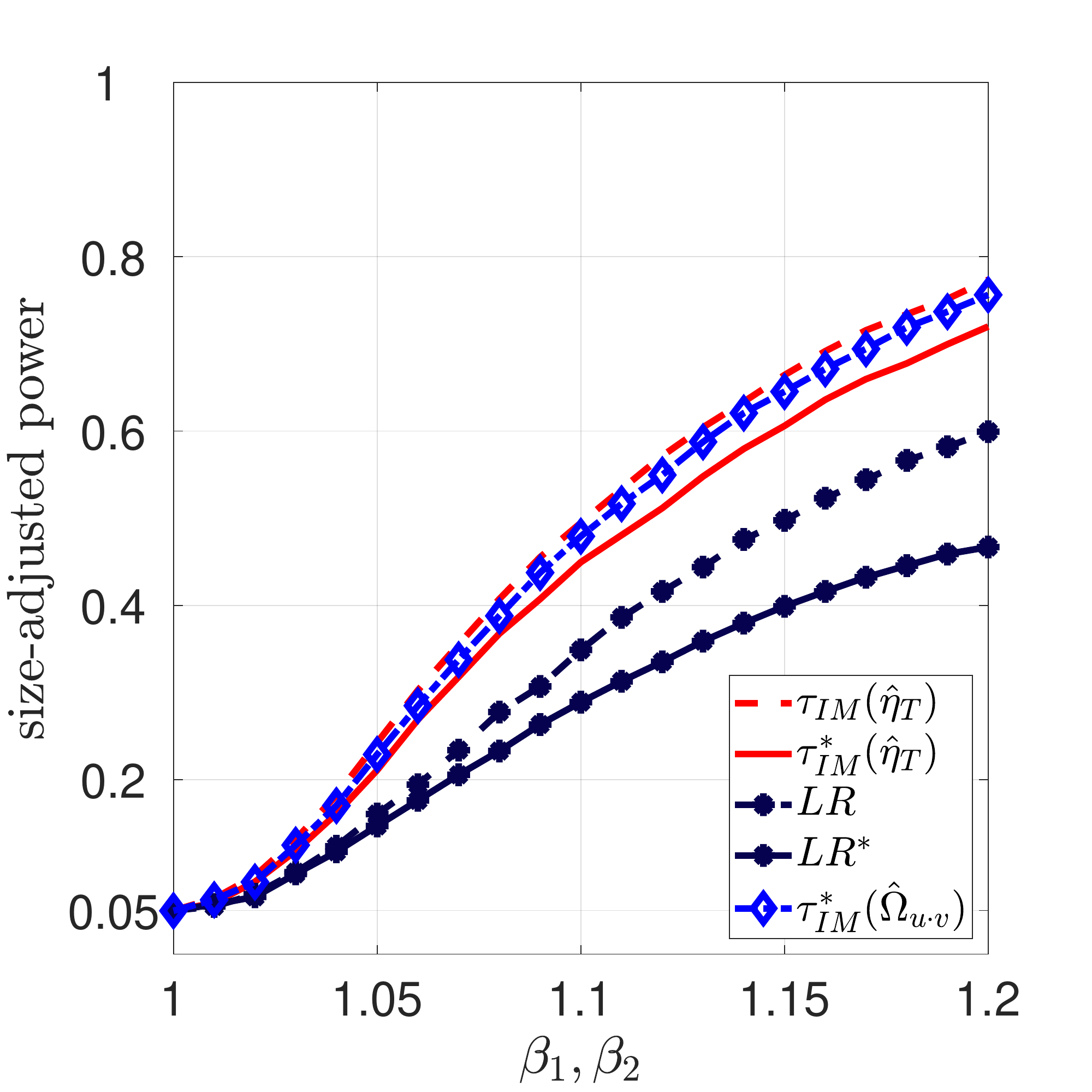}
		\end{subfigure}
		
		\vspace{2ex}
		
		\begin{subfigure}{0.4\textwidth}
			\centering
			\caption*{$T=250$, $\phi=0$}
			\vspace{-1ex}
			\includegraphics[trim={0cm 0cm 1cm 1cm},width=\textwidth,clip]{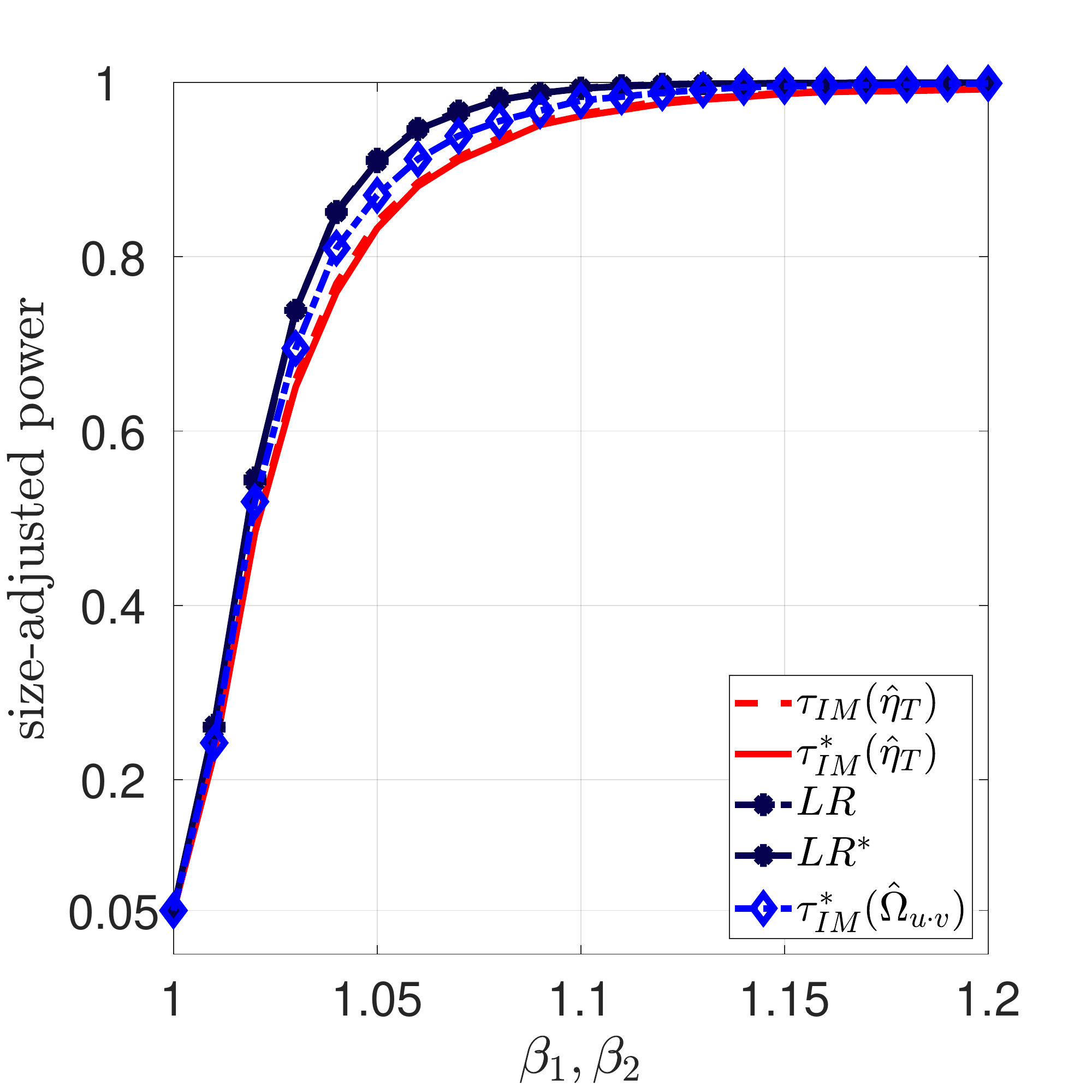}
		\end{subfigure}\begin{subfigure}{0.4\textwidth}
			\centering
			\caption*{$T=250$, $\phi=0.9$}
			\vspace{-1ex}
			\includegraphics[trim={0cm 0cm 1cm 1cm},width=\textwidth,clip]{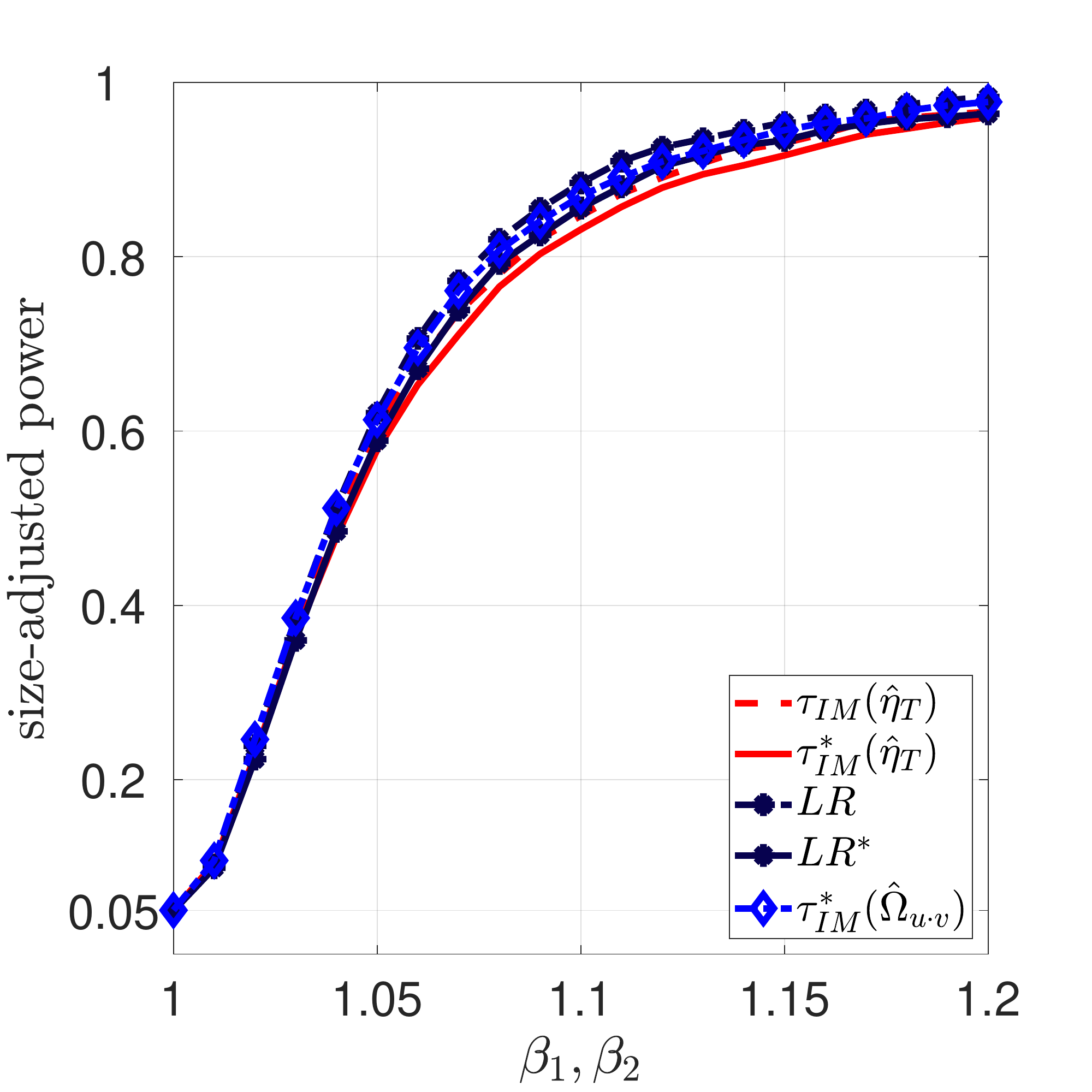}
		\end{subfigure}
		
	\end{center}
	\vspace{-2ex}
	\caption{Size-adjusted power of the tests for $\text{H}_0: \beta_1=1,\ \beta_2=1$ at 5\% level for $\rho_1,\rho_2=0.6$. Notes: Superscript~``$*$'' signifies the use of bootstrap critical values. Results for the (bootstrap-assisted) self-normalized test and the IM-OLS Wald-type bootstrap tests coincide with those in Figure~\ref{fig:power}.}
	\label{fig:powerLR_rho106}
\end{figure}

\begin{figure}[!t]
	\begin{center}
		\begin{subfigure}{0.4\textwidth}
			\centering
			\caption*{$T=75$, $\phi=0$}
			\vspace{-1ex}
			\includegraphics[trim={0cm 0cm 1cm 1cm},width=\textwidth,clip]{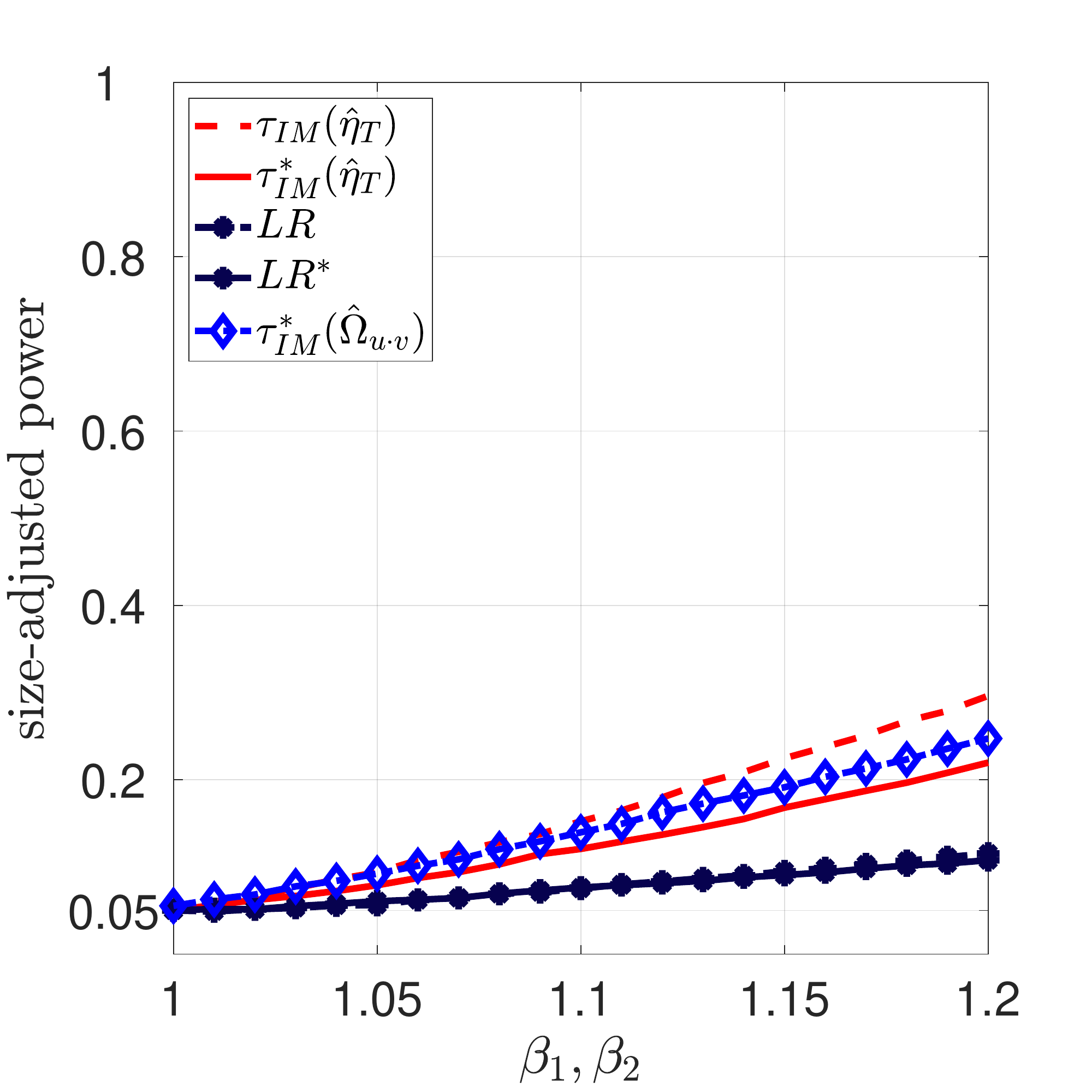}
		\end{subfigure}\begin{subfigure}{0.4\textwidth}
			\centering
			\caption*{$T=75$, $\phi=0.9$}
			\vspace{-1ex}
			\includegraphics[trim={0cm 0cm 1cm 1cm},width=\textwidth,clip]{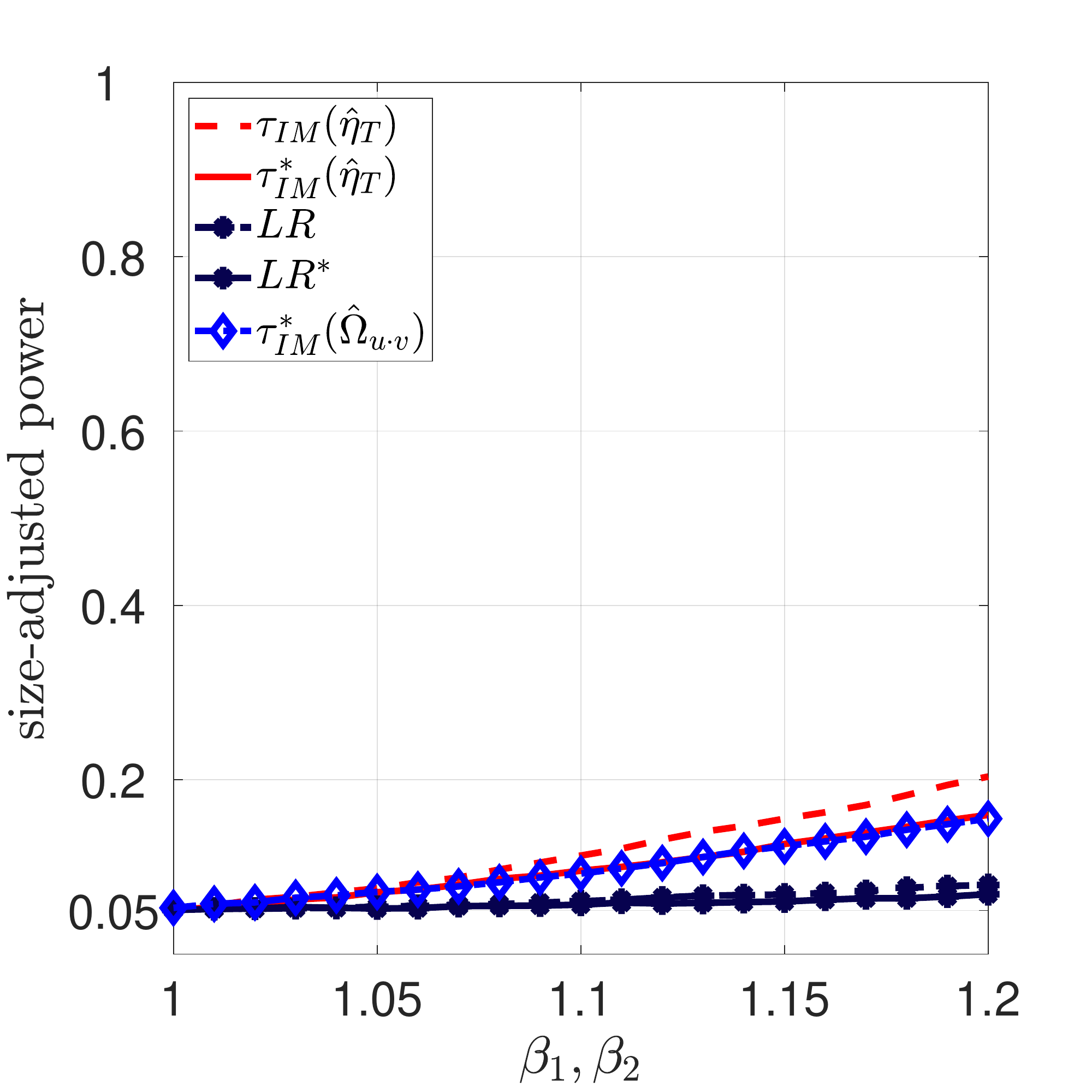}
		\end{subfigure}
		
		\vspace{2ex}
		
		\begin{subfigure}{0.4\textwidth}
			\centering
			\caption*{$T=100$, $\phi=0$}
			\vspace{-1ex}
			\includegraphics[trim={0cm 0cm 1cm 1cm},width=\textwidth,clip]{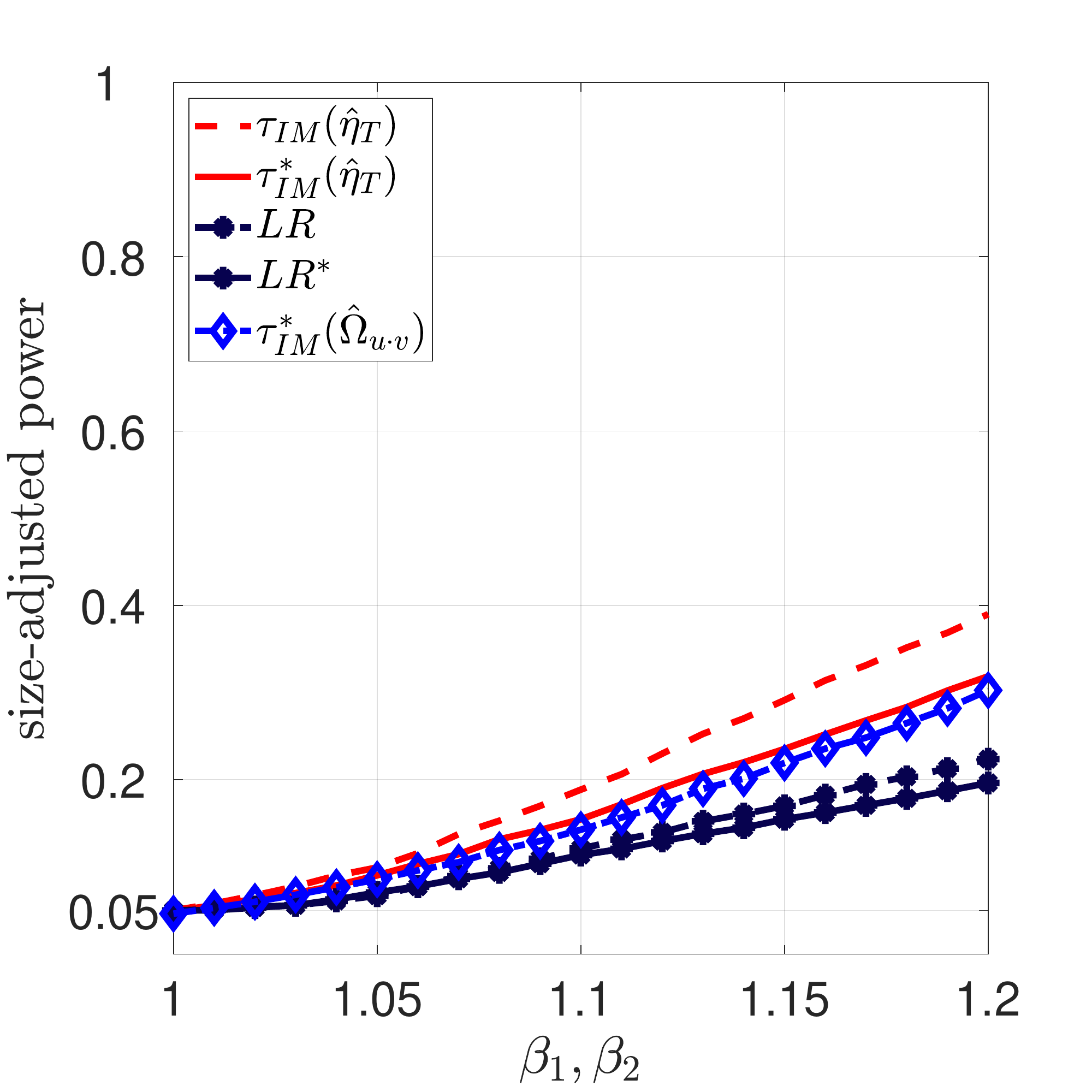}
		\end{subfigure}\begin{subfigure}{0.4\textwidth}
			\centering
			\caption*{$T=100$, $\phi=0.9$}
			\vspace{-1ex}
			\includegraphics[trim={0cm 0cm 1cm 1cm},width=\textwidth,clip]{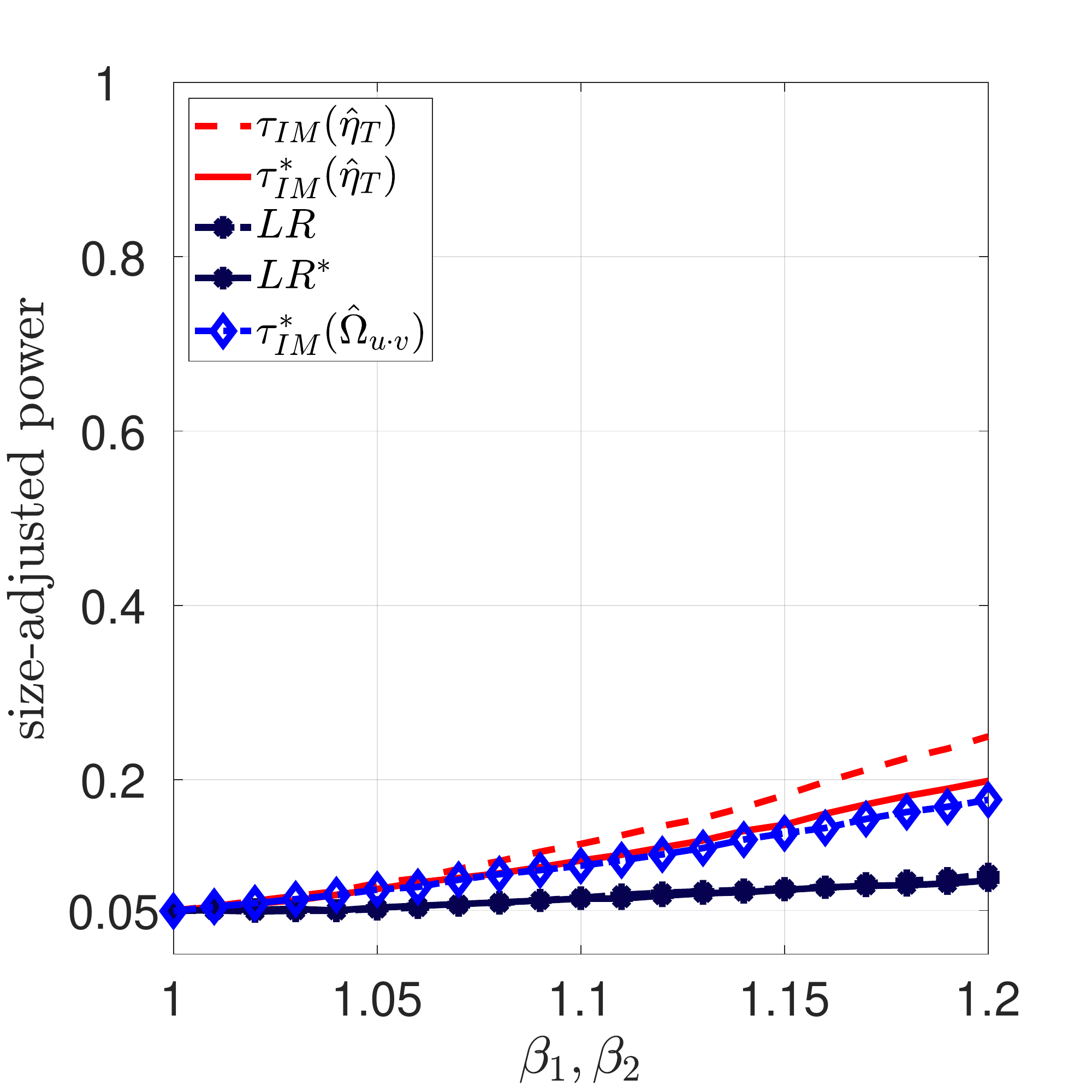}
		\end{subfigure}
		
	\end{center}
	\vspace{-2ex}
	\caption{Size-adjusted power of the tests for $\text{H}_0: \beta_1=1,\ \beta_2=1$ at 5\% level for $\rho_1,\rho_2=0.9$. Note: Superscript~``$*$'' signifies the use of bootstrap critical values. Results for the IM-OLS based Wald-type bootstrap test are based on the Bartlett kernel.}
	\label{fig:powerLR_rho109}
\end{figure}

\newpage
\clearpage

\section{Asymptotic Critical Values}\label{app:critvalsdeter}

To simulate asymptotically valid critical values, we approximate standard Brownian motions with normalized sums of $10{,}000$ i.i.d.~standard normal random variables and approximate the corresponding integrals accordingly. The fact that the numerator of the limiting null distribution of the self-normalized test statistic given in~\eqref{eq:Geta} is invariant to $\Pi$ and only depends on the number of linearly independent restrictions under the null hypothesis justifies to set $\Pi = I_{2m}$ and $R_2=[I_s,0_{s\times (2m-s)}]$.\footnote{Note that the dependence between numerator and denominator in~\eqref{eq:Geta} implies that simply drawing from a chi-square distribution for the numerator is invalid.} We tabulate critical values based on $10{,}000$ replications for various choices of $m$ and $s$ and different deterministic regressors in~\eqref{eq:y} in Table~\ref{tab:critvals}.

\begin{table}[!ht]
	\centering
	\adjustbox{max width=0.85\textwidth}{\begin{threeparttable}
			\caption{Asymptotic critical values for $\tau_{\scriptstyle\text{IM}}(\hat \eta_T)$}
			\label{tab:critvals}
			\begin{tabular}{ccccccccccc}
				\toprule[1pt]\midrule[0.3pt]
				\multicolumn{1}{c}{}&\multicolumn{1}{c}{$m=1$}&\multicolumn{2}{c}{$m=2$}&\multicolumn{3}{c}{$m=3$}&\multicolumn{4}{c}{$m=4$}\\
				\cmidrule(lr){2-2}
				\cmidrule(lr){3-4}
				\cmidrule(lr){5-7}
				\cmidrule(lr){8-11}
				$\%$ & $s=1$ & $s=1$ & $s=2$ & $s=1$ & $s=2$ & $s=3$ & $s=1$ & $s=2$ & $s=3$ & $s=4$ \\
				\midrule
				\multicolumn{11}{l}{Panel A: No deterministic regressors}\\
				\midrule
				90.0 & 36.63 & 66.33 & 122.32 & 94.04 & 172.00 & 240.58 & 131.68 & 232.77 & 318.25 & 402.61 \\
				95.0 & 56.58 & 96.51 & 167.23 & 140.69 & 231.79 & 313.46 & 189.15 & 309.06 & 407.17 & 510.60 \\
				97.5 & 79.24 & 131.79 & 216.99 & 191.68 & 290.47 & 390.38 & 256.38 & 390.07 & 504.08 & 630.19 \\
				99.0 & 120.10 & 189.69 & 286.97 & 266.16 & 375.30 & 494.00 & 355.25 & 505.21 & 645.89 & 767.61 \\
				\midrule
				\multicolumn{11}{l}{Panel B: Intercept ($d_t=1$)}\\
				\midrule
				90.0 & 64.13 & 94.15 & 168.58 & 126.65 & 221.45 & 305.36 & 162.08 & 278.05 & 382.30 & 481.15 \\
				95.0 & 95.81 & 140.55 & 233.15 & 187.03 & 297.11 & 396.56 & 236.54 & 372.79 & 487.71 & 596.15 \\
				97.5 & 136.10 & 190.23 & 292.64 & 245.93 & 375.55 & 488.35 & 325.56 & 458.37 & 587.85 & 720.31 \\
				99.0 & 187.13 & 263.92 & 381.78 & 338.59 & 474.31 & 602.27 & 421.68 & 582.89 & 719.98 & 872.07 \\
				\midrule
				\multicolumn{11}{l}{Panel C: Intercept and linear time trend ($d_t=[1,t]'$)}\\
				\midrule
				90.0 & 90.44 & 122.19 & 209.54 & 152.66 & 261.47 & 363.17 & 180.25 & 311.22 & 434.29 & 545.37 \\
				95.0 & 134.19 & 171.46 & 283.33 & 219.51 & 354.08 & 460.37 & 258.75 & 423.39 & 546.31 & 688.21 \\
				97.5 & 183.51 & 231.09 & 357.66 & 294.26 & 433.33 & 569.84 & 342.56 & 524.25 & 686.44 & 810.09 \\
				99.0 & 243.72 & 304.08 & 460.98 & 409.03 & 556.42 & 713.24 & 478.05 & 680.76 & 821.12 & 977.09 \\
				\midrule
				\multicolumn{11}{l}{Panel D: Intercept, linear and square time trend ($d_t=[1,t,t^2]'$)}\\
				\midrule
				90.0 & 115.13 & 138.49 & 245.91 & 175.40 & 302.95 & 418.77 & 205.72 & 352.96 & 479.57 & 608.35 \\
				95.0 & 166.35 & 200.65 & 331.26 & 255.74 & 402.90 & 530.94 & 303.58 & 465.28 & 621.70 & 764.20 \\
				97.5 & 217.42 & 268.86 & 401.63 & 348.51 & 509.51 & 637.29 & 390.59 & 589.05 & 762.57 & 902.89 \\
				99.0 & 290.63 & 357.58 & 513.85 & 472.49 & 646.48 & 800.81 & 527.20 & 754.26 & 923.59 & 1070.32 \\
				\midrule
				\multicolumn{11}{l}{Panel E: Intercept, linear, square and cubic time trend ($d_t=[1,t,t^2,t^3]'$)}\\
				\midrule
				90.0 & 137.70 & 166.87 & 292.13 & 197.84 & 340.61 & 465.58 & 229.38 & 392.80 & 533.60 & 680.84 \\
				95.0 & 198.48 & 237.82 & 379.15 & 288.65 & 446.27 & 590.05 & 334.55 & 509.11 & 684.33 & 858.04 \\
				97.5 & 263.30 & 308.64 & 467.71 & 391.70 & 565.65 & 720.19 & 438.56 & 645.41 & 853.07 & 1004.50 \\
				99.0 & 352.56 & 406.48 & 587.03 & 539.71 & 726.07 & 903.53 & 592.44 & 846.82 & 1052.20 & 1222.78 \\
				\midrule[0.3pt]\bottomrule[1pt]
			\end{tabular}
			\begin{tablenotes}
				\item Notes: $s$ is the number of linearly independent restrictions under the null hypothesis on the coefficients corresponding to the $m\geq s$ integrated regressors.
			\end{tablenotes}
	\end{threeparttable}}
\end{table}

\newpage
\clearpage

\section{Proofs of Main Results}\label{app:mainproofs}
\setlength{\parindent}{0mm}
{\bf Proof of Theorem~\ref{thm:SN}.} \citeasnoun[Lemma~2]{VoWa14} show that
%
\begin{align*}
	T^{-1/2} \sum_{t=2}^{\floor{rT}} \Delta\hat{S}_t^u \overset{w}{\longrightarrow} \Omega_{u \cdot v}^{1/2}\left(W_{u\cdot v}(r) - g(r)'\mathcal{Z}\right),\quad 0\leq r \leq 1,
\end{align*}
as $T\rightarrow \infty$. The continuous mapping theorem thus yields
\begin{align*}
	\hat \eta_T = T^{-1} \sum_{t=2}^T \left( T^{-1/2} \sum_{s=2}^t \Delta\hat{S}_s^u \right)^2 \overset{w}{\longrightarrow} \Omega_{u \cdot v} \int_0^1 \left(W_{u\cdot v}(r) - g(r)'\mathcal{Z}\right)^2 dr,
\end{align*}
as $T\rightarrow \infty$. The final result now follows with standard arguments from~\eqref{eq:G1}. \hfill$\square$\\

The proof of the remaining results relies on the following lemma.
\begin{lemma}\label{lem:hat-tilde}
	Under Assumptions~\ref{ass:w}, \ref{ass:cumulants} and \ref{ass:q}, it holds that
	\begin{align}\label{eq:maxw}
		\max_{q+1\leq t \leq T} \vert \hat w_t - w_t \vert_F = O_\mP(T^{-1/2})
	\end{align}
	and 
	\begin{align}\label{eq:diffPhi}
		q^{1/2}\sum_{j=1}^q \vert \hat \Phi_j(q) - \tilde \Phi_j(q)\vert_F= O_\mP(q^3/T)=o_\mP(1),
	\end{align}
	where $\tilde \Phi_1(q),\ldots,\tilde \Phi_q(q)$ denote the solution of the sample Yule-Walker equations in the regression of $w_t$ on $w_{t-1},\ldots,w_{t-q}$, $t=q+1,\ldots,T$.
\end{lemma}
\begin{proof}
	We provide a proof in Online Appendix~\ref{app:auxproofs}.
\end{proof}
Two key ingredients in the proof of Lemma~\ref{lem:hat-tilde}, which are also useful hereafter, are the following: First, it holds under Assumptions~\ref{ass:w} and~\ref{ass:q} that
\begin{align}\label{eq:supPhi}
	q^{3/2} \sup_{1\leq j \leq q}\vert \tilde \Phi_j(q) - \Phi_j(q)\vert_F = q^{3/2} O_\mP((\ln(T)/T)^{1/2}) = O_\mP(1),
\end{align}
compare \citeasnoun[Remark 3.3.]{MeKr15}, where $\Phi_1(q),\ldots,\Phi_q(q)$ denote the finite predictor coefficients, \ie, the solution of the population Yule-Walker equations based on the true moments. Second, under Assumption~\ref{ass:w} with $k\geq 3/2$, there exist constants $q_0\in \mN$ and $c< \infty$ such that
\begin{align}\label{eq:Baxter}
	\sum_{j=1}^q (1+j)^k \vert \Phi_j(q) - \Phi_j \vert_F \leq c \sum_{j=q+1}^\infty (1+j)^k \vert \Phi_j\vert_F,
\end{align}
for all $q\geq q_0$ and the right-hand side converges to zero as $q\rightarrow \infty$, see \citeasnoun[Lemma 3.1]{MeKr15}.\footnote{This result is known as the generalized Baxter's inequality, see \citeasnoun{Ba62} and \citeasnoun[p.\,269]{HaDe88}.}\\

We use Lemma~\ref{lem:hat-tilde} to prove the following two lemmas. Subsequently, we use the following two lemmas to prove Proposition~\ref{prop:Wstar}.
\begin{lemma}\label{lem:moments}
	It holds under Assumptions~\ref{ass:w} and \ref{ass:q} that
	\begin{align*}
		\mE^*\left(\vert \ve_t^*\vert_F^a\right) = (T-q)^{-1}\sum_{t=q+1}^T \vert \hat \ve_t(q) - \bar{\hat{\ve}}_T(q)\vert_F^a = O_\mP(1),
	\end{align*}
	in $\mP$, for the $a>2$ from Assumption~\ref{ass:w}.
\end{lemma}
\begin{proof}
	We provide a proof in Online Appendix~\ref{app:auxproofs}.
\end{proof}
\begin{lemma}\label{lem:Sigmastar}
	It holds under Assumptions~\ref{ass:w}, \ref{ass:cumulants} and \ref{ass:q} that
	\begin{align*}
		\mE^*\left(\ve_t^* \ve_t^{*\prime}\right) = (T-q)^{-1}\sum_{t=q+1}^T \left(\hat \ve_t(q) - \bar{\hat{\ve}}_T(q)\right)\left(\hat \ve_t(q) - \bar{\hat{\ve}}_T(q)\right)' = \Sigma + o_\mP(1),
	\end{align*}
	in $\mP$.
\end{lemma}
\begin{proof}
	We provide a proof in Online Appendix~\ref{app:auxproofs}.
\end{proof}

{\bf Proof of Proposition~\ref{prop:Wstar}.} Given Lemma~\ref{lem:moments} and Lemma~\ref{lem:Sigmastar}, the result follows immediately from \citeasnoun{Ei87}, as in \citeasnoun[p.\,714]{CPS06}.\footnote{For more details we refer to the pre-print of \citeasnoun{PSU10}, which is available on \href{https://www.stephansmeekes.nl/research}{https://www.stephansmeekes.nl/research} (Accessed: March 24, 2022).}\hfill$\square$\\

{\bf Proof of Theorem~\ref{Thm:Bstar}.} Using similar arguments as \citeasnoun[p.\,670]{PSU10}, it follows that
%
%
\begin{align*}
	B_T^*(r)= T^{-1/2} \sum_{t=1}^{\floor{rT}} w_t^* = \left(I - \sum_{j=1}^q \hat \Phi_j(q)\right)^{-1} W_T^*(r) + T^{-1/2} (\bar w_0^* - \bar w_{\floor{rT}}^*),
\end{align*}
where $\bar w_{t-1}^*\coloneqq \left(I - \sum_{j=1}^q \hat \Phi_j(q)\right)^{-1}\sum_{i=1}^q\left(\sum_{j=i}^q \hat \Phi_j(q)\right) w_{t-i}^*$. It thus remains to show that
\begin{align}\label{eq:ThmBstar-1}
	I - \sum_{j=1}^q \hat \Phi_j(q) \overset{p}{\longrightarrow} \Phi(1)
\end{align}
and
\begin{align}\label{eq:ThmBstar-2}
	\mP^*\left( \max_{0\leq t\leq T} \vert T^{-1/2} \bar w_t^* \vert_F > \delta \right)=o_\mP(1).
\end{align}
We first show~\eqref{eq:ThmBstar-1}. From Lemma~\ref{lem:hat-tilde},~\eqref{eq:supPhi} and~\eqref{eq:Baxter}, we obtain
\begin{align*}
	\vert I - \sum_{j=1}^q \hat \Phi_j(q) - \Phi(1)\vert_F &\leq \sum_{j=1}^q \vert \hat \Phi_j(q) - \tilde\Phi_j(q) \vert_F + \sum_{j=1}^q \vert \tilde \Phi_j(q) - \Phi_j(q) \vert_F \\
	&\quad + \sum_{j=1}^q \vert \Phi_j(q) - \Phi_j \vert_F + \sum_{j=q+1}^\infty \vert \Phi_j \vert_F\\
	& \leq \sum_{j=1}^q \vert \hat \Phi_j(q) - \tilde\Phi_j(q) \vert_F + q\sup_{1\leq j \leq q}\vert \tilde \Phi_j(q)-\Phi_j(q)\vert_F \\
	&\quad+ c \sum_{j=q+1}^\infty \vert \Phi_j\vert_F + \sum_{j=q+1}^\infty \vert \Phi_j \vert_F\\
	& = o_\mP(1) + o_\mP(1) + o(1) + o(1) = o_\mP(1).
\end{align*}
To prove~\eqref{eq:ThmBstar-2}, we note that it follows from strict stationarity of $\{\bar w_t^*\}_{t\in \mZ}$ and Markov's inequality, that
\begin{align*}
	\mP^*\left( \max_{0\leq t\leq T} \vert T^{-1/2} \bar w_t^* \vert_F > \delta \right) 
	& \leq \sum_{t=0}^T \mP^*\left( \vert T^{-1/2} \bar w_t^* \vert_F > \delta \right)\\
	&\leq (T+1) \mP^*\left( \vert T^{-1/2} \bar w_t^* \vert_F > \delta \right)\\
	&\leq \delta^{-a}(T^{1-a/2}+T^{-a/2}) \mE^*\left(\vert \bar w_t^* \vert_F^a \right),
\end{align*}
with the $a>2$ from Assumption~\ref{ass:w}, compare \citeasnoun[p.\,486]{Pa02}. Similarly as in \citeasnoun[p.\,671]{PSU10}, we obtain
\begin{align*}
	\mE^*\left(\vert \bar w_t^* \vert_F^a \right) \leq c\left(m+1\right)^{a/2-1} \left(\sum_{j=0}^\infty \vert \bar{\hat\Psi}_j(q)\vert_F^2\right)^{a/2} \mE^*\left(\vert \ve_t^* \vert_F^a \right),
\end{align*}
for some constant $c$ and $\bar{\hat\Psi}_j(q) \coloneqq  \sum_{i=j+1}^\infty \hat \Psi_i(q)$, where the matrices $\hat \Psi_j$ are determined by the power series expansion of the inverse of $I-\sum_{j=1}^q \hat \Phi_j(q)z^j$. As discussed in \citeasnoun[p.\,671]{PSU10} it follows that $\sum_{j=0}^\infty \vert \bar{\hat\Psi}_j(q)\vert_F^2=O_\mP(1)$ if we can show that $\sum_{j=1}^q j^{1/2}\vert \hat\Psi_j(q)\vert_F =  O_\mP(1)$, which in turn holds if $\sum_{j=1}^q j^{1/2}\vert \hat\Phi_j(q)\vert_F =  O_\mP(1)$. Using again Lemma~\ref{lem:hat-tilde},~\eqref{eq:supPhi} and~\eqref{eq:Baxter}, we obtain
\begin{align*}
	\sum_{j=1}^q j^{1/2}\vert \hat\Phi_j(q)\vert_F 
	&\leq \sum_{j=1}^q j^{1/2}\vert \hat\Phi_j(q) - \tilde \Phi_j(q)\vert_F + \sum_{j=1}^q j^{1/2}\vert \tilde\Phi_j(q) - \Phi_j(q)\vert_F \\
	& \quad + \sum_{j=1}^q j^{1/2}\vert \Phi_j(q) - \Phi_j\vert_F + \sum_{j=1}^q j^{1/2}\vert \Phi_j\vert_F\\
	&\leq q^{1/2} \sum_{j=1}^q \vert \hat\Phi_j(q) - \tilde \Phi_j(q)\vert_F + q^{1/2}\sum_{j=1}^q \vert \tilde\Phi_j(q) - \Phi_j(q)\vert_F\\
	&\quad + \sum_{j=1}^q (1+j)\vert \Phi_j(q) - \Phi_j\vert_F + \sum_{j=1}^q j^{1/2}\vert \Phi_j\vert_F\\
	&\leq q^{1/2} \sum_{j=1}^q \vert \hat\Phi_j(q) - \tilde \Phi_j(q)\vert_F + q^{3/2} \sup_{1\leq j \leq q} \vert \tilde\Phi_j(q) - \Phi_j(q)\vert_F\\
	&\quad + \sum_{j=1}^q (1+j)\vert \Phi_j(q) - \Phi_j\vert_F + \sum_{j=1}^q j^{1/2}\vert \Phi_j\vert_F\\
	& = o_\mP(1) + O_\mP(1) + o(1) + O(1) = O_\mP(1).
\end{align*}
This completes the proof, since $\mE^*\left(\vert \ve_t^* \vert_F^a \right)= O_\mP(1)$ by Lemma~\ref{lem:moments} for the $a>2$ from Assumption~\ref{ass:w}.~\hfill$\square$\\

{\bf Proof of Theorem~\ref{Thm:IMstar}.} The first result in Theorem~\ref{Thm:IMstar} follows from the bootstrap invariance principle result in Theorem~\ref{Thm:Bstar} and similar arguments as used in \citeasnoun[Proof of Theorem~2]{VoWa14}. The second result then follows from the bootstrap invariance principle and similar arguments as used in the proof of Theorem~\ref{thm:SN}.~\hfill$\square$

\section{Proofs of Auxiliary Results}\label{app:auxproofs}
\setlength{\parindent}{0mm}
{\bf Proof of Lemma~\ref{lem:hat-tilde}.} The solution of the sample Yule-Walker equations in the regression of $w_t$ on $w_{t-1},\ldots,w_{t-q}$, $t=q+1,\ldots,T$, can be written in compact form as a ($(m+1)\times q(m+1)$)-dimensional matrix
\begin{align*}
	\boldsymbol{\tilde \Phi}(q)\coloneqq [\tilde \Phi_1(q),\ldots,\tilde \Phi_q(q)] = \tilde \Gamma \tilde G^{-1},
\end{align*}
with the ($(m+1)\times q(m+1)$)-dimensional matrix $\tilde \Gamma \coloneqq  \left[\tilde \Gamma(1),\ldots,\tilde\Gamma(q)\right]$, the ($q(m+1)\times q(m+1)$)-dimensional matrix $\tilde G \coloneqq  \left(\tilde \Gamma(s-r) \right)_{r,s=1,\ldots,q}$ and the ($(m+1)\times (m+1)$)-dimensional empirical autocovariance matrix of $w_1,\ldots,w_T$ at lag $-q+1\leq h \leq q$, given by
\begin{align*}
	\tilde \Gamma(h) \coloneqq  T^{-1} \sum_{t=\max\{1,1-h\}}^{\min\{T,T-h\}}(w_{t+h} - \bar{w}_T)(w_t - \bar{w}_T)',
\end{align*}
where $\bar{w}_T\coloneqq T^{-1}\sum_{t=1}^T w_t$. Analogously, the solution of the sample Yule-Walker equations in the regression of $\hat w_t$ on $\hat w_{t-1},\ldots,\hat w_{t-q}$, $t=q+1,\ldots,T$, can be written in compact form as 
\begin{align*}
	\boldsymbol{\hat \Phi}(q)\coloneqq [\hat \Phi_1(q),\ldots,\hat \Phi_q(q)] = \hat \Gamma \hat G^{-1},
\end{align*}
with $\hat \Gamma \coloneqq  \left[\hat \Gamma(1),\ldots,\hat\Gamma(q)\right]$, $\hat G \coloneqq  \left(\hat \Gamma(s-r) \right)_{r,s=1,\ldots,q}$ and $\hat \Gamma(h)$ the empirical autocovariance matrix of $\hat w_1,\ldots,\hat w_T$ at lag $-q+1\leq h \leq q$. Taking the difference of $\boldsymbol{\tilde \Phi}(q)$ and $\boldsymbol{\hat \Phi}(q)$, adding and subtracting $\hat \Gamma \tilde G^{-1}$ and using
\begin{align*}
	\tilde G^{-1} - \hat G^{-1}=\tilde G^{-1}(\hat G - \tilde G)\hat G^{-1}
\end{align*}
leads to
\begin{align*}
	\boldsymbol{\hat\Phi}(q) - \boldsymbol{\tilde\Phi}(q) = \hat \Gamma \tilde G^{-1}(\tilde G - \hat G)\hat G^{-1} - (\tilde \Gamma - \hat \Gamma) \tilde G^{-1}
\end{align*}
Hence, we have to consider $\tilde G - \hat G$ in more detail ($\tilde \Gamma - \hat \Gamma$ works similarly). A typical block element of $\tilde G - \hat G$ is
\begin{align*}
	\tilde \Gamma (h) - \hat \Gamma(h) & = T^{-1} \sum_{t=\max\{1,1-h\}}^{\min\{T,T-h\}}(w_{t+h} - \bar{w}_T)(w_t - \bar{w}_T)'\\
	&\quad -T^{-1} \sum_{t=\max\{1,1-h\}}^{\min\{T,T-h\}}(\hat w_{t+h} - \bar{\hat w}_T)(\hat w_t - \bar{\hat w}_T)'	\\
	& = T^{-1} \sum_{t=\max\{1,1-h\}}^{\min\{T,T-h\}}(w_{t+h} - \bar{w}_T)(w_t - \hat w_t - (\bar{w}_T- \bar{\hat w}_T))'	\\
	& \quad - T^{-1} \sum_{t=\max\{1,1-h\}}^{\min\{T,T-h\}}(\hat w_{t+h} - w_{t+h} - (\bar{\hat w}_T - \bar{w}_T))(\hat w_t - \bar{\hat w}_T)'	\\
	&= A_1(h)-A_2(h),
\end{align*}
with an obvious definition for $A_1(h)$ and $A_2(h)$. Let us consider $A_1(h)$ in more detail ($A_2(h)$ works similarly). Using $\hat w_t=[\hat u_t,v_t']'$ and $\hat u_t = y_t - x_t'\hat\beta_{\scriptstyle\text{IM}}$ together with the model equations~\eqref{eq:y} and~\eqref{eq:x}, we get
\begin{align}\label{eq:Lem1tmp1}
	w_t-\hat w_t=\begin{bmatrix} u_t-\hat u_t \\ 0_{m\times 1}	\end{bmatrix}=\begin{bmatrix} y_t - x_t'\beta-(y_t - x_t'\hat\beta_{\scriptstyle\text{IM}}) \\ 0_{m\times 1}	\end{bmatrix}=\begin{bmatrix}  x_t'(\hat\beta_{\scriptstyle\text{IM}}-\beta) \\ 0_{m\times 1}	\end{bmatrix}.
\end{align}
Before we continue, note that the last equality implies 
\begin{align*}
	\max_{1\leq t \leq T}\vert  \hat w_t - w_t \vert_F \leq T^{-1/2}\max_{1\leq t \leq T}\vert T^{-1/2} x_t \vert_F\vert T\left(\hat\beta_{\scriptstyle\text{IM}} - \beta\right)\vert_F = O_\mP(T^{-1/2}),
\end{align*}
since $\max_{1\leq t \leq T}\vert T^{-1/2} x_t \vert_F=\sup_{0\leq r \leq 1}\vert T^{-1/2} x_{\floor{rT}}\vert_{F}$ converges by Assumption~\ref{ass:FCLT} and the continuous mapping theorem to $\sup_{0\leq r \leq 1}\vert B_v(r)\vert_{F}=O_\mP(1)$ and $\hat\beta_{\scriptstyle\text{IM}}$ is rate-$T$ consistent. This proves~\eqref{eq:maxw}.

We now proceed with the proof of~\eqref{eq:diffPhi}. From~\eqref{eq:Lem1tmp1} we obtain
\begin{align*}
	A_1(h) 
	& = T^{-1} \sum_{t=\max\{1,1-h\}}^{\min\{T,T-h\}}(w_{t+h} - \bar{w}_T)\begin{bmatrix}  (x_t-\bar{x}_T)'(\hat\beta_{\scriptstyle\text{IM}}-\beta) \\ 0_{m\times 1}	\end{bmatrix}'	\\
	& = T^{-1} \sum_{t=\max\{1,1-h\}}^{\min\{T,T-h\}}(w_{t+h} - \bar{w}_T)[(x_t-\bar{x}_T)'(\hat\beta_{\scriptstyle\text{IM}}-\beta),0_{1\times m}]	\\
	& = \tilde\Gamma_{w,x}(h)(\hat\beta_{\scriptstyle\text{IM}}-\beta)e_1',
\end{align*}
where $\tilde\Gamma_{w,x}(h)\coloneqq T^{-1} \sum_{t=\max\{1,1-h\}}^{\min\{T,T-h\}}(w_{t+h} - \bar{w}_T)(x_t-\bar{x}_T)'$ is ($(m+1)\times m$)-dimensional and $e_1\coloneqq(1,0_{1\times m})'$ is the first $(m+1)$-dimensional unit vector. Denoting the part of $\tilde G - \hat G$ that consists of block-entries $A_1(h)$ by $\tilde G_1 - \hat G_1$, we get
\begin{align*}
	\tilde G_1 - \hat G_1  =
	\begin{bmatrix}
		\tilde\Gamma_{w,x}(s-r)(\hat\beta_{\scriptstyle\text{IM}}-\beta)e_1'
	\end{bmatrix}_{r,s=1,\ldots,q}
	= \tilde\Gamma_{w,x} (I_q\otimes ((\hat\beta_{\scriptstyle\text{IM}}-\beta)e_1')),
\end{align*}
where $\tilde\Gamma_{w,x}\coloneqq (\tilde\Gamma_{w,x}(s-r))_{r,s=1,\ldots,q}$ is ($q(m+1)\times qm$)-dimensional. For the second factor we have
\begin{align*}
	\vert I_q\otimes ((\hat\beta_{\scriptstyle\text{IM}}-\beta)e_1')\vert_F & = \sqrt{\text{tr}\left((I_q\otimes ((\hat\beta_{\scriptstyle\text{IM}}-\beta)e_1'))'I_q\otimes ((\hat\beta_{\scriptstyle\text{IM}}-\beta)e_1')\right)}	\\
	& = \sqrt{\text{tr}\left(I_q\otimes (e_1(\hat\beta_{\scriptstyle\text{IM}}-\beta)'(\hat\beta_{\scriptstyle\text{IM}}-\beta)e_1'\right))}	\\
	& = \sqrt{\text{tr}\left(I_q\otimes \text{diag}\left(\sum_{i=1}^m (\hat\beta_{\scriptstyle\text{IM},i}-\beta_i)^2,0,\ldots,0\right)\right)}	\\
	& = \sqrt{q\sum_{i=1}^m (\hat\beta_{\scriptstyle\text{IM},i}-\beta_i)^2}	\\
	& = q^{1/2}T^{-1} \vert T(\hat\beta_{\scriptstyle\text{IM}}-\beta)\vert_F	\\
	& = O_P(q^{1/2}T^{-1}).
\end{align*}
Next, let us consider $\tilde\Gamma_{w,x}$ in more detail. Recall that $w_t=(u_t,v_t')'$ and $x_t=\sum_{k=1}^tv_k$. To avoid lengthy index notation, w.l.o.g. we can assume that $m=1$ and consider the second element of $w_t$ only (the first element works similarly). We thus consider the scalar quantity
\begin{align*}
	\tilde\Gamma_{w,x}(h)=T^{-1} \sum_{t=\max\{1,1-h\}}^{\min\{T,T-h\}}(v_{t+h} - \bar{v}_T)(\sum_{k=1}^t v_k-T^{-1}\sum_{i=1}^T\sum_{j=1}^iv_j).
\end{align*}
Taking the expectation of the squared Frobenius norm of the corresponding ($q\times q$)-dimensional matrix $\tilde\Gamma_{w,x}$ and combining the sums over $r$ and $s$, leads to
\begin{align*}
	& \mE(\vert\tilde\Gamma_{w,x}\vert_F^2)	\\
	&= \sum_{r,s=1}^q T^{-2} \sum_{t_1,t_2=\max\{1,1-(s-r)\}}^{\min\{T,T-(s-r)\}}\mE\left[(v_{t_1+s-r} - \bar{v}_T)(\sum_{k_1=1}^{t_1} v_{k_1}-T^{-1}\sum_{i_1=1}^T\sum_{j_1=1}^{i_1}v_{j_1})\right.	\\
	& \qquad\qquad\qquad\qquad\qquad\times\left.(v_{t_2+s-r} - \bar{v}_T)(\sum_{k_2=1}^{t_2} v_{k_2}-T^{-1}\sum_{i_2=1}^T\sum_{j_2=1}^{i_2}v_{j_2})\right]	\\
	&= T^{-2}\sum_{h=-q+1}^{q-1}(q-\vert h\vert)\sum_{t_1,t_2=\max\{1,1-h\}}^{\min\{T,T-h\}}\mE\left[(v_{t_1+h} - \bar{v}_T)(\sum_{k_1=1}^{t_1} v_{k_1}-T^{-1}\sum_{i_1=1}^T\sum_{j_1=1}^{i_1}v_{j_1})\right.	\\
	& \qquad\qquad\qquad\qquad\qquad\times\left.(v_{t_2+h} - \bar{v}_T)(\sum_{k_2=1}^{t_2} v_{k_2}-T^{-1}\sum_{i_2=1}^T\sum_{j_2=1}^{i_2}v_{j_2})\right].
\end{align*}
Note that the last expectation is of the form $\mE(ABCD)$ with $\mE(A)=\mE(B)=\mE(C)=\mE(D)=0$. Hence, by using common rules for joint cumulants of centered random variables \citeaffixed{Br81}{see, \eg,}, we get
\begin{align*}
	& \mE(ABCD)	\\
	& = \text{cum}(A,B,C,D)+\mE(AB)\mE(CD)+\mE(AC)\mE(BD)+\mE(AD)\mE(BC)	\\
	& = \text{cum}(A,B,C,D)+\text{Cov}(A,B)\text{Cov}(C,D)+\text{Cov}(A,C)\text{Cov}(B,D)\\
	& \quad +\text{Cov}(A,D)\text{Cov}(B,C),
\end{align*}
where $\text{Cov}(\cdot,\cdot)$ denotes the covariance of two random variables. Hence, the first term corresponding to the fourth-order cumulant becomes
\begin{align*}
	& T^{-2}\sum_{h=-q+1}^{q-1}(q-\vert h\vert) \sum_{t_1,t_2=\max\{1,1-h\}}^{\min\{T,T-h\}}\text{cum}\left(v_{t_1+h} - \bar{v}_T,\sum_{k_1=1}^{t_1} v_{k_1}-T^{-1}\sum_{i_1=1}^T\sum_{j_1=1}^{i_1}v_{j_1},\right.	\\
	& \qquad\qquad\qquad\qquad\qquad\left.v_{t_2+h} - \bar{v}_T,\sum_{k_2=1}^{t_2} v_{k_2}-T^{-1}\sum_{i_2=1}^T\sum_{j_2=1}^{i_2}v_{j_2}\right)
\end{align*}
leading to $2^4=16$ terms when exapnding the cumulant. Exemplarily, for the absolute value of the first one (the others work similarly), we get from the common calculation rules for cumulants
\begin{align*}
	& \vert T^{-2}\sum_{h=-q+1}^{q-1}(q-\vert h\vert)  \sum_{t_1,t_2=\max\{1,1-h\}}^{\min\{T,T-h\}}\text{cum}\left(v_{t_1+h},\sum_{k_1=1}^{t_1} v_{k_1},v_{t_2+h},\sum_{k_2=1}^{t_2} v_{k_2}\right)\vert	\\
	&\leq  T^{-2}\sum_{h=-q+1}^{q-1}\vert q-\vert h\vert\vert \sum_{t_1,t_2=\max\{1,1-h\}}^{\min\{T,T-h\}}\sum_{k_1=1}^{t_1}\sum_{k_2=1}^{t_2} \vert \text{cum}\left(v_{t_1+h},v_{k_1},v_{t_2+h},v_{k_2}\right)\vert	\\
	&\leq \frac{q}{T^2}\sum_{h=-q+1}^{q-1} \sum_{t_1,t_2=1}^{T}\sum_{k_1,k_2=1}^T \vert \text{cum}\left(v_{t_1+h},v_{k_1},v_{t_2+h},v_{k_2}\right)\vert.
\end{align*}
By combining the sums over $t_1$ and $t_2$ and those over $k_1$ and $k_2$, respectively, the above term becomes
\begin{align*}
	& \frac{q}{T^2}\sum_{h=-q+1}^{q-1} \sum_{l=-(T-1)}^{T-1}\sum_{s=\max\{1,1-l\}}^{\min\{T,T-l\}}\sum_{i=-(T-1)}^{T-1}\sum_{j=\max\{1,1-i\}}^{\min\{T,T-i\}}\vert \text{cum}\left(v_{s+l+h},v_{j+i},v_{s+h},v_{j}\right)\vert	\\
	&\leq \frac{q}{T^2}\sum_{h=-q+1}^{q-1} \sum_{i,l=-(T-1)}^{T-1}\sum_{s,j=1}^T\vert \text{cum}\left(v_{s+l+h},v_{j+i},v_{s+h},v_{j}\right)\vert	\\
	&\leq \frac{q}{T^2}\sum_{h=-q+1}^{q-1} \sum_{i,l=-(T-1)}^{T-1} \sum_{k=-(T-1)}^{T-1}\sum_{r=\max\{1,1-k\}}^{\min\{T,T-k\}}\vert \text{cum}\left(v_{r+k+l+h},v_{r+i},v_{r+k+h},v_{r}\right)\vert	\\
	&\leq \frac{q}{T}\sum_{h=-q+1}^{q-1} \sum_{i,l=-(T-1)}^{T-1} \sum_{k=-(T-1)}^{T-1}\vert \text{cum}\left(v_{k+l+h},v_{i},v_{k+h},v_{0}\right)\vert,
\end{align*}
where we also combined the sums over $s$ and $j$ and made use of the (strict) stationarity of $\{v_t\}_{t\in\mathbb{Z}}$. Finally, combining the sums over $h$ and $k$, we get the bound
\begin{align*}
	& \frac{q}{T}\sum_{r=-(T+q-2)}^{T+q-2} (2q-1) \sum_{i,l=-(T-1)}^{T-1} \vert \text{cum}\left(v_{r+l},v_{i},v_{r},v_{0}\right)\vert	\\
	&\leq 2\frac{q^2}{T} \sum_{j=-(2T+q-3)}^{2T+q-3}\sum_{r=-(T+q-2)}^{T+q-2} \sum_{i=-(T-1)}^{T-1} \vert \text{cum}\left(v_{j},v_{i},v_{r},v_{0}\right)\vert\\
	&\leq 2\frac{q^2}{T}\sum_{j,i,r=-\infty}^{\infty} \vert \text{cum}\left(v_{j},v_{i},v_{r},v_{0}\right)\vert=O(q^2 T^{-1})
\end{align*}
due to the summability condition imposed on the fourth order cumulants in Assumption~\ref{ass:cumulants} and, hence, vanishes for $T\rightarrow \infty$. However, the leading term is the term corresponding to $\text{Cov}(A,B)\text{Cov}(C,D)$. That is, we have to consider
\begin{align*}
	& T^{-2}\sum_{h=-q+1}^{q-1}(q-\vert h\vert)\sum_{t_1,t_2=\max\{1,1-h\}}^{\min\{T,T-h\}}\text{Cov}(v_{t_1+h} - \bar{v}_T,\sum_{k_1=1}^{t_1} v_{k_1}-T^{-1}\sum_{i_1=1}^T\sum_{j_1=1}^{i_1}v_{j_1})	\\
	& \qquad\qquad\qquad\qquad\qquad\times \text{Cov}(v_{t_2+h} - \bar{v}_T,\sum_{k_2=1}^{t_2} v_{k_2}-T^{-1}\sum_{i_2=1}^T\sum_{j_2=1}^{i_2}v_{j_2})	\\
	&= \sum_{h=-q+1}^{q-1}(q-\vert h\vert) \left(T^{-1}\sum_{t=\max\{1,1-h\}}^{\min\{T,T-h\}}\text{Cov}(v_{t+h} - \bar{v}_T,\sum_{k=1}^{t} v_{k}-T^{-1}\sum_{i=1}^T\sum_{j=1}^{i}v_{j})\right)^2.
\end{align*}
Hence, we have to compute
\begin{align*}
	T^{-1}\sum_{t=\max\{1,1-h\}}^{\min\{T,T-h\}}\text{Cov}(v_{t+h} - \bar{v}_T,\sum_{k=1}^{t} v_{k}-T^{-1}\sum_{i=1}^T\sum_{j=1}^{i}v_{j}).
\end{align*}
This leads to four terms to consider, which can be treated with similar arguments. For the first term we get
\begin{align*}
	T^{-1}\sum_{t=\max\{1,1-h\}}^{\min\{T,T-h\}}\text{Cov}(v_{t+h},\sum_{k=1}^{t} v_{k}) = T^{-1}\sum_{t=\max\{1,1-h\}}^{\min\{T,T-h\}}\sum_{k=1}^{t}\gamma_v(t+h-k),
\end{align*}
where $\gamma_v(h)$ is the covariance of the one-dimensional process (still assumed for notational brevity) $\{v_t\}_{t\in\mZ}$ at lag $h$. \Eg,~for $h\geq 0$, this can be exactly computed to equal
\begin{align*}
	T^{-1}\sum_{t=1}^{T-h}\sum_{k=1}^{t}\gamma_v(t+h-k)=T^{-1}\sum_{j=h}^{T-1}(T-j)\gamma_v(j)\leq \sum_{j=h}^{T-1}\gamma_v(j)
\end{align*}
and its absolute vale can be bounded by $\sum_{j=-\infty}^{\infty}\vert \gamma_v(j)\vert <\infty$ due to the second-order cumulant condition imposed in Assumption~\ref{ass:cumulants}. Similar arguments yield the same bound for $h<0$ and for the other three terms. Hence, the term of $\mE(\vert \tilde\Gamma_{w,x}\vert_F^2)$ that corresponds to $\text{Cov}(A,B)\text{Cov}(C,D)$ is of order $O(q^2)$.

In total we thus have $\mE(\vert \tilde\Gamma_{w,x}\vert_F^2)=O(q^2)$. Note that we have proven this result for $m=1$. However, as $m$ is fixed, the result also holds for $m>1$. Therefore, we obtain for the ($q(m+1)\times qm$)-dimensional matrix $\tilde\Gamma_{w,x}$ that $\vert \tilde\Gamma_{w,x}\vert_F=O_\mP(q)$. It follows that $\vert \tilde G_1-\hat G_1\vert_F=O_\mP(q)O_P(q^{1/2}T^{-1})=O_\mP(q^{3/2}T^{-1})$ and similarly also $\vert \tilde G-\hat G\vert_F=O_\mP(q)O_\mP(q^{1/2}T^{-1})=O_\mP(q^{3/2}T^{-1})$.

Further, we have to consider $\tilde G^{-1}$. In the following, let $\mu_{\min}(A)$ and $\mu_{\max}(A)$ denote the smallest and largest eigenvalue of a matrix $A$, respectively and define $G \coloneqq  \left( \Gamma(s-r) \right)_{r,s=1,\ldots,q}\in\mR^{q(m+1)\times q(m+1)}$. Similar to the above, using the fourth-order cumulant condition from Assumption~\ref{ass:cumulants}, we can show that $\vert \tilde G-G\vert_F=O_\mP(qT^{-1/2})=o_\mP(1)$. Then, to show boundedness in probability of $\tilde G^{-1}$ (similar for $\hat G^{-1}$), for all $\epsilon>0$, we have to find a $K<\infty$ and a $T_0<\infty$ both large enough such that for all $T>T_0$, it holds that
\begin{align*}
	\mP(\vert \tilde G^{-1}\vert_2>K)<\epsilon,
\end{align*}
where $\vert A \vert_2$ denotes the spectral norm of a matrix $A$. Let $\epsilon>0$. Then, due to positive semi-definiteness of $\tilde G$ by construction and invertibility, see \citeasnoun[Lemma~3.4 and Remark~3.2]{MeKr15}, we have positive definiteness of $\tilde G$ and, consequently, of $\tilde G^{-1}$. Hence, we get
\begin{align*}
	\mP(\vert \tilde G^{-1}\vert_2>K) =&\mP(\mu_{\max}(\tilde G^{-1})>K)=\mP(\mu_{\min}^{-1}(\tilde G)>K)	\\
	=& \mP(\mu_{\min}(\tilde G)<\frac{1}{K},\vert \tilde G-G\vert_2\geq \delta)	\\
	&+ \mP(\mu_{\min}(\tilde G)<\frac{1}{K},\vert \tilde G-G\vert_2<\delta).
\end{align*}
Further, as $\vert \tilde G-G\vert_2 \leq \vert \tilde G-G\vert_F=o_\mP(1)$, for any $\delta>0$, we can choose $T_0$ large enough to have $\mP(\vert \tilde G-G\vert_2\geq \delta)\leq \epsilon$. Then the first term on the last right-hand side can be bounded by $\mP(\vert \tilde G-G\vert_2\geq \delta)\leq\epsilon$. For the second term, as $\tilde G$, $G$ and hence also $\tilde G-G$ are symmetric with real-valued entries, these matrices are Hermitian such that Weyl's theorem \citeaffixed{HoJo12}{see, \eg, Theorem~4.3.1 in} applies, leading to the inequality $\mu_{\min}(G)+\mu_{\min}(\tilde G-G)\leq\mu_{\min}(\tilde G)$. It follows that
\begin{align*}
	& \mP(\mu_{\min}(\tilde G)<\frac{1}{K},\vert \tilde G-G\vert_2<\delta)	\\
	&\leq \mP(\mu_{\min}(G)+\mu_{\min}(\tilde G-G)<\frac{1}{K},\vert \tilde G-G\vert_2<\delta)	\\
	&= \mP(\mu_{\min}(G)<\frac{1}{K}-\mu_{\min}(\tilde G-G),\vert \tilde G-G\vert_2<\delta).
\end{align*}
From symmetry of $\tilde G-G$ we get that the eigenvalues of $(\tilde G-G)'(\tilde G-G)$ are exactly the squared eigenvalues of $\tilde G-G$. Hence, the bound
\begin{align*}
	\vert \tilde G-G\vert_2=\sqrt{\mu_{\max}((\tilde G-G)'(\tilde G-G))}<\delta
\end{align*}
implies also $\mu_{\min}(\tilde G-G)\geq -\delta$ such that the last right-hand side can be bounded by
\begin{align}\label{eq:muG}
	\mP(\mu_{\min}(G)<\frac{1}{K}+\delta,\vert \tilde G-G\vert_2<\delta)\leq \mP(\mu_{\min}(G)<\frac{1}{K}+\delta).
\end{align}
Next, note that $\mu_{\min}(G)\geq \tilde c$ for some constant $\tilde c>0$ by Assumption~\ref{ass:w}. Therefore, the right-hand side in~\eqref{eq:muG} becomes zero if we choose $\delta<\tilde c/2$ small enough and $K>2/\tilde c$ large enough, such that $\frac{1}{K}+\delta<\tilde c$. This completes the proof of $\vert \tilde G^{-1}\vert_2=O_\mP(1)$. Furthermore, from Assumption~\ref{ass:w} we also get $\vert \tilde G\vert_2=O_\mP(1)$ and similarly $\vert \hat G\vert_2=O_\mP(1)$ and $\vert \hat G\vert_2=O_\mP(1)$. Altogether, we get
\begin{align*}
	&\vert \boldsymbol{\tilde \Phi}(q) - \boldsymbol{\hat \Phi}(q)\vert_2 \leq \vert \tilde \Gamma - \hat \Gamma\vert_2 \vert \tilde G^{-1}\vert_2 + \vert \hat \Gamma\vert_2 \vert \tilde G^{-1}\vert_2\vert \tilde G - \hat G\vert_2\vert \hat G^{-1}\vert_2	\\
	&\leq \vert \tilde G - \hat G\vert_2 \left(\vert \tilde G^{-1}\vert_2 + \vert \hat G\vert_2 \vert \tilde G^{-1}\vert_2\vert \hat G^{-1}\vert_2\right)	\\
	&\leq \vert \tilde G - \hat G\vert _{F} \left(\vert \tilde G^{-1}\vert_2 + \vert \hat G\vert_2 \vert \tilde G^{-1}\vert_2\vert \hat G^{-1}\vert_2\right)	\\
	&= O_\mP(q^{3/2}T^{-1})\left(O_\mP(1)+O_\mP(1)O_\mP(1)O_\mP(1)\right)	\\
	&= O_\mP(q^{3/2}T^{-1}).
\end{align*}
Since $\boldsymbol{\tilde \Phi}(q) - \boldsymbol{\hat \Phi}(q)$ is ($(m+1)\times q(m+1)$)-dimensional and $m$ is fixed, it holds that \citeaffixed{Ge07}{see, \eg}
\begin{align*}
	\vert \boldsymbol{\tilde \Phi}(q) - \boldsymbol{\hat \Phi}(q)\vert _{F} \leq \sqrt{m+1}\  \vert \boldsymbol{\tilde \Phi}(q) - \boldsymbol{\hat \Phi}(q)\vert_2 = O_\mP(q^{3/2}T^{-1}).
\end{align*}
This implies that
\begin{align*}
	q^{1/2}\sum_{j=1}^q \vert \hat \Phi_j(q) - \tilde \Phi_j(q)\vert_F \leq q^{3/2} \vert \boldsymbol{\tilde \Phi}(q) - \boldsymbol{\hat \Phi}(q)\vert _{F} = O_\mP(q^3T^{-1}),
\end{align*}
which is $o_\mP(1)$ since $q^{3}T^{-1}= o(1)$ by Assumption~\ref{ass:q}.~\hfill~$\square$

In the proofs of Lemma~\ref{lem:moments} and~\ref{lem:Sigmastar} we repeatedly use the fact that by convexity, $\vert \sum_{i=1}^k z_i \vert^a \leq k^{a-1} \sum_{i=1}^k \vert z_i\vert^a$, for all $a,k\geq 1$.\\

{\bf Proof of Lemma~\ref{lem:moments}.} Let $\tilde \ve_t(q) \coloneqq   w_t - \sum_{j=1}^q \tilde \Phi_j(q) w_{t-j}$, $t=q+1,\ldots,T$, denote the Yule-Walker residuals in the regression of $w_t$ on $w_{t-1},\ldots,w_{t-q}$, $t=q+1,\ldots,T$ and define $\bar{\tilde{\ve}}_T(q)\coloneqq (T-q)^{-1}\sum_{t=q+1}^T \tilde{\ve}_{t}(q)$.\footnote{In the following, $a$ denotes the fixed $a>2$ from Assumption~\ref{ass:w}. However, the results also hold for $1\leq \tilde a < a$.} For $q+1\leq t \leq T$ we have
\begin{align*}
	&\vert \hat \ve_t(q) - \bar{\hat{\ve}}_T(q)\vert_F^a\\
	&\leq \left(\vert \hat \ve_t(q) - \tilde \ve_t(q)\vert_F + \vert \tilde \ve_t(q) - \bar{\tilde{\ve}}_T(q) \vert_F + \vert \bar{\hat{\ve}}_T(q) - \bar{\tilde{\ve}}_T(q) \vert_F \right)^a\\
	& \leq \left(\vert \hat \ve_t(q) - \tilde \ve_t(q)\vert_F + \vert \tilde \ve_t(q) - \bar{\tilde{\ve}}_T(q) \vert_F + (T-q)^{-1} \sum_{t=q+1}^T \vert \hat \ve_t(q) - \tilde \ve_t(q) \vert_F \right)^a\\
	& \leq 3^{a-1} \left(\vert \hat \ve_t(q) - \tilde \ve_t(q)\vert_F^a + \vert \tilde \ve_t(q) - \bar{\tilde{\ve}}_T(q) \vert_F^a + \left((T-q)^{-1} \sum_{t=q+1}^T \vert \hat \ve_t(q) - \tilde \ve_t(q) \vert_F\right)^a \right).
\end{align*}
Hence,
\begin{align*}
	&(T-q)^{-1} \sum_{t=q+1}^T \vert \hat \ve_t(q) - \bar{\hat{\ve}}_T(q)\vert_F^a\\ 
	&\leq 3^{a-1}\left((T-q)^{-1} \sum_{t=q+1}^T \vert \hat \ve_t(q) - \tilde \ve_t(q) \vert_F^a + (T-q)^{-1} \sum_{t=q+1}^T \vert \tilde \ve_t(q) - \bar{\tilde{\ve}}_T(q) \vert_F^a\right. \\
	& \hspace{2cm}+ \left.\left((T-q)^{-1} \sum_{t=q+1}^T \vert \hat \ve_t(q) - \tilde \ve_t(q) \vert_F\right)^a\right)\\
	& = 3^{a-1}\left(F_{T,a} + (T-q)^{-1} \sum_{t=q+1}^T \vert \tilde \ve_t(q) - \bar{\tilde{\ve}}_T(q) \vert_F^a + (F_{T,1})^a\right),
\end{align*}
where $F_{T,a}\coloneqq (T-q)^{-1} \sum_{t=q+1}^T \vert \hat \ve_t(q) - \tilde \ve_t(q) \vert_F^a$. 

We now consider $F_{T,a}$ in more detail. For $q+1\leq t \leq T$ we have
\begin{align*}
	&\vert \hat \ve_t(q) - \tilde \ve_t(q) \vert_F\\
	&= \vert\hat w_t - \sum_{j=1}^q \hat \Phi_j(q) \hat w_{t-j} - (w_t - \sum_{j=1}^q \tilde \Phi_j(q) w_{t-j})\vert_F\\
	& = \vert\hat w_t - w_t - \sum_{j=1}^q \left( \hat \Phi_j(q) - \tilde \Phi_j(q) + \tilde \Phi_j(q)\right)\left(\hat w_{t-j} - w_{t-j} + w_{t-j}\right) + \sum_{j=1}^q \tilde \Phi_j(q) w_{t-j}\vert_F\\
	& = \vert\hat w_t - w_t - \sum_{j=1}^q \left( \hat \Phi_j(q) - \tilde \Phi_j(q)\right)\left(\hat w_{t-j} - w_{t-j} + w_{t-j}\right) - \sum_{j=1}^q \tilde \Phi_j(q) \left(\hat w_{t-j} - w_{t-j}\right)\vert_F\\
	& \leq \max_{1\leq t \leq T} \vert\hat w_t - w_t\vert_F + \sum_{j=1}^q \vert \hat \Phi_j(q) - \tilde \Phi_j(q)\vert_F \vert \hat w_{t-j} - w_{t-j}\vert_F \\
	&\hspace{0.5cm} + \sum_{j=1}^q \vert\tilde \Phi_j(q)\vert_F \vert\hat w_{t-j} - w_{t-j}\vert_F + \sum_{j=1}^q \vert\hat \Phi_j(q) - \tilde \Phi_j(q)\vert_F \vert w_{t-j}\vert_F\\
	& \leq \max_{1\leq t \leq T} \vert\hat w_t - w_t\vert_F + \max_{1\leq t \leq T} \vert\hat w_t - w_t\vert_F \sum_{j=1}^q \vert \hat \Phi_j(q) - \tilde \Phi_j(q)\vert_F\\
	& \hspace{0.5cm} + \max_{1\leq t \leq T} \vert\hat w_t - w_t\vert_F \sum_{j=1}^q \vert \tilde \Phi_j(q)\vert_F + \sqrt{m+1} q \vert \boldsymbol{\hat \Phi}(q) - \boldsymbol{\tilde \Phi}(q)\vert_F\ 
	q^{-1} \sum_{j=1}^q \vert w_{t-j}\vert_F,
\end{align*}
where we have used that 
\begin{align*}
	\vert \hat \Phi_j(q) - \tilde \Phi_j(q)\vert_F &\leq \vert \boldsymbol{\hat \Phi}(q) - \boldsymbol{\tilde \Phi}(q)\vert_F \vert [0,I_{m+1},0]'\vert_F\\
	&= \sqrt{m+1}\vert \boldsymbol{\hat \Phi}(q) - \boldsymbol{\tilde \Phi}(q)\vert_{F}.
\end{align*}
Therefore,
\begin{align*}
	&\vert \hat \ve_t(q) - \tilde \ve_t(q) \vert_F^a\\
	&\leq 4^{a-1} \left( \left(\max_{1\leq t \leq T} \vert\hat w_t - w_t\vert_F\right)^a + \left(\max_{1\leq t \leq T} \vert\hat w_t - w_t\vert_F\right)^a \left( \sum_{j=1}^q \vert\hat \Phi_j(q) - \tilde \Phi_j(q)\vert_F \right)^a\right.\\
	&\hspace{0.5cm}\left. + \left(\max_{1\leq t \leq T} \vert\hat w_t - w_t\vert_F\right)^a \left(\sum_{j=1}^q \vert \tilde \Phi_j(q)\vert_F\right)^a \right.\\
	&\hspace{0.5cm} \left.+ \left(\sqrt{m+1} q \vert \boldsymbol{\hat \Phi}(q) - \boldsymbol{\tilde \Phi}(q)\vert_F\right)^a \left(q^{-1} \sum_{j=1}^q \vert w_{t-j}\vert_F\right)^a\right).
\end{align*}
It follows that
\begin{align*}
	F_{T,a}& \leq 4^{a-1} \left( \left(\max_{1\leq t \leq T} \vert\hat w_t - w_t\vert_F\right)^a + \left(\max_{1\leq t \leq T} \vert\hat w_t - w_t\vert_F\right)^a \left( \sum_{j=1}^q \vert\hat \Phi_j(q) - \tilde \Phi_j(q)\vert_F \right)^a\right.\\
	&\hspace{2cm}\left. + \left(\max_{1\leq t \leq T} \vert\hat w_t - w_t\vert_F\right)^a \left(\sum_{j=1}^q \vert \tilde \Phi_j(q)\vert_F\right)^a \right.\\
	& \hspace{2cm}\left. + \left(\sqrt{m+1} q \vert \boldsymbol{\hat \Phi}(q) - \boldsymbol{\tilde \Phi}(q)\vert_{F}\right)^a (T-q)^{-1} \sum_{t=q+1}^T \left( q^{-1} \sum_{j=1}^q \vert w_{t-j}\vert_F\right)^a\right).
\end{align*}
From Lemma~\ref{lem:hat-tilde} we have $\max_{1\leq t \leq T} \vert\hat w_t - w_t\vert_F = O_\mP(T^{-1/2})$ and $\sum_{j=1}^q \vert \hat \Phi_j(q) - \tilde \Phi_j(q)\vert_F= O_\mP(q^{5/2}T^{-1})$. Moreover, the proof of Lemma~\ref{lem:hat-tilde} shows that $q \vert \boldsymbol{\hat \Phi}(q) - \boldsymbol{\tilde \Phi}(q)\vert_{F}=O_\mP(q^{5/2} T^{-1})$. Further, by~\eqref{eq:supPhi},~\eqref{eq:Baxter} and Assumption~\ref{ass:w} we have
\begin{align*}
	\sum_{j=1}^q \vert \tilde \Phi_j(q)\vert_F &\leq \sum_{j=1}^q \vert \tilde \Phi_j(q) - \Phi_j(q)\vert_F + \sum_{j=1}^q \vert \Phi_j(q) - \Phi_j\vert_F + \sum_{j=1}^q \vert \Phi_j\vert_F\\
	& \leq q \sup_{1\leq j \leq q} \vert \tilde \Phi_j(q) - \Phi_j(q)\vert_F + c \sum_{j=q+1}^\infty \vert \Phi_j\vert_F + \sum_{j=1}^\infty \vert \Phi_j\vert_F\\
	& = O_\mP(1).
\end{align*}
Finally, note that
\begin{align*}
	(T-q)^{-1} \sum_{t=q+1}^T \left( q^{-1} \sum_{j=1}^q \vert w_{t-j}\vert_F\right)^a&\leq (T-q)^{-1} \sum_{t=q+1}^T q^{-a} q^{a-1} \sum_{j=1}^q \vert w_{t-j}\vert_F^a\\
	& = (T-q)^{-1} q^{-1} \sum_{t=q+1}^T \sum_{j=1}^q \vert w_{t-j}\vert_F^a\\
	& \leq (T-q)^{-1} \sum_{t=1}^{T-1} \vert w_t \vert_F^a,
\end{align*}
where the last inequality follows from the fact that each element in the double sum occurs at most $q$ times, \ie, $\sum_{t=q+1}^T \sum_{j=1}^q \vert w_{t-j} \vert_F^a \leq q \sum_{t=1}^{T-1} \vert w_t \vert_F^a$. From $\sup_{t\in \mZ}\mE\left( \vert  w_t \vert_F^a \right)< \infty$ (by stationarity of $\{w_t\}_{t\in \mZ}$) and Markov's inequality, it follows that $(T-q)^{-1} \sum_{t=1}^{T-1} \vert w_t \vert_F^a=O_\mP(1)$. In total, we thus have $F_{T,a} = O_{\mP}((q^{5/2}T^{-1})^a)=o_{\mP}(1)$. Therefore,
\begin{align*}
	(T-q)^{-1} \sum_{t=q+1}^T \vert \hat \ve_t(q) - \bar{\hat{\ve}}_T(q)\vert_F^a \leq 3^{a-1} \left((T-q)^{-1} \sum_{t=q+1}^T \vert \tilde \ve_t(q) - \bar{\tilde{\ve}}_T(q) \vert_F^a + o_\mP(1)\right).
\end{align*}
It thus remains to show that $(T-q)^{-1} \sum_{t=q+1}^T \vert \tilde \ve_t(q) - \bar{\tilde{\ve}}_T(q) \vert_F^a = O_\mP(1)$. We now follow \citeasnoun[Proof of Lemma~3.2]{Pa02} and \citeasnoun[Proof of Lemma~2]{PSU10}. Define $\ve_t(q)\coloneqq  \ve_t + \sum_{j=q+1}^\infty \Phi_j w_{t-j}$ and note that
\begin{align*}
	&(T-q)^{-1} \sum_{t=q+1}^T \vert \tilde \ve_t(q) - \bar{\tilde{\ve}}_T(q) \vert_F^a \\&= (T-q)^{-1} \sum_{t=q+1}^T \vert \tilde \ve_t(q) - \ve_t(q) + \ve_t(q) - \ve_t + \ve_t - \bar{\tilde{\ve}}_T(q) \vert_F^a\\
	&\leq 4^{a-1}\left(A_{T,a}+B_{T,a}+C_{T,a}+D_{T,a}\right),
\end{align*}
where
\begin{align*}
	A_{T,a} &\coloneqq  (T-q)^{-1} \sum_{t=q+1}^T \vert \ve_t \vert_F^a,\\
	B_{T,a} &\coloneqq  (T-q)^{-1} \sum_{t=q+1}^T \vert \ve_t(q) - \ve_t \vert_F^a = (T-q)^{-1} \sum_{t=q+1}^T \vert \sum_{j=q+1}^\infty \Phi_j w_{t-j} \vert_F^a,\\
	C_{T,a} &\coloneqq  (T-q)^{-1} \sum_{t=q+1}^T \vert \tilde\ve_t(q) - \ve_t(q) \vert_F^a,\\
	D_{T,a} &\coloneqq  (T-q)^{-1} \sum_{t=q+1}^T \vert \bar{\tilde{\ve}}_T(q) \vert_F^a = \vert \bar{\tilde{\ve}}_T(q) \vert_F^a = \vert (T-q)^{-1} \sum_{t=q+1}^T \tilde \ve_t(q) \vert_F^a.
\end{align*}
We first consider $B_{T,a}$. Note that $\mE\left(\vert B_{T,a}\vert_F\right) \leq \sup_{t\in\mZ} \mE\left(\vert \ve_t(q)-\ve_t\vert_F^a\right)$. Using Minkowski's inequality, we have
\begin{align*}
	\mE\left(\vert \ve_t(q)-\ve_t\vert_F^a\right)
	&= \mE\left(\vert \sum_{j=q+1}^\infty \Phi_j w_{t-j} \vert_F^a\right)
	= \left(\left[ \mE\left(\vert \sum_{j=q+1}^\infty \Phi_j w_{t-j} \vert_F^a\right) \right]^{1/a}\right)^a\\
	&\leq \left( \sum_{j=q+1}^\infty \left[\mE\left( \vert \Phi_j w_{t-j} \vert_F^a \right)\right]^{1/a} \right)^a
	\leq \left( \sum_{j=q+1}^\infty \left[\mE\left( \vert \Phi_j \vert_F^a\vert  w_{t-j} \vert_F^a \right)\right]^{1/a} \right)^a\\
	&\leq \left( \sum_{j=q+1}^\infty \vert \Phi_j \vert_F \left[\mE\left( \vert  w_{t-j} \vert_F^a \right)\right]^{1/a} \right)^a
	\leq \left( \sum_{j=q+1}^\infty \vert \Phi_j \vert_F \left[\sup_{t\in \mZ}\mE\left( \vert  w_t \vert_F^a \right)\right]^{1/a} \right)^a\\
	& \leq \sup_{t\in \mZ}\mE\left( \vert  w_t \vert_F^a \right) \left( \sum_{j=q+1}^\infty \vert \Phi_j \vert_F \right)^a.
\end{align*}
From Assumption~\ref{ass:w} we have $\sup_{t\in \mZ}\mE\left( \vert  w_t \vert_F^a \right)<\infty$ and $\sum_{j=q+1}^\infty \vert \Phi_j \vert_F = o(1)$. Markov's inequality thus yields $B_{T,a}=o_\mP(1)$. Analogously, $\mE\left(\vert A_{T,a}\vert_F\right) \leq \sup_{t\in\mZ} \mE\left(\vert \ve_t\vert_F^a\right)$. Using Minkowski's inequality, we have as above
\begin{align*}
	\mE\left(\vert \ve_t\vert_F^a\right) 
	&= \mE\left(\vert w_t - \sum_{j=1}^\infty \Phi_j w_{t-j}\vert_F^a\right)
	\leq 2^{a-1}\left(\mE\left(\vert w_t \vert_F^a \right)+ \mE\left(\vert \sum_{j=1}^\infty \Phi_j w_{t-j}\vert_F^a\right)\right)\\
	&\leq 2^{a-1}\left(\sup_{t\in\mZ} \mE\left(\vert w_t\vert_F^a\right)+ \mE\left(\vert \sum_{j=1}^\infty \Phi_j w_{t-j}\vert_F^a\right)\right)\\
	& \leq 2^{a-1} \sup_{t\in\mZ} \mE\left(\vert w_t\vert_F^a\right) \left(1+\left( \sum_{j=1}^\infty \vert \Phi_j \vert_F\right)^a\right)
	< \infty.
\end{align*}
Using Markov's inequality we conclude that $A_{T,a}=O_\mP(1)$. We now turn to $C_{T,a}$. By definition,
\begin{align*}
	\tilde \ve_t(q) &= w_t - \sum_{j=1}^q \tilde \Phi_j(q) w_{t-j} 
	= \ve_t(q) - \sum_{j=1}^q \left(\tilde \Phi_j(q) - \Phi_j \right) w_{t-j}\\
	&= \ve_t(q) - \sum_{j=1}^q \left(\tilde \Phi_j(q) - \Phi_j(q) \right) w_{t-j} - \sum_{j=1}^q \left(\Phi_j(q) - \Phi_j \right) w_{t-j}.
\end{align*}
Hence,
\begin{align*}
	\vert \tilde \ve_t(q) - \ve_t(q) \vert_F^a \leq 2^{a-1} \left( \vert \sum_{j=1}^q \left(\tilde \Phi_j(q) - \Phi_j(q) \right) w_{t-j} \vert_F^a + \vert \sum_{j=1}^q \left(\Phi_j(q) - \Phi_j \right) w_{t-j} \vert_F^a \right).
\end{align*}
It follows that $C_{T,a} = 2^{a-1} \left( C_{1T,a} + C_{2T,a}\right)$, where
\begin{align*}
	C_{1T,a} &\coloneqq  (T-q)^{-1} \sum_{t=q+1}^T \vert \sum_{j=1}^q \left(\tilde \Phi_j(q) - \Phi_j(q) \right) w_{t-j} \vert_F^a,\\
	C_{2T,a} &\coloneqq  (T-q)^{-1} \sum_{t=q+1}^T \vert \sum_{j=1}^q \left(\Phi_j(q) - \Phi_j \right) w_{t-j} \vert_F^a.
\end{align*}
We consider both terms separately. First note that
\begin{align*}
	C_{1T,a} &\leq q^{a-1} (T-q)^{-1} \sum_{t=q+1}^T \sum_{j=1}^q \vert \tilde \Phi_j(q) - \Phi_j(q) \vert_F^a \vert w_{t-j} \vert_F^a\\
	& \leq q^{a-1} \left( \sup_{1\leq j \leq q} \vert \tilde \Phi_j(q) - \Phi_j(q) \vert_F \right)^a (T-q)^{-1} \sum_{t=q+1}^T \sum_{j=1}^q \vert w_{t-j} \vert_F^a\\
	& \leq \left( q \sup_{1\leq j \leq q} \vert \tilde \Phi_j(q) - \Phi_j(q) \vert_F \right)^a (T-q)^{-1} \sum_{t=1}^{T-1} \vert w_t \vert_F^a,
\end{align*}
where the third inequality follows again from the fact that $\sum_{t=q+1}^T \sum_{j=1}^q \vert w_{t-j} \vert_F^a \leq q \sum_{t=1}^{T-1} \vert w_t \vert_F^a$. As $(T-q)^{-1} \sum_{t=1}^{T-1} \vert w_t \vert_F^a=O_\mP(1)$ it follows from~\eqref{eq:supPhi} that $C_{1T,a} = o_\mP(1)$. Moreover, using Minkowski's inequality, we obtain
\begin{align*}
	\mE\left(\vert C_{2T,a} \vert_F\right) &= (T-q)^{-1} \sum_{t=q+1}^T \mE\left( \vert \sum_{j=1}^q \left(\Phi_j(q) - \Phi_j \right) w_{t-j} \vert_F^a \right)\\
	& = (T-q)^{-1} \sum_{t=q+1}^T \left(\left[\mE\left( \vert \sum_{j=1}^q \left(\Phi_j(q) - \Phi_j \right) w_{t-j} \vert_F^a \right)\right]^{1/a}\right)^a\\
	& \leq (T-q)^{-1} \sum_{t=q+1}^T \left(\sum_{j=1}^q \left[ \mE\left( \vert \left(\Phi_j(q) - \Phi_j \right) w_{t-j} \vert_F^a \right) \right]^{1/a}\right)^a\\
	& \leq (T-q)^{-1} \sum_{t=q+1}^T \sup_{t\in \mZ}\mE\left( \vert  w_t \vert_F^a \right) \left( \sum_{j=1}^q \vert \Phi_j(q) - \Phi_j \vert_F \right)^a\\
	& = \sup_{t\in \mZ}\mE\left( \vert  w_t \vert_F^a \right) \left( \sum_{j=1}^q \vert \Phi_j(q) - \Phi_j \vert_F \right)^a.
\end{align*}
From~\eqref{eq:Baxter} and Markov's inequality it follows that $C_{2T,a} = o_\mP(1)$. In total we thus have $C_{T,a} = o_\mP(1)$. Finally, we consider $D_{T,a}$. It holds that $(T-q)^{-1} \sum_{t=q+1}^T \tilde \ve_t(q) = D_{1T} + D_{2T} + D_{3T}$, where
\begin{align*}
	D_{1T} &\coloneqq  (T-q)^{-1} \sum_{t=q+1}^T \ve_t,\\
	D_{2T} &\coloneqq  (T-q)^{-1} \sum_{t=q+1}^T (\ve_t(q) - \ve_t),\\
	D_{3T} &\coloneqq  (T-q)^{-1} \sum_{t=q+1}^T (\tilde\ve_t(q) - \ve_t(q)).
\end{align*}
By Chebyshev's weak law of large numbers \cite[p.\,25]{Wh01}, $D_{1T} \overset{p}{\longrightarrow} \mE\left(\ve_t\right) = 0$, \ie, $D_{1T}=o_\mP(1)$. Moreover, $\vert D_{2T}\vert_F\leq B_{T,1} = o_\mP(1)$ and $\vert D_{3T} \vert_F \leq C_{T,1} = o_{\mP}(1)$. By the continuous mapping theorem we thus have $D_T=o_\mP(1)$. This completes the proof.~\hfill~$\square$\\

{\bf Proof of Lemma~\ref{lem:Sigmastar}.} It follows from Assumption~\ref{ass:cumulants} that 
\begin{align*}
	\vert(T-q)^{-1} \sum_{t=q+1}^T \ve_t\ve_t' - \Sigma\vert_F = o_\mP(1).	
\end{align*}
Therefore,
%
\begin{align*}
	\vert \mE^*\left(\ve_t^* \ve_t^{*\prime}\right) - \Sigma\vert_F \leq \vert \mE^*\left(\ve_t^* \ve_t^{*\prime}\right) - (T-q)^{-1} \sum_{t=q+1}^T \ve_t\ve_t'\vert_F
	+ o_\mP(1).
\end{align*}
Moreover,
\begin{align*}
	&\vert \mE^*\left(\ve_t^* \ve_t^{*\prime}\right) - (T-q)^{-1} \sum_{t=q+1}^T \ve_t\ve_t'\vert_F \\&= \vert (T-q)^{-1} \sum_{t=q+1}^T \left(\hat \ve_t(q) - \bar{\hat{\ve}}_T(q)\right)\left(\hat \ve_t(q) - \bar{\hat{\ve}}_T(q)\right)' - \ve_t\ve_t'\vert_F \\
	& = \vert(T-q)^{-1} \sum_{t=q+1}^T \left(\left[\left(\hat \ve_t(q) - \bar{\hat{\ve}}_T(q)\right) - \ve_t\right]\left(\hat \ve_t(q) - \bar{\hat{\ve}}_T(q)\right)'\right. \\
	&\hspace{4cm} \left.+ \ve_t \left[\left(\hat \ve_t(q) - \bar{\hat{\ve}}_T(q)\right)-\ve_t\right]'\right)\vert_F\\
	& \leq E_{1T} + E_{2T},
\end{align*}
where
\begin{align*}
	E_{1T} & = (T-q)^{-1} \sum_{t=q+1}^T \vert\hat \ve_t(q) - \bar{\hat{\ve}}_T(q) - \ve_t\vert_F\vert\hat \ve_t(q) - \bar{\hat{\ve}}_T(q)\vert_F,\\
	E_{2T} & = (T-q)^{-1} \sum_{t=q+1}^T  \vert\hat \ve_t(q) - \bar{\hat{\ve}}_T(q)-\ve_t\vert_F\vert\ve_t\vert_F.
\end{align*}
The Cauchy-Schwarz inequality yields
\begin{align*}
	E_{1T} &\leq \left((T-q)^{-1} \sum_{t=q+1}^T \vert\hat \ve_t(q) - \bar{\hat{\ve}}_T(q) - \ve_t\vert_F^2
	(T-q)^{-1} \sum_{t=q+1}^T \vert\hat \ve_t(q) - \bar{\hat{\ve}}_T(q)\vert_F^2\right)^{1/2},\\
	E_{2T} &\leq \left((T-q)^{-1} \sum_{t=q+1}^T \vert\hat \ve_t(q) - \bar{\hat{\ve}}_T(q)-\ve_t\vert_F^2 (T-q)^{-1} \sum_{t=q+1}^T \vert\ve_t\vert_F^2\right)^{1/2}.
\end{align*}
From the proof of Lemma~\ref{lem:moments} we have $(T-q)^{-1} \sum_{t=q+1}^T \vert\ve_t\vert_F^2 = A_{T,2} = O_\mP(1)$ and $(T-q)^{-1} \sum_{t=q+1}^T \vert\hat \ve_t(q) - \bar{\hat{\ve}}_T(q)\vert_F^2 = O_\mP(1)$. It thus remains to show that $(T-q)^{-1} \sum_{t=q+1}^T \vert\hat \ve_t(q) - \bar{\hat{\ve}}_T(q) - \ve_t\vert_F^2 = o_\mP(1)$. To this end note that
\begin{align*}
	&\vert\hat \ve_t(q) - \bar{\hat{\ve}}_T(q) - \ve_t\vert_F\\ 
	&\leq \vert \hat \ve_t(q) - \tilde \ve_t(q)\vert_F 
	+ \vert \tilde \ve_t(q) - (w_t - \sum_{j=1}^q \Phi_j(q)w_{t-j})\vert_F\\
	&\quad + \vert w_t - \sum_{j=1}^q \Phi_j(q)w_{t-j} - \ve_t \vert_F 
	+ \vert \bar{\hat{\ve}}_T(q) \vert_F\\
	& = \vert \hat \ve_t(q) - \tilde \ve_t(q)\vert_F 
	+ \vert \sum_{j=1}^q \left(\tilde\Phi_j(q) - \Phi_j(q)\right)w_{t-j}\vert_F\\
	&\quad + \vert \sum_{j=1}^q \left(\Phi_j(q) - \Phi_j\right)w_{t-j}\vert_F
	+ \vert \sum_{j=q+1}^\infty \Phi_j w_{t-j}\vert_F + \vert \bar{\hat{\ve}}_T(q)\vert_F.
\end{align*}
Hence,
\begin{align*}
	&\vert\hat \ve_t(q) - \bar{\hat{\ve}}_T(q) - \ve_t\vert_F^2\\
	&\leq 5\left( 
	\vert \hat \ve_t(q) - \tilde \ve_t(q)\vert_F^2
	+ \vert \sum_{j=1}^q \left(\tilde\Phi_j(q) - \Phi_j(q)\right)w_{t-j}\vert_F^2\right.\\
	&\hspace{1cm}\left. + \vert \sum_{j=1}^q \left(\Phi_j(q) - \Phi_j\right)w_{t-j}\vert_F^2
	+  \vert \sum_{j=q+1}^\infty \Phi_j w_{t-j}\vert_F^2
	+ \vert \bar{\hat{\ve}}_T(q) \vert_F^2
	\right).
\end{align*}
In the notation of the proof of Lemma~\ref{lem:moments}, we obtain
\begin{align*}
	(T-q)^{-1} \sum_{t=q+1}^T \vert\hat \ve_t(q) - \bar{\hat{\ve}}_T(q) - \ve_t\vert_F^2
	&\leq 5\left( F_{T,2} + C_{1T,2} + C_{2T,2} + B_{T,2} + \vert \bar{\hat{\ve}}_T(q) \vert_F^2 \right)\\
	& = 5\vert \bar{\hat{\ve}}_T(q) \vert_F^2 + o_\mP(1).
\end{align*}
From $\hat \ve_t(q) = \hat \ve_t(q) - \tilde \ve_t(q) + \tilde \ve_t(q) - \ve_t(q) + \ve_t(q) - \ve_t + \ve_t$, with $\ve_t(q)$ as defined in the proof of Lemma~\ref{lem:moments}, it follows that
\begin{align*}
	\vert \bar{\hat{\ve}}_T(q) \vert_F
	&\leq \vert (T-q)^{-1} \sum_{t=q+1}^T \hat \ve_t(q) - \tilde \ve_t(q) \vert_F
	+ \vert (T-q)^{-1} \sum_{t=q+1}^T \tilde \ve_t(q) - \ve_t(q) \vert_F\\
	&\quad+ \vert (T-q)^{-1} \sum_{t=q+1}^T \ve_t(q) - \ve_t \vert_F
	+ \vert (T-q)^{-1} \sum_{t=q+1}^T \ve_t \vert_F\\
	& \leq F_{T,1} + C_{T,1} + B_{T,1} + \vert D_{1T}\vert_F = o_\mP(1).
\end{align*}
This completes the proof.~\hfill~$\square$

\bibliographystyle{ifac}

\end{NoHyper}

\end{document}